# A Sub-Cell Based Indicator for Troubled Zones in RKDG Schemes and a Novel Class of Hybrid RKDG+HWENO Schemes

By

Dinshaw S. Balsara[1], Christoph Altmann[1,2], Claus-Dieter Munz[2], Michael Dumbser[2]

(dbalsara@nd.edu, iagaltmann@iag.uni-stuttgart.de, munz@iag.uni-stuttgart.de, michael.dumbser@iag.uni-stuttgart.de)

[1]Center for Astrophysics, Department of Physics, University of Notre Dame, Notre Dame, IN 46556, U.S.A.

[2]Institute for Aerodynamics & Gas Dynamics, University of Stuttgart, Stuttgart, 70569, Germany

## Abstract

Runge Kutta Discontinuous Galerkin (RKDG) schemes can provide highly accurate solutions for a large class of important scientific problems. Using them for problems with shocks and other discontinuities requires that one has a strategy for detecting the presence of these discontinuities. Strategies that are based on total variation diminishing (TVD) limiters can be problem-independent and scale-free but they can indiscriminately clip extrema, resulting in degraded accuracy. Those based on total variation bounded (TVB) limiters are neither problem-independent nor scale-free. In order to get past these limitations we realize that the solution in RKDG schemes can carry meaningful sub-structure within a zone that may not need to be limited. To make this sub-structure visible, we take a sub-cell approach to detecting zones with discontinuities, known as troubled zones. A monotonicity preserving (MP) strategy is applied to distinguish between meaningful sub-structure and shocks. The strategy does not indiscriminately clip extrema and is, nevertheless, scale-free and problem-independent. It, therefore, overcomes some of the limitations of previously-used strategies for detecting troubled zones. The moments of the troubled zones can then be corrected using a weighted essentially non-oscillatory (WENO) or Hermite WENO (HWENO) approach. In the course of doing this work it was also realized that the most significant variation in the solution is contained in the solution variables and their first moments. Thus the additional moments can be reconstructed using the variables and their first moments, resulting in a very substantial savings in computer memory. We call such schemes hybrid RKDG+HWENO schemes. It is shown that such schemes can attain the same formal accuracy as RKDG schemes, making them attractive, low-storage alternatives to RKDG schemes. Particular attention has been paid to the reconstruction of cross-terms in multidimensional problems and explicit, easy to implement formulae have been catalogued for third and fourth order of spatial accuracy. The utility of hybrid RKDG+WENO schemes has been illustrated with several stringent test problems in one and two dimensions. It is shown that their accuracy is usually competitive with the accuracy of RKDG schemes of the same order. Because of their compact stencils and low storage, hybrid RKDG+HWENO schemes could be very useful for large-scale parallel adaptive mesh refinement calculations.

# 1 Introduction

Discontinuous Galerkin (DG) methods were first introduced by Reed and Hill [24] for solving linear hyperbolic problems associated with neutron transfer. Cockburn and Shu [7], Cockburn, Lin and Shu [8] formulated the method for nonlinear hyperbolic problems and its application to the case of systems of conservation laws in one and multiple dimensions was carried out in Cockburn, Hou and Shu [9] and Cockburn and Shu [10]. The basic idea of Cockburn, *et al* was a simple yet elegant one. They used a DG discretization in space along with the total variation diminishing (TVD) preserving Runge-Kutta time discretization from Shu and Osher [26] to arrive at a methodology they called Runge-Kutta Discontinuous Galerkin (RKDG). The method was explicit and non-linearly stable. Exact or approximate Riemann solvers were used to obtain interface fluxes and a limiting strategy that was based on total variation bounded (TVB) limiters was used to achieve non-linear stability in the presence of strong shocks. A recent review of RKDG methods is available in Cockburn, Karniadakis and Shu [11].

The RKDG method has several desirable features. All the moments of the solution are explicitly carried, thus making a well-defined solution available at any point in the computational domain. It provides a simple strategy for obtaining the update equations for all the moments, making it easy to arrive at a computer implementation of the method. The RKDG method can be formulated on logically rectangular meshes or on triangular and tetrahedral meshes using a constructive strategy that is independent of the mesh used. It can be formulated for all orders, a fact that has been used to advantage by Biswas, Devine and Flaherty [5] to design hp-adaptive RKDG formulations for the Euler equations and by Hesthaven and Warburton [16] to achieve very high orders of accuracy for Maxwell's equations.

A potential advantage of RKDG schemes stems from the fact that the high order of accuracy of RKDG schemes can be achieved by using rather small stencils. The small stencils give RKDG methods an advantage over the essentially non-oscillatory (ENO) schemes of Shu and Osher [26], [27] and the weighted ENO (WENO) schemes of Liu,

Osher and Chan [20], Jiang and Shu [18], Balsara and Shu [2] and Hu and Shu [17]. Recently Qiu and Shu [21], [22] formulated Hermite WENO (HWENO) schemes which also rely on smoothness indicators to pick the best stencil, just like the WENO schemes. The WENO schemes all have stencil widths that increase with increasing order, which might be held to be a disadvantage of the WENO schemes. Because WENO schemes reconstruct all the moments of the solution, they, however, have the advantage of having a low storage requirement. RKDG schemes, on the other hand, have storage requirements that increase dramatically with increasing orders, especially in multiple dimensions. The RKDG and WENO methods have complementing strengths even when one assesses their performance on practical problems. RKDG formulations can be very robust even in the presence of problems with strong shocks. WENO schemes offer high formal accuracy and yet some of the higher order WENO schemes can lose robustness and even display spurious oscillations that need to be controlled, as was discussed in Balsara and Shu [2] and Shi, Hu and Shu [25]. It is, therefore, interesting to ask whether there might be schemes that draw on the strengths of both approaches? In the course of carrying out this work it was realized that an affirmative answer could be found for the above question. We call such schemes hybrid RKDG+HWENO schemes. The hybrid RKDG+HWENO schemes have the intrinsic robustness and smaller stencils of RKDG schemes and yet they draw on ideas from WENO schemes to reconstruct some of the higher moments, thus enhancing the order of accuracy of the underlying RKDG scheme. The schemes rely on evolving the lower moments with sufficiently high accuracy while reconstructing the higher moments. The utilitarian justification of such a division stems from the fact that the majority of the variation of the flow is contained in the first few moments. Since only the lower moments require fixed storage in computer memory, the new schemes might have some attractive features for large-scale, high-performance parallel computations. In this work we give explicit formulae for reconstructing the second and third moments using the zeroth and first moments. For one and two dimensional problems, these formulae are furnished in a format that is easy to implement in numerical codes. It is thus possible to formulate hybrid RKDG+HWENO schemes of third and fourth order accuracy.

Despite their many attractive features, the development of limiters for RKDG schemes has presented something of a challenge. Several approaches have been tried. Early attempts by Cockburn and Shu [10] consisted of using a TVB strategy to detect troubled zones. The TVB strategy suffers from the fact that it is neither scale-free nor problem-independent. Biswas, Devine and Flaherty [5] designed a limiter that adaptively limited the moments, going from the highest moment to the lowest. Burbeau, Sagaut and Bruneau [6] formulated a problem-independent limiter that limits not just the variable but also its successive moments, while making some allowance for the existence of local extrema. Krivodonova et al [19] presented a limiting strategy that is based on a strong superconvergence at the outflow boundaries for the Euler equations. Qiu and Shu [21] showed that the problem of non-linear limiting can be broken up into two conceptual parts. The first part consisted of using a TVB limiter to detect zones that could be classified as troubled zones because they had a discontinuity passing through them. The second part consisted of using WENO schemes to supply the moments in the troubled zones. In subsequent work, Qiu and Shu [22], [23] replaced the second part with a HWENO scheme for supplying the moments in the troubled zones. The formulation by Qiu and Shu furnishes a logical strategy for rebuilding the moments in zones that are already identified as troubled. It can even be used in the hybrid RKDG+HWENO schemes mentioned in the previous paragraph. However, their strategy for detecting troubled zones is still based on the TVB limiter, which suffers from the dual defects that it is neither scale-free nor problem-independent.

In order to ensure the end-to-end success of RKDG schemes and the previously mentioned hybrid RKDG+HWENO schemes we still need a scale-free, problem-independent strategy for detecting troubled zones. Such a strategy has to be built very carefully because it should be able to tell the difference between a genuine discontinuity in the solution and a smoothly varying extremum. Higher order RKDG schemes can carry a considerable amount of meaningful sub-structure within each zone. A good detector of troubled zones should be able to tell apart this meaningful sub-structure from a genuine discontinuity. If a bad detector is used, a great many zones will be flagged as troubled zones and the WENO scheme will be needlessly called upon to supply moments in those zones. Practical experimentation has shown that while this will not degrade the formal order of the RKDG scheme it will, nevertheless, degrade the intrinsic accuracy of the scheme. For very higher order RKDG schemes, this degradation of intrinsic accuracy can be by several orders of magnitude. We realize, therefore, that as the order of accuracy of the RKDG scheme is increased we have to be willing to carefully examine the solution for meaningful sub-scale structure and distinguish between smooth local extrema at the subscale level within each zone and gross discontinuities in the flow. A good detector of troubled zones should also have free parameters that allow us to decide how large and rapidly-varying a local extremum we are willing to tolerate before labeling it a spurious oscillation associated, for example, with an upstream shock. Experience has shown that such parameters are usually determined by the intrinsic properties of the hyperbolic system being considered. To give an example, because the Euler system has a convex flux and does not generate any compound shocks, it will permit such parameters to be set liberally. The magnetohydrodynamic (MHD) system, which has a non-convex flux and can generate compound shocks, will begin to show spurious oscillations if the same parameters are set too liberally. In their examination of very high order WENO schemes, Balsara and Shu [2] found that the monotonicity preserving (MP) limiter of Suresh and Huynh [29] provided such a scale-free, problem-independent strategy for stabilizing WENO schemes. The MP limiter uses a five-point stencil to detect local extrema, instead of a three-point stencil that is used by a TVD limiter. In this work we make two extensions: a) We reformulate the MP limiter, showing that there are some further opportunities for providing extra space at well-defined extrema. b) We apply the reformulated MP limiter to suitably chosen sub-cells of a cell that is undergoing update via an RKDG or hybrid RKDG+HWENO scheme. The above-mentioned advances enable us to efficiently detect troubled zones. The detector of troubled zones described in this paper is scale-free and problem-independent in the following sense: For a given hyperbolic system and a chosen order of the RKDG or hybrid RKDG+HWENO scheme we choose the free parameters in the MP limiter once and once only. Once that choice is made, we require that all problems with any shock strength can be treated without changing the parameters.

In this paper we focus on developing the present ideas very thoroughly for one and two-dimensional problems. Extensions to three dimensions will be the topic of a later paper. In Section 2 we briefly catalogue RKDG schemes and then introduce the hybrid RKDG+HWENO schemes. In Section 3 we present our strategy for flagging troubled zones. In section 4 we provide an extensive accuracy analysis. In Section 5 we present several test problems. Conclusions are presented in Section 5.

## 2 RKDG and hybrid RKDG+HWENO Schemes

We divide this section into three sub-sections. In the first sub-section we provide a very brief introduction to RKDG schemes with the intent of setting up a common notation. In the second sub-section we present the hybrid RKDG+HWENO schemes in one dimension. In the third sub-section we present two-dimensional extensions of hybrid RKDG+HWENO schemes.

## 2.a Introduction to RKDG Schemes

The RKDG schemes have been presented in detail in Cockburn and Shu [10]. Following Cockburn and Shu [10], we present just enough detail here to set up a common notation for the rest of this paper. Thus consider the system of conservation laws:

$$\frac{\partial \mathbf{u}}{\partial t} + \mathrm{div}\, \mathbf{f(u)} = 0 \qquad (2.1)$$

Any specific component of the vector of conserved variables $\mathbf{u}$ can be denoted by u (x,t). Eqn. (2.1) is discretized over the physical domain $\Omega$ using the discontinuous Galerkin method. The approximate solution to u (x,t) is sought in the finite element space of discontinuous functions

$$V = \{ v_h \in L^\infty(\Omega) : v_h|_K \in V(K), \forall K \in \mathcal{T}_h \} \qquad (2.2)$$

where $T_h$ is a triangulation of the domain Ω and V(K) is the local space. To take an example, consider subdividing the real line into intervals labeled by "i" so that each such zone has a local coordinate that goes from [-1/2, 1/2] . Then, for a fourth order accurate representation, u (x,t) can be written in a set of modal bases, $P_0$ (x), $P_1$ (x), $P_2$ (x) and $P_3$ (x) as:

$$u(x,t) = u_0(t)\,P_0(x) + u_1(t)\,P_1(x) + u_2(t)\,P_2(x) + u_3(t)\,P_3(x)$$
$$\text{where } P_0(x) = 1\ ;\ P_1(x) = x\ ;\ P_2(x) = x^2 - \frac{1}{12}\ ;\ P_3(x) = x^3 - x\frac{3}{20} \qquad (2.3)$$

The modal bases in eq. (2.3) are just the Legendre polynomials, suitably scaled to the local coordinates of the zone being considered. For logically rectangular meshes in multiple dimensions, one can use tensor products of the above bases. The bases used in eq. (2.3) are orthogonal but not orthonormal and result in the mass matrix:

$$M = \text{diag}\left[\,1,\ \frac{1}{12}\ ,\ \frac{1}{180}\ ,\ \frac{1}{2800}\ \right] \qquad (2.4)$$

The terms in eqn. (2.3) can also be viewed as the modes in a functional expansion. One can also take the view that they represent the zeroth through third moments of the solution u (x,t) within the zone. Since it will be useful to be able to calculate these moments, we briefly explain that in the Appendix. The equations for the evolution of $u_i(t)$ can be obtained by using a smooth test function v(x) to make a weak formulation of eqn. (2.1) over the domain K as follows:

$$\frac{d}{dt}\int_K u(x,t)\,v(x)\,dx + \sum_{e\in K}\int_e \mathbf{f}\,(u(x,t))\bullet n_{e,K}\,v(x)\,d\Gamma - \int_K \mathbf{f}\,(u(x,t))\bullet \text{grad}\,v(x)\,dx = 0 \qquad (2.5)$$

Here $n_{e,K}$ denotes the outward unit normal to the edge "e" of the domain K. The smooth test functions v(x) are usually drawn from the set of bases functions, eqn. (2.2). The integrals in eqn. (2.5) are replaced by discrete sums using quadrature rules of the appropriate accuracy. We take our example from eqn. (2.3) and explicitly write out the evolution equations for the zeroth through third moments as follows:

$$\frac{d\,u_0(t)}{d\,t} + \left[f\,(1/2) - f\,(-1/2)\right] = 0 \tag{2.6}$$

$$\frac{1}{12}\;\frac{d\,u_1(t)}{d\,t} + \left[P_1(1/2)\,f\,(1/2) - P_1(-1/2)\,f\,(-1/2)\right] - \left[\int_{-1/2}^{1/2} f\,(u\,(x,t))\,P_1^{/}(x)\,dx\right] = 0 \tag{2.7}$$

$$\frac{1}{180}\;\frac{d\,u_2(t)}{d\,t} + \left[P_2(1/2)\,f\,(1/2) - P_2(-1/2)\,f\,(-1/2)\right] - \left[\int_{-1/2}^{1/2} f\,(u\,(x,t))\,P_2^{/}(x)\,dx\right] = 0 \tag{2.8}$$

$$\frac{1}{2800}\;\frac{d\,u_3(t)}{d\,t} + \left[P_3(1/2)\,f\,(1/2) - P_3(-1/2)\,f\,(-1/2)\right] - \left[\int_{-1/2}^{1/2} f\,(u\,(x,t))\,P_3^{/}(x)\,dx\right] = 0 \tag{2.9}$$

The fluxes f(1/2) and f(–1/2) can be obtained by solving a Riemann problem at the zone boundaries. Notice that because of our choice of a one-dimensional formulation the surface integrals at the zone boundaries have been much simplified. Typically, simple flux functions such as the ones by Harten, Lax and van Leer flux or (local) Lax-Friedrichs flux are used. The integrals in eqns. (2.7) to (2.9) are evaluated using Gauss quadrature points of the appropriate order and such quadrature points and their weights are catalogued in texts by Stroud and Secrest [28] or Abromowitz and Stegun [1]. Notice that eqn. (2.6) is our familiar expression of a conservation law while the higher moments in eqns. (2.7) to (2.9) represent evolutionary equations for the moments and so do not have to be in a conservation law form. This completes our description of the RKDG scheme.

## 2.b Hybrid RKDG+HWENO Schemes

We now describe the hybrid RKDG+HWENO schemes. We preface our description by pointing out that if the application scientist has sufficient amount of computer memory to retain all the moments up to the desired level of accuracy then s/he should do so. In that case, RKDG schemes will serve the purpose optimally because none of the data associated with the higher moments will need to be destroyed at the end of a timestep and reconstructed at the start of the next timestep. However, most large applications almost never have a sufficient amount of computer memory. Furthermore, large parallel applications usually use a certain style of processing data where the computation is domain-decomposed into a large number of sub-domains that are processed using a substantially smaller number of processors. A prominent example of such an application could be large, parallel adaptive mesh refinement calculations where the solution is stored over many thousands of sub-domains and those sub-domains are processed using a few hundred to a thousand processors, see Balsara and Norton [3]. In such calculations, one wants to store a small amount of solution-specific data in each of the sub-domains. However, one can assign a substantially larger amount of local data on each processor and use it for processing each of the sub-domains. We build the hybrid RKDG+HWENO algorithm around the premise that we are willing to store the variables as well as their slopes for each of the sub-domains. However, for the third order hybrid RKDG+HWENO scheme we would like to reconstruct the second moment in eqn. (2.3) and use the third order quadratures in eqns. (2.6) to (2.8) to evolve the terms for an entire timestep, after which we store just the variables and their slopes back to the computer's main memory. Similarly, for the fourth order hybrid RKDG+HWENO scheme we would like to reconstruct the second and third moments in eqn. (2.3) and use the fourth order quadratures in eqns. (2.6) to (2.9) to evolve the terms for an entire timestep, after which we store just the variables and their slopes back to the computer's main memory. In the next three sub-sections we describe our reconstruction strategy for third and fourth order hybrid RKDG+HWENO schemes as well as our treatment of cross-terms for multi-dimensional problems.

## 2.b.i Reconstruction Strategy for Third Order Hybrid RKDG+HWENO Scheme

In this sub-section we reconstruct the second moment in the zone “i” where we take $u_i$ as representing the mean value in that zone and $v_i$ as representing the first moment (i.e. the slope) in the same zone. In terms of eqn. (2.3), $u_i$ and $v_i$ are the first two coefficients in the modal expansion. If the zone “i” has been identified as being a troubled zone then we assume that we have used the strategy given in Qiu and Shu [21] to obtain a corrected value for the first moment $v_i$ . We also base our scheme on using the values $u_{i-1}$ , $u_i$ , $v_i$ and $u_{i+1}$ to reconstruct the second moment in the zone “i”. Following the HWENO philosophy of Qiu and Shu [22], we wish to find two quadratic polynomials : The first quadratic, $p_0(x)$, covers the stencil formed by zones “i” and “i–1” and reproduces the zeroth and first moments, $u_i$ and $v_i$ , in zone “i” and the zeroth moment “$u_{i-1}$” in zone “i–1”. The second quadratic, $p_1(x)$, covers the stencil formed by zones “i” and “i+1” and reproduces the zeroth and first moments, $u_i$ and $v_i$ , in zone “i” and the zeroth moment “$u_{i+1}$” in zone “i+1”. We wish to obtain the solution at preferred Gauss quadrature points within the zone “i” so that we can use the value of the solution at those quadrature points to obtain the second moment, i.e. the third coefficient in the modal expansion that is given in eqn. (2.3). The Gauss quadrature points for third order accurate integration in the local coordinates of zone “i” that span [-0.5, 0.5] are given by $x_G = -\sqrt{3/20}$ , $x_G = 0$ and $x_G = \sqrt{3/20}$ . The corresponding weights can be found from Stroud and Secrest [28]. (Note that the standard Gauss formulae in Stroud and Secrest’s text use the domain [-1,1] so a normalization by a factor of 0.5 is needed for the weights.) Following the HWENO philosophy of Qiu and Shu [22], we choose the large stencil formed by zones “i–1”, “i” and “i+1” and find the cubic polynomial Q(x) which reproduces the zeroth and first moments, $u_i$ and $v_i$ , in zone “i” , the zeroth moment “$u_{i-1}$” in zone “i–1” and the zeroth moment “$u_{i+1}$” in zone “i+1”. The HWENO strategy consists of trying to find positive linear weights, $\gamma_k^G$ ( also known as optimal weights) for each Gauss quadrature point $x_G$ that satisfy the condition:

$$Q(x_G) = \sum_{k=0}^{1} \gamma_k^G \; p_k(x_G) \tag{2.10}$$

For the third order hybrid RKDG+WENO we come up with the extremely simple result that:

$$\gamma_k^G = \frac{1}{2} \text{ for all Gauss quadrature points "G" and both polynomials "k".} \tag{2.11}$$

Once the optimal weights are known, we combine the solution from the each of the two small stencils at each of the Gauss quadrature points using smoothness measures that keep track of how rapidly the solution is varying in that stencil. The smoothness measures for each stencil evaluate the quality of the data available on that stencil and are, therefore, used to determine the fraction of the solution that will be used from that stencil. Since the procedure favors the stencil with the smoothest solution, it introduces an upwind bias into the scheme. The construction of the smoothness measures follows Jiang and Shu [18]. To complete the present scheme we simply need to provide closed-form expressions for the smoothness measures and explicit evaluations of the polynomials $p_0(x)$ and $p_1(x)$ at each of the Gauss quadrature points $x_G$ . The smoothness measure for the stencil spanned by the polynomial $p_0(x)$ is given by:

$$\beta_0 = (\, 13\, (\, u_i - u_{i-1} - v_i)^2 \, + 3\, v_i^2 \,\,) \,/\, 3 \tag{2.12}$$

The smoothness measure for the stencil spanned by the polynomial $p_1(x)$ is given by:

$$\beta_1 = (\, 13\, (\, u_{i+1} - u_i - v_i)^2 \, + 3\, v_i^2 \,\,) \,/\, 3 \tag{2.13}$$

We only need to provide explicit evaluations of the polynomials $p_0(x)$ and $p_1(x)$ at two of the three Gauss quadrature points, since the third can be obtained by symmetry considerations. Thus at $x_G = 0$ the polynomials, with coefficients that are evaluated with 16 digit accuracy, are:

$$p_0(x_G) = 1.0833333333333333\, u_i - 0.08333333333333333\, u_{i-1} - 0.08333333333333333\, v_i$$
$$p_1(x_G) = 1.0833333333333333\, u_i - 0.08333333333333333\, u_{i+1} + 0.08333333333333333\, v_i$$

(2.14)

At $x_G = \sqrt{3/20}$ the polynomials, with coefficients that are evaluated with 16 digit accuracy, are:

$$p_0(x_G) = 0.9333333333333333\, u_i + 0.0666666666666666\, u_{i-1} + 0.4539650012874083\, v_i$$
$$p_1(x_G) = 0.9333333333333333\, u_i + 0.0666666666666666\, u_{i+1} + 0.3206316679540702\, v_i$$
(2.15)

The solution $u(x_G)$ at a Gauss quadrature point $x_G$ can be constructed by first building the non-linear weights, $\omega_k^G$, as follows:

$$\omega_k^G = \frac{\overline{\omega}_k^G}{\sum_k \overline{\omega}_k^G} \; ; \quad \overline{\omega}_k^G = \frac{\gamma_k^G}{(\beta_k + \varepsilon)^2} \tag{2.16}$$

We take $\varepsilon = 10^{-12}$ in eqn. (2.16). The solution $u(x_G)$ is then given by:

$$u(x_G) = \sum_k \omega_k^G \, p_k(x_G) \tag{2.17}$$

The solution at all the Gauss quadrature points can then be used in conjunction with the modal basis to reconstruct the second moment, as shown in the Appendix. This completes our description of the third order hybrid RKDG+HWENO scheme.

We make several observations below:

1) It is known that for schemes that are better than second order the best flagging strategies for detecting troubled zones rely on characteristic variables. The hybrid

RKDG+HWENO schemes are well-matched with such a flagging strategy because for a small addition to the cost of the flagging strategy one can also obtain an additional one or two orders of accuracy. Moreover, this advantage is obtained without any significant increase in storage.

2) For multiple dimensions we can still obtain the quadratic terms using this strategy. We will present further ideas for the reconstruction of cross-terms in Sub-section 2.c.i.

3) While the method is described here as if it is being applied to a scalar problem, it extends naturally to systems as long as characteristic variables are used. We have found it worthwhile to use characteristic variables in all the tests reported here.

4) Qiu and Shu [22] suggested that one can correct the first moment in the central zone "i" using a HWENO scheme that uses the uncorrected first moments in zones "i+1" and "i−1". We favor the WENO correction strategies for obtaining the slopes from Qiu and Shu [21]. In this work we always use the corrected first moment $v_i$ to reconstruct the higher moments. Once such a corrected first moment is available we can use the approach in this section to reconstruct the second moment.

5) When using Runge-Kutta timestepping there are two ways in which this reconstruction can be used on a sub-domain. In the first approach, we reconstruct the second moment in each fractional timestep. In the second, we reconstruct the second moment at the start of a timestep on a sub-domain and evolve it for the current timestep. The latter requires a larger layer of ghost boundaries than the former. Experimentation has shown that both strategies produce comparable results, hence we reconstruct the second moment in each fractional timestep for all the test problems shown here.

6) The ADER approaches that were first developed by Titarev and Toro [31] and applied to DG schemes by Dumbser and Munz [12], [13] would prove especially useful here because they would ameliorate the problems associated with Runge-Kutta timestepping that are mentioned in the previous point. Such methods have also been extended to magnetohydrodynamics by Taube et al [14].

7) When the slopes are available, the present method has a small advantage over the WENO-based limiters in Qiu and Shu [21] because it generates positive weights for the Gauss quadrature points instead of the Gauss-Lobatto points. This advantage also extends to the fourth order hybrid RKDG+HWENO scheme described in the next sub-section.

8) Notice that in a well-resolved calculation most zones will have smooth flow. Thus the first moment will not go through any limiting and the scheme described above will be a third order scheme with a compact stencil that extends over just the immediate neighbors of the zone being considered.

9) The present set of ideas has some parallels with the work of Takewaki, Nishiguchi and Yabe [30] though the underlying RKDG scheme is different as is the reconstruction strategy for higher moments. The present schemes are also upwind and conservative.

10) The ideas described here are also potentially very useful for unstructured meshes.

11) Because the second moment is reconstructed, it is possible that the present scheme might permit a higher Courant number than the limiting Courant numbers cataloged in Cockburn, Karniadakis and Shu [11] for the corresponding RKDG scheme of the same order. In this work though, we use Courant numbers for the third order hybrid RKDG+HWENO scheme that are the same as those used for p=2 RKDG.

12) For notational consistency we will also refer to the third order accurate hybrid RKDG+HWENO scheme as the p=2 RKDG+HWENO scheme.

## 2.b.ii Reconstruction Strategy for Fourth Order Hybrid RKDG+HWENO Scheme

In this sub-section we reconstruct the second and third moments in the zone "i" where we take $u_i$ as representing the mean value in that zone and $v_i$ as representing the first moment in the same zone. If the zone "i" has been identified as being a troubled zone then we assume that we have used the strategy given in Qiu and Shu [21] to obtain a corrected value for the first moment $v_i$ . We also base our scheme on using the values $u_{i-2}$ , $u_{i-1}$ , $u_i$ , $v_i$ , $u_{i+1}$ and $u_{i+2}$ to reconstruct the second and third moments in the zone "i". Again following the HWENO philosophy of Qiu and Shu [22], we wish to find three cubic polynomials : The first cubic, $p_0(x)$, covers the stencil formed by zones "i", "i–1" and "i–2" and reproduces the zeroth and first moments, $u_i$ and $v_i$ , in zone "i" , the zeroth moment "$u_{i-1}$" in zone "i–1" and the zeroth moment "$u_{i-2}$" in zone "i–2". The second cubic, $p_1(x)$, covers the stencil formed by zones "i–1", "i" and "i+1" and reproduces the zeroth and first moments, $u_i$ and $v_i$ , in zone "i" , the zeroth moment "$u_{i-1}$" in zone "i–1" and the zeroth moment "$u_{i+1}$" in zone "i+1". The third cubic, $p_2(x)$, covers the stencil formed by zones "i", "i+1" and "i+2" and reproduces the zeroth and first moments, $u_i$ and $v_i$ , in zone "i" , the zeroth moment "$u_{i+1}$" in zone "i+1" and the zeroth moment "$u_{i+2}$" in zone "i+2". We wish to obtain the solution at preferred Gauss quadrature points within the zone "i" so that we can use the value of the solution at those quadrature points to obtain the second and third moments, i.e. the third and fourth coefficients in the modal expansion that is given in eqn. (2.3). The Gauss quadrature points for fourth order accurate integration in the local coordinates of the zone "i" that span [-0.5, 0.5] are given by $x_G = -0.43056815579702623$ , $x_G = -0.1699905217924281$ , $x_G = 0.1699905217924281$ and $x_G = 0.43056815579702623$ . The corresponding weights can be found from Stroud and Secrest [28] and can be used after normalizing by a factor of 0.5. Following the HWENO philosophy of Qiu and Shu [22], we choose the large stencil formed by zones "i–2", "i–1", "i", "i+1" and "i+2" and find the quintic polynomial Q(x) which reproduces the zeroth and first moments, $u_i$ and $v_i$ , in zone "i" , the zeroth moment "$u_{i-2}$" in zone "i–2" , the zeroth moment "$u_{i-1}$" in zone "i–1", the zeroth moment "$u_{i+1}$" in zone "i+1" and the zeroth moment "$u_{i+2}$" in zone "i+2". The HWENO strategy consists of trying to find the optimal weights, $\gamma_k^G$ for each Gauss quadrature point $x_G$ that satisfy the condition in eqn. (2.10). For the third order hybrid RKDG+WENO we find the optimal weights for $x_G = 0.1699905217924281$ to be:

$$\gamma_0^G = 0.1086206894023039 \; ; \; \gamma_1^G = 0.5497704011732130 \; ; \\ \gamma_2^G = 0.3416089094244832 \tag{2.18}$$

The corresponding optimal weights for $x_G = 0.43056815579702623$ are:

$$\gamma_0^G = 0.1493553825751190 \; ; \; \gamma_1^G = 0.6186539367058337 \; ; \\ \gamma_2^G = 0.2319906807190475 \tag{2.19}$$

The optimal weights at the two other Gauss quadrature points can be obtained by symmetry. The smoothness measure for the stencil spanned by the polynomial $p_0(x)$ is given by:

$$\begin{aligned}\beta_0 = (&170995\,u_i^2 + 274900\,u_{i-1}^2 - 117196\,u_{i-1}\,u_{i-2} \\ &+ 13291\,u_{i-2}^2 + u_i\,(-432604\,u_{i-1} + 90614\,u_{i-2} - 252060\,v_i) \\ &+ 316320\,u_{i-1}\,v_i - 64260\,u_{i-2}\,v_i + 98460\,v_i^2)\,/\,4332\end{aligned} \tag{2.20}$$

The smoothness measure for the stencil spanned by the polynomial $p_1(x)$ is given by:

$$\begin{aligned}\beta_1 = (&6292\,u_i^2 + 13291\,u_{i-1}^2 - 20290\,u_{i-1}\,u_{i+1} + 13291\,u_{i+1}^2 \\ &- 6292\,u_i\,(u_{i-1} + u_{i+1}) + 46740\,u_{i-1}\,v_i - 46740\,u_{i+1}\,v_i \\ &+ 48060\,v_i^2)\,/\,1452\end{aligned} \tag{2.21}$$

The smoothness measure for the stencil spanned by the polynomial $p_2(x)$ is given by:

$$\begin{aligned}\beta_2 = (&170995\,u_i^2 + 274900\,u_{i+1}^2 + 13291\,u_{i+2}^2 + 64260\,u_{i+2}\,v_i \\ &+ 98460\,v_i^2 - 4\,u_{i+1}\,(29299\,u_{i+2} + 79080\,v_i) \\ &+ u_i\,(-432604\,u_{i+1} + 90614\,u_{i+2} + 252060\,v_i))\,/\,4332\end{aligned} \tag{2.22}$$

At $x_G = 0.1699905217924281$ the polynomials, with coefficients that are evaluated with 16 digit accuracy, are:

$$\begin{aligned}p_0(x_G) &= 1.117962827992853\,u_i - 0.1391382520457803\,u_{i-1} + 0.02117542405292723\,u_{i-2} \\ &\quad + 0.07320311785250228\,v_i \\ p_1(x_G) &= 1.054436555834071\,4\,u_i - 0.01786082337901523\,4\,u_{i-1} - 0.03657573245505612\,u_{i+1} \\ &\quad + 0.188705430868469\,v_i \\ p_2(x_G) &= 1.085457985913413\,7\,u_i - 0.09579846260652769\,u_{i+1} + 0.01034047669311408\,1\,u_{i+2} \\ &\quad + 0.2451080310127276\,v_i\end{aligned} \tag{2.23}$$

At $x_G = 0.43056815579702623$ the polynomials, with coefficients that are evaluated with 16 digit accuracy, are:

$$\begin{aligned}
p_0(x_G) &= 0.7972876757209407\ u_i + 0.23626456455437267\ u_{i-1} - 0.03355224027531344\ u_{i-2} \\
&\quad + 0.599728239800772\ v_i \\
p_1(x_G) &= 0.897944396546881\ u_i + 0.044101733886668416\ u_{i-1} + 0.057953869566450496\ u_{i+1} \\
&\quad + 0.4167160201172442\ v_i \\
p_2(x_G) &= 0.8213466482174043\ u_i + 0.2041859345590879\ u_{i+1} - 0.02553258277649225\ u_{i+2} \\
&\quad + 0.27744738679092285\ v_i
\end{aligned} \tag{2.24}$$

The polynomials at the two other Gauss quadrature points can be obtained by symmetry. The solution can be obtained at the Gauss quadrature points by using eqns. (2.16) and (2.17). The second and third moments can be reconstructed using the Appendix. This completes our description of the fourth order hybrid RKDG+HWENO scheme.

We make a few observations below:

1) Notice that there is a difference in the way eqns. (2.6) and (2.7) are used in the third and fourth order RKDG+WENO schemes. For the third order scheme we use third order accurate spatial quadrature and third order accurate Runge-Kutta timestepping. For the fourth order scheme the spatial quadrature and timestepping have to be upgraded to fourth order accuracy.

2) It is tempting to try and use both the variable and slope in each of the zones “i–1”, “i” and “i+1” to obtain a RKDG+HWENO scheme that has an even more compact stencil. Such an effort results in negative values for the optimal weights and is, therefore, unproductive. Following the idea of Friedrichs [15], we even tried assigning equal linear weights of 0.5 to each of the stencils and used the smoothness measures to combine the two stencils non-linearly. The resulting scheme was still found to be unstable. Thus there seems to be a limit to the level of compactness that can be achieved.

3) For notational consistency we will also refer to the fourth order accurate hybrid RKDG+HWENO scheme as p=3 RKDG+HWENO scheme.

## 2.c)Multidimensional Reconstruction Strategy for Higher Order Hybrid RKDG+HWENO Schemes

The above Sub-sections have provided considerable detail on the reconstruction strategy in one dimension. However, it is interesting to develop the same scheme for multiple dimensions. In this section we focus on two dimensions, but the same methods translate well to three dimensions. We will denote the variable in zone (i,j) as $u_{i,j}$ and its x and y-directional slopes as $v_{i,j}$ and $w_{i,j}$ respectively. A general third order accurate representation in two dimensions is given by:

$$\begin{aligned} u(x,y,t) = {} & u_{i,j}(t)\,P_0(x) + v_{i,j}(t)\,P_1(x) + w_{i,j}(t)\,P_1(y) \\ & + u_{xx,i,j}(t)\,P_2(x) + u_{yy,i,j}(t)\,P_2(y) + u_{xy,i,j}(t)\,P_1(x)\,P_1(y) \end{aligned} \tag{2.25}$$

Retaining just the first line in the above test function yields a second order scheme while inclusion of the second line in (2.25) yields a third order scheme. The multidimensional reconstruction problem consists of finding a good representation for $u_{xx,i,j}(t)$, $u_{yy,i,j}(t)$ and $u_{xy,i,j}(t)$. In practice, $u_{xx,i,j}(t)$ and $u_{yy,i,j}(t)$ can be found using a dimensionally swept strategy in each direction, as shown in the previous sub-section, so the problem reduces to finding a good representation for the cross-term $u_{xy,i,j}(t)$. The strategy for obtaining $u_{xy,i,j}(t)$ from various compact stencils is presented in sub-section 2.c.i as is the strategy for limiting the cross-term.

A fourth order accurate representation in two dimensions is given by:

$$\begin{aligned} u(x,y,t) = {} & u_{i,j}(t)\,P_0(x) + v_{i,j}(t)\,P_1(x) + w_{i,j}(t)\,P_1(y) \\ & + u_{xx,i,j}(t)\,P_2(x) + u_{yy,i,j}(t)\,P_2(y) + u_{xy,i,j}(t)\,P_1(x)\,P_1(y) \\ & + u_{xxx,i,j}(t)\,P_3(x) + u_{yyy,i,j}(t)\,P_3(y) + u_{xxy,i,j}(t)\,P_2(x)\,P_1(y) + u_{xyy,i,j}(t)\,P_1(x)\,P_2(y) \end{aligned} \tag{2.26}$$

As before, $u_{xx,i,j}(t)$, $u_{yy,i,j}(t)$, $u_{xxx,i,j}(t)$ and $u_{yyy,i,j}(t)$ can be found using a dimensionally swept strategy in each direction, as shown in the previous sub-section, so the problem reduces to finding a good representation for the cross-terms $u_{xy,i,j}(t)$, $u_{xxy,i,j}(t)$ and $u_{xyy,i,j}(t)$. The strategy for obtaining $u_{xy,i,j}(t)$, $u_{xxy,i,j}(t)$ and $u_{xyy,i,j}(t)$ from various compact stencils is presented in sub-section 2.c.ii as is the strategy for limiting the cross-terms. Schemes of fifth and higher order can be constructed by repeating the same procedure with a larger basis set.

**2.c.i) Multidimensional Reconstruction Strategy for the Cross-Terms for Third Order Hybrid RKDG+HWENO Scheme**

Fig. 1 shows the nine zones around zone (i,j) that are used to carry out spatially third order accurate reconstruction. The maximally compact eight stencils that yield a third order accurate interpolant are also shown. The stencils are labeled $S_1$ through $S_8$ . The reconstruction in stencil $S_1$ is based on requiring the polynomial in eqn. (2.25) to satisfy the following constraints:

$$\iint_{[-3/2,-1/2]\times[-1/2,1/2]} u(x,y,t)\,dx\,dy = u_{i-1,j}(t) \quad ; \quad \iint_{[-1/2,1/2]\times[-3/2,-1/2]} u(x,y,t)\,dx\,dy = u_{i,j-1}(t) \; ;$$
$$\iint_{[-3/2,-1/2]\times[-3/2,-1/2]} u(x,y,t)\,dx\,dy = u_{i-1,j-1}(t) \qquad (2.27)$$

As a result, the coefficients $u_{xx,i,j}(t)$ , $u_{yy,i,j}(t)$ and $u_{xy,i,j}(t)$ for stencil $S_1$ are given by:

$$\begin{aligned} u_{xx,i,j}(t) &= u_{i-1,j}(t) - u_{i,j}(t) + v_{i,j}(t) \\ u_{yy,i,j}(t) &= u_{i,j-1}(t) - u_{i,j}(t) + w_{i,j}(t) \\ u_{xy,i,j}(t) &= u_{i-1,j-1}(t) - u_{i,j-1}(t) - u_{i-1,j}(t) + u_{i,j}(t) \end{aligned} \qquad (2.28)$$

The remaining stencils are given below and the time level "(t)" that is explicited in eqn. (2.28) is dropped. Thus for $S_2$ we have:

$$\begin{aligned} u_{xx,i,j} &= u_{i+1,j} - u_{i,j} - v_{i,j} \\ u_{yy,i,j} &= u_{i,j-1} - u_{i,j} + w_{i,j} \\ u_{xy,i,j} &= u_{i,j-1} - u_{i+1,j-1} - u_{i,j} + u_{i+1,j} \end{aligned} \qquad (2.29)$$

$S_3$ gives:

$$\begin{aligned} u_{xx,i,j} &= u_{i-1,j} - u_{i,j} + v_{i,j} \\ u_{yy,i,j} &= u_{i,j+1} - u_{i,j} - w_{i,j} \\ u_{xy,i,j} &= u_{i-1,j} - u_{i,j} - u_{i-1,j+1} + u_{i,j+1} \end{aligned} \tag{2.30}$$

$S_4$ gives:

$$\begin{aligned} u_{xx,i,j} &= u_{i+1,j} - u_{i,j} - v_{i,j} \\ u_{yy,i,j} &= u_{i,j+1} - u_{i,j} - w_{i,j} \\ u_{xy,i,j} &= u_{i,j} - u_{i+1,j} - u_{i,j+1} + u_{i+1,j+1} \end{aligned} \tag{2.31}$$

$S_5$ gives:

$$\begin{aligned} u_{xx,i,j} &= \frac{1}{2}\left(u_{i-1,j-1} - 2\,u_{i,j-1} + u_{i+1,j-1}\right) \\ u_{yy,i,j} &= u_{i,j-1} - u_{i,j} + w_{i,j} \\ u_{xy,i,j} &= \frac{1}{2}\left(u_{i-1,j-1} - u_{i+1,j-1} + 2\,v_{i,j}\right) \end{aligned} \tag{2.32}$$

$S_6$ gives:

$$\begin{aligned} u_{xx,i,j} &= u_{i+1,j} - u_{i,j} - v_{i,j} \\ u_{yy,i,j} &= \frac{1}{2}\left(u_{i+1,j-1} - 2\,u_{i+1,j} + u_{i+1,j+1}\right) \\ u_{xy,i,j} &= \frac{1}{2}\left(u_{i+1,j-1} - u_{i+1,j+1} + 2\,w_{i,j}\right) \end{aligned} \tag{2.33}$$

$S_7$ gives:

$$
\begin{aligned}
u_{xx,i,j} &= \frac{1}{2}\left(u_{i-1,j+1} - 2\,u_{i,j+1} + u_{i+1,j+1}\right) \\
u_{yy,i,j} &= u_{i,j+1} - u_{i,j} - w_{i,j} \\
u_{xy,i,j} &= \frac{1}{2}\left(u_{i-1,j+1} - u_{i+1,j+1} + 2\,v_{i,j}\right)
\end{aligned}
\qquad (2.34)
$$

$S_8$ gives:

$$
\begin{aligned}
u_{xx,i,j} &= u_{i-1,j} - u_{i,j} + v_{i,j} \\
u_{yy,i,j} &= \frac{1}{2}\left(u_{i-1,j-1} - 2\,u_{i-1,j} + u_{i-1,j+1}\right) \\
u_{xy,i,j} &= \frac{1}{2}\left(u_{i-1,j-1} - u_{i-1,j+1} + 2\,w_{i,j}\right)
\end{aligned}
\qquad (2.35)
$$

The $S_1$ to $S_4$ stencils catalogued above can be viewed as stencils that permit us to select upwind solutions that are diagonally aligned, while the $S_5$ to $S_8$ stencils permit us to select solutions that are upwinded in mesh-aligned directions. Unlike WENO, the goal in reconstructing higher order terms is not to improve the accuracy of the solution by one or more orders. As a result, there is no need to seek out optimal weights for all the stencils. We, therefore, assign equal weights to all the stencils and construct smoothness measures for each stencil by using :

$$
\beta = \iint_{[-1/2,1/2]\times[-1/2,1/2]} \left[ \left( \frac{\partial^2\, u\,(x,y,t)}{\partial x^2} \right)^2 + \left( \frac{\partial^2\, u\,(x,y,t)}{\partial y^2} \right)^2 + \left( \frac{\partial^2\, u\,(x,y,t)}{\partial x\, \partial y} \right)^2 \right] dx\, dy \qquad (2.36)
$$

The smoothness measures are easily constructed using the coefficients in eqs. (2.28) to (2.35). It is important to realize that the $u_{xx,i,j}$ and $u_{yy,i,j}$ terms are more easily reconstructed using the expressions in sub-section 2.b.i. In fact, the expressions in Sub-section 2.b.i are very suitable for reconstruction in the characteristic variables. Since characteristics-based reconstruction is desired in the design of higher order schemes, we construct the $u_{xx,i,j}$ and $u_{yy,i,j}$ terms by limiting in the characteristic variables and then projecting back into the space of conserved variables. For that reason, even though the reconstruction presented here can yield expressions for $u_{xx,i,j}$ and $u_{yy,i,j}$ terms, we only need the reconstruction for the construction of the cross-term, $u_{xy,i,j}$ . It is, therefore, interesting to ask whether the construction of the computationally costly smoothness measures described in eqn. (2.36) can be dispensed with entirely? We have indeed found that by sequentially applying a min-mod limiter to all the $u_{xy,i,j}$ cross-terms for each of the eight stencils $S_1$ to $S_8$ , the evaluation of the smoothness measures can be avoided entirely. Numerical experimentation has shown that this causes only a small degradation in the quality of the solution at a much-reduced cost. This completes our description of the reconstruction of cross-terms for the third order hybrid RKDG+HWENO scheme.

We make a few observations below:

1) The idea of limiting in each of the modes has already been explored in Burbeau, Sagaut and Bruneau [6]. The style and sequence in which the limiting is applied is, however, different here.

2) In this work we have chosen to limit the $u_{xy,i,j}$ cross-term in the space of conserved variables. This is the least expensive strategy and seems to work well for the problems explored here.

3) It is also possible to formulate the limiter in characteristic variables. The large number of projections into the space of characteristic variables may make the procsess computationally costly. It is, however, conceivable that for small systems the computational complexity is not prohibitive. When making such a projection it may be interesting to use diagonally oriented eigenvectors for stencils $S_1$ to $S_4$ and mesh-aligned eigenvectors for stencils $S_5$ to $S_8$ .

4) The stencils $S_1$ to $S_8$ are unique only in that they are the most compact choices. It is certainly possible to choose other sets of stencils. Just as $S_5$ to $S_8$ represent upwinded stencils along the cardinal directions of the mesh and $S_1$ to $S_4$ represent upwinded stencils along the primary inter-cardinal directions of the mesh, it is certainly possible to choose stencils that represent the secondary inter-cardinal directions of the mesh.

5) The strategy described here goes over unchanged in three dimensions. In three dimensions, one only needs the cross terms in xy, yz, zx which can all be obtained by cyclic rotation of the indices in the formulae given here.

**2.c.ii) Multidimensional Reconstruction Strategy for the Cross-Terms for Fourth Order Hybrid RKDG+HWENO Scheme**

Fig. 2 shows the twenty-one zones around zone (i,j) that are used to carry out spatially fourth order accurate reconstruction. The maximally compact eight stencils that yield a fourth order accurate interpolant are also shown. The stencils are labeled $S_1$ through $S_8$ . The reconstruction in stencil $S_1$ is based on requiring the polynomial in eqn. (2.26) to satisfy the following constraints:

$$\iint_{[-3/2,-1/2]\times[-1/2,1/2]} u\,(x, y, t)\,dx\,dy = u_{i-1,j}(t) \quad ; \quad \iint_{[-1/2,1/2]\times[-3/2,-1/2]} u\,(x, y, t)\,dx\,dy = u_{i,j-1}(t)\ ;$$
$$\iint_{[-3/2,-1/2]\times[-3/2,-1/2]} u\,(x, y, t)\,dx\,dy = u_{i-1,j-1}(t) \quad ; \quad \iint_{[-5/2,-3/2]\times[-1/2,1/2]} u\,(x, y, t)\,dx\,dy = u_{i-2,j}(t) \quad ;$$
$$\iint_{[-5/2,-3/2]\times[-3/2,-1/2]} u\,(x, y, t)\,dx\,dy = u_{i-2,j-1}(t) \quad ; \quad \iint_{[-1/2,1/2]\times[-5/2,-3/2]} u\,(x, y, t)\,dx\,dy = u_{i,j-2}(t)\ ;$$
$$\iint_{[-3/2,-1/2]\times[-5/2,-3/2]} u\,(x, y, t)\,dx\,dy = u_{i-1,j-2}(t) \qquad (2.37)$$

As before, the time level "(t)" that is explicited in eqns. (2.26) and (2.37) is dropped. As a result, the coefficients $u_{xx,i,j}$ , $u_{yy,i,j}$, $u_{xy,i,j}$, $u_{xxx,i,j}$, $u_{yyy,i,j}$, $u_{xxy,i,j}$ and $u_{xyy,i,j}$ for stencil $S_1$ are given by:

$$u_{xx,i,j} = \frac{1}{38}\left(-11\,u_{i-2,j} + 82\,u_{i-1,j} - 71\,u_{i,j} + 60\,v_{i,j}\right)$$
$$u_{yy,i,j} = \frac{1}{38}\left(-71\,u_{i,j} - 11\,u_{i,j-2} + 82\,u_{i,j-1} + 60\,w_{i,j}\right)$$
$$u_{xy,i,j} = \frac{1}{2}\left(u_{i-2,j} - 5\,u_{i-1,j} + 4\,u_{i,j} - u_{i-1,j-2} + u_{i,j-2} - u_{i-2,j-1} + 6\,u_{i-1,j-1} - 5\,u_{i,j-1}\right)$$
$$u_{xxx,i,j} = -\frac{5}{19}\left(u_{i-2,j} - 4\,u_{i-1,j} + 3\,u_{i,j} - 2\,v_{i,j}\right)$$
$$u_{yyy,i,j} = -\frac{5}{19}\left(3\,u_{i,j} + u_{i,j-2} - 2\left(2\,u_{i,j-1} + w_{i,j}\right)\right)$$
$$u_{xxy,i,j} = \frac{1}{2}\left(u_{i-2,j} - 2\,u_{i-1,j} + u_{i,j} - u_{i-2,j-1} + 2\,u_{i-1,j-1} - u_{i,j-1}\right)$$
$$u_{xyy,i,j} = \frac{1}{2}\left(-u_{i-1,j} + u_{i,j} - u_{i-1,j-2} + u_{i,j-2} + 2\,u_{i-1,j-1} - 2\,u_{i,j-1}\right)$$

(2.38)

$S_2$ gives:

$$u_{xx,i,j} = \frac{1}{38}\left(-71\,u_{i,j} + 82\,u_{i+1,j} - 11\,u_{i+2,j} - 60\,v_{i,j}\right)$$
$$u_{yy,i,j} = \frac{1}{38}\left(-71\,u_{i,j} - 11\,u_{i,j-2} + 82\,u_{i,j-1} + 60\,w_{i,j}\right)$$
$$u_{xy,i,j} = \frac{1}{2}\left(u_{i+2,j-1} - 4\,u_{i,j} + 5\,u_{i+1,j} - u_{i+2,j} - u_{i,j-2} + u_{i+1,j-2} + 5\,u_{i,j-1} - 6\,u_{i+1,j-1}\right)$$
$$u_{xxx,i,j} = \frac{5}{19}\left(3\,u_{i,j} - 4\,u_{i+1,j} + u_{i+2,j} + 2\,v_{i,j}\right)$$
$$u_{yyy,i,j} = -\frac{5}{19}\left(3\,u_{i,j} + u_{i,j-2} - 2\left(2\,u_{i,j-1} + w_{i,j}\right)\right)$$
$$u_{xxy,i,j} = \frac{1}{2}\left(-u_{i+2,j-1} + u_{i,j} - 2\,u_{i+1,j} + u_{i+2,j} - u_{i,j-1} + 2\,u_{i+1,j-1}\right)$$
$$u_{xyy,i,j} = \frac{1}{2}\left(-u_{i,j} + u_{i+1,j} - u_{i,j-2} + u_{i+1,j-2} + 2\,u_{i,j-1} - 2\,u_{i+1,j-1}\right)$$

(2.39)

$S_3$ gives:

$$u_{xx,i,j} = \frac{1}{38}\left(-11\,u_{i-2,j} + 82\,u_{i-1,j} - 71\,u_{i,j} + 60\,v_{i,j}\right)$$
$$u_{yy,i,j} = \frac{1}{38}\left(-71\,u_{i,j} + 82\,u_{i,j+1} - 11\,u_{i,j+2} - 60\,w_{i,j}\right)$$
$$u_{xy,i,j} = \frac{1}{2}\left(-u_{i-2,j} + 5\,u_{i-1,j} - 4\,u_{i,j} + u_{i-2,j+1} - 6\,u_{i-1,j+1} + 5\,u_{i,j+1} + u_{i-1,j+2} - u_{i,j+2}\right)$$
$$u_{xxx,i,j} = -\frac{5}{19}\left(u_{i-2,j} - 4\,u_{i-1,j} + 3\,u_{i,j} - 2\,v_{i,j}\right)$$
$$u_{yyy,i,j} = \frac{5}{19}\left(3\,u_{i,j} - 4\,u_{i,j+1} + u_{i,j+2} + 2\,w_{i,j}\right)$$
$$u_{xxy,i,j} = \frac{1}{2}\left(-u_{i-2,j} + 2\,u_{i-1,j} - u_{i,j} + u_{i-2,j+1} - 2\,u_{i-1,j+1} + u_{i,j+1}\right)$$
$$u_{xyy,i,j} = \frac{1}{2}\left(-u_{i-1,j} + u_{i,j} + 2\,u_{i-1,j+1} - 2\,u_{i,j+1} - u_{i-1,j+2} + u_{i,j+2}\right)$$

(2.40)

$S_4$ gives:

$$u_{xx,i,j} = \frac{1}{38}\left(-71\,u_{i,j} + 82\,u_{i+1,j} - 11\,u_{i+2,j} - 60\,v_{i,j}\right)$$
$$u_{yy,i,j} = \frac{1}{38}\left(-71\,u_{i,j} + 82\,u_{i,j+1} - 11\,u_{i,j+2} - 60\,w_{i,j}\right)$$
$$u_{xy,i,j} = \frac{1}{2}\left(4\,u_{i,j} - 5\,u_{i+1,j} + u_{i+2,j} - 5\,u_{i,j+1} + 6\,u_{i+1,j+1} - u_{i+2,j+1} + u_{i,j+2} - u_{i+1,j+2}\right)$$
$$u_{xxx,i,j} = \frac{5}{19}\left(3\,u_{i,j} - 4\,u_{i+1,j} + u_{i+2,j} + 2\,v_{i,j}\right)$$
$$u_{yyy,i,j} = \frac{5}{19}\left(3\,u_{i,j} - 4\,u_{i,j+1} + u_{i,j+2} + 2\,w_{i,j}\right)$$
$$u_{xxy,i,j} = \frac{1}{2}\left(-u_{i,j} + 2\,u_{i+1,j} - u_{i+2,j} + u_{i,j+1} - 2\,u_{i+1,j+1} + u_{i+2,j+1}\right)$$
$$u_{xyy,i,j} = \frac{1}{2}\left(-u_{i,j} + u_{i+1,j} + 2\,u_{i,j+1} - 2\,u_{i+1,j+1} - u_{i,j+2} + u_{i+1,j+2}\right)$$

(2.41)

$S_5$ gives:

$$u_{xx,i,j} = \frac{1}{2}\left(u_{i-1,j} - 2\,u_{i,j} + u_{i+1,j}\right)$$

$$u_{yy,i,j} = \frac{1}{76}\Big(-11\,u_{i-1,j} - 120\,u_{i,j} - 11\,u_{i+1,j} - 11\,u_{i-1,j-2}$$
$$- 11\,u_{i+1,j-2} + 22\,u_{i-1,j-1} + 120\,u_{i,j-1} + 22\,u_{i+1,j-1} + 120\,w_{i,j}\Big)$$

$$u_{xy,i,j} = \frac{1}{4}\left(-3\,u_{i-1,j} + 3\,u_{i+1,j} - u_{i-1,j-2} + u_{i+1,j-2} + 4\,u_{i-1,j-1} - 4\,u_{i+1,j-1}\right)$$

$$u_{xxx,i,j} = -\frac{5}{11}\left(u_{i-1,j} - u_{i+1,j} + 2\,v_{i,j}\right)$$

$$u_{yyy,i,j} = -\frac{5}{38}\Big(u_{i-1,j} + 4\,u_{i,j} + u_{i+1,j} + u_{i-1,j-2} +$$
$$u_{i+1,j-2} - 2\,u_{i-1,j-1} - 4\,u_{i,j-1} - 2\,u_{i+1,j-1} - 4\,w_{i,j}\Big)$$

$$u_{xxy,i,j} = \frac{1}{2}\left(u_{i-1,j} - 2\,u_{i,j} + u_{i+1,j} - u_{i-1,j-1} + 2\,u_{i,j-1} - u_{i+1,j-1}\right)$$

$$u_{xyy,i,j} = \frac{1}{4}\left(-u_{i-1,j} + u_{i+1,j} - u_{i-1,j-2} + u_{i+1,j-2} + 2\,u_{i-1,j-1} - 2\,u_{i+1,j-1}\right)$$

(2.42)

$S_6$ gives:

$$u_{xx,i,j} = \frac{1}{76}\Big(-11\,u_{i+2,j-1} - 120\,u_{i,j} + 120\,u_{i+1,j} - 11\,u_{i,j+1} + 22\,u_{i+1,j+1} - 11\,u_{i+2,j+1} - 11\,u_{i,j-1} + 22\,u_{i+1,j-1} - 120\,v_{i,j}\Big)$$

$$u_{yy,i,j} = \frac{1}{2}\Big(-2\,u_{i,j} + u_{i,j+1} + u_{i,j-1}\Big)$$

$$u_{xy,i,j} = \frac{1}{4}\Big(u_{i+2,j-1} - 3\,u_{i,j+1} + 4\,u_{i+1,j+1} - u_{i+2,j+1} + 3\,u_{i,j-1} - 4\,u_{i+1,j-1}\Big)$$

$$u_{xxx,i,j} = \frac{5}{38}\Big(u_{i+2,j-1} + 4\,u_{i,j} - 4\,u_{i+1,j} + u_{i,j+1} - 2\,u_{i+1,j+1} + u_{i+2,j+1} + u_{i,j-1} - 2\,u_{i+1,j-1} + 4\,v_{i,j}\Big)$$

$$u_{yyy,i,j} = \frac{5}{11}\Big(u_{i,j+1} - u_{i,j-1} - 2\,w_{i,j}\Big)$$

$$u_{xxy,i,j} = \frac{1}{4}\Big(-u_{i+2,j-1} + u_{i,j+1} - 2\,u_{i+1,j+1} + u_{i+2,j+1} - u_{i,j-1} + 2\,u_{i+1,j-1}\Big)$$

$$u_{xyy,i,j} = \frac{1}{2}\Big(2\,u_{i,j} - 2\,u_{i+1,j} - u_{i,j+1} + u_{i+1,j+1} - u_{i,j-1} + u_{i+1,j-1}\Big)$$

(2.43)

$S_7$ gives:

$$u_{xx,i,j} = \frac{1}{2}\left(u_{i-1,j} - 2\,u_{i,j} + u_{i+1,j}\right)$$

$$u_{yy,i,j} = \frac{1}{76}\left(-11\,u_{i-1,j} - 120\,u_{i,j} - 11\,u_{i+1,j} + 22\,u_{i-1,j+1}\right.$$

$$\left. + 120\,u_{i,j+1} + 22\,u_{i+1,j+1} - 11\,u_{i-1,j+2} - 11\,u_{i+1,j+2} - 120\,w_{i,j}\right)$$

$$u_{xy,i,j} = \frac{1}{4}\left(3\,u_{i-1,j} - 3\,u_{i+1,j} - 4\,u_{i-1,j+1} + 4\,u_{i+1,j+1} + u_{i-1,j+2} - u_{i+1,j+2}\right)$$

$$u_{xxx,i,j} = -\frac{5}{11}\left(u_{i-1,j} - u_{i+1,j} + 2\,v_{i,j}\right)$$

$$u_{yyy,i,j} = \frac{5}{38}\left(u_{i-1,j} + 4\,u_{i,j} + u_{i+1,j} - 2\,u_{i-1,j+1} - \right.$$

$$\left. 4\,u_{i,j+1} - 2\,u_{i+1,j+1} + u_{i-1,j+2} + u_{i+1,j+2} + 4\,w_{i,j}\right)$$

$$u_{xxy,i,j} = \frac{1}{2}\left(-u_{i-1,j} + 2\,u_{i,j} - u_{i+1,j} + u_{i-1,j+1} - 2\,u_{i,j+1} + u_{i+1,j+1}\right)$$

$$u_{xyy,i,j} = \frac{1}{4}\left(-u_{i-1,j} + u_{i+1,j} + 2\,u_{i-1,j+1} - 2\,u_{i+1,j+1} - u_{i-1,j+2} + u_{i+1,j+2}\right)$$

(2.44)

$S_8$ gives:

$$u_{xx,i,j} = \frac{1}{76}\Big(120\, u_{i-1,j} - 120\, u_{i,j} - 11\, u_{i-2,j+1} + 22\, u_{i-1,j+1}$$
$$- 11\, u_{i,j+1} - 11\, u_{i-2,j-1} + 22\, u_{i-1,j-1} - 11\, u_{i,j-1} + 120\, v_{i,j}\Big)$$
$$u_{yy,i,j} = \frac{1}{2}\left(-2\, u_{i,j} + u_{i,j+1} + u_{i,j-1}\right)$$
$$u_{xy,i,j} = \frac{1}{4}\left(u_{i-2,j+1} - 4\, u_{i-1,j+1} + 3\, u_{i,j+1} - u_{i-2,j-1} + 4\, u_{i-1,j-1} - 3\, u_{i,j-1}\right)$$
$$u_{xxx,i,j} = \frac{5}{38}\Big(4\, u_{i-1,j} - 4\, u_{i,j} - u_{i-2,j+1} + 2\, u_{i-1,j+1} -$$
$$u_{i,j+1} - u_{i-2,j-1} + 2\, u_{i-1,j-1} - u_{i,j-1} + 4\, v_{i,j}\Big)$$
$$u_{yyy,i,j} = \frac{5}{11}\left(u_{i,j+1} - u_{i,j-1} - 2\, w_{i,j}\right)$$
$$u_{xxy,i,j} = \frac{1}{4}\left(u_{i-2,j+1} - 2\, u_{i-1,j+1} + u_{i,j+1} - u_{i-2,j-1} + 2\, u_{i-1,j-1} - u_{i,j-1}\right)$$
$$u_{xyy,i,j} = \frac{1}{2}\left(2\, u_{i-1,j} - 2\, u_{i,j} - u_{i-1,j+1} + u_{i,j+1} - u_{i-1,j-1} + u_{i,j-1}\right)$$

(2.45)

As in the previous sub-section, we assign equal optimal weights to all the stencils. The smoothness measures that we use for limiting $u_{xy,i,j}$ are given by eqn. (2.36), this time applied to the polynomials that we have constructed in this section. The smoothness measures that we use for limiting $u_{xxy,i,j}$ and $u_{xyy,i,j}$ are given by:

$$\beta = \iint_{[-1/2,1/2]\times[-1/2,1/2]} \left[ \left( \frac{\partial^3\, u\,(x,y,t)}{\partial x^3} \right)^2 + \left( \frac{\partial^3\, u\,(x,y,t)}{\partial y^3} \right)^2 + \left( \frac{\partial^3\, u\,(x,y,t)}{\partial x^2\, \partial y} \right)^2 + \left( \frac{\partial^3\, u\,(x,y,t)}{\partial x\, \partial y^2} \right)^2 \right] dx\, dy$$

(2.45)

This completes our description of the reconstruction of cross-terms for the fourth order hybrid RKDG+HWENO scheme.

# 3 A Sub-Cell Based Algorithm for Flagging Troubled Zones

Both RKDG and hybrid RKDG-HWENO schemes really do not need a lot of limiting. For smooth portions of the flow the schemes are very stable and the higher moments do not need any limiting. Some practitioners have even reported that they have successfully solved problems with mild shocks by using RKDG schemes without any resort to limiting. However, for most problems with strong shocks, some amount of limiting is indeed needed. The MP limiter which we develop here follows the philosophy of Suresh and Huynh [29] but is a little different in details. It permits a monotonicity property to be applied in a controlled way and does not clip all local extrema, as would a TVD limiter. It parametrizes the amount of solution-dependent curvature we permit in a bonafide local extremum and the rapidity with which this curvature may vary. Both these features are very desirable when it comes to preserving some forms of local extrema, especially when used with RKDG schemes that do not seem to need as much limiting as comparable TVD schemes.

It is important to realize two important features associated with local extrema in RKDG schemes: 1) The higher order RKDG schemes can retain meaningful sub-cell structure that may not need to be limited. 2) It is also important to realize that spurious extrema in second order finite volume schemes with piecewise-linear slopes only show up at the zone boundaries. In contrast, when dealing with higher order RKDG schemes we have to realize that such spurious extrema will not necessarily show up only at a zone boundary but may also manifest themselves within the interior of a zone. Applying any form of discontinuity detector at just the zone boundary, as is done by Biswas, Devine and Flaherty [5] and Burbeau, Sagaut and Bruneau [6], would fail to distinguish between the two types of extrema catalogued earlier in this paragraph. A higher order scheme that retains higher moments simply forces us to take a sub-cell based approach to limitng. The schemes presented in this paper have all been formulated in a modal basis set. We observe though that a nodal formulation of RKDG schemes makes it even easier to see the need for a sub-cell based limiting strategy.

The number of sub-cells that each zone should be divided into is not rigidly determined but should be sufficient to distinguish between the different types of extrema that may form. The number of sub-cells used will not determine the cost of the scheme because all limiters for higher order schemes operate in characteristic variables and the cost of building eigenvectors and projecting the solution and its moments into characteristic variables far outweighs the cost of taking a sub-cell approach. As a practical matter, we do not divide the piecewise linear representation of second order RKDG schemes into subcells at all because we know from the above discussion that the extrema will show up at the zone boundaries. For third order schemes, which retain piecewise parabolic data, we subdivide each zone into three sub-cells of equal size and apply the modified MP algorithim within each of them. For fourth order schemes, which retain piecewise cubic data, we subdivide each zone into four sub-cells of equal size and apply the modified MP algorithim within each of them. The entire zone being considered is flagged as a troubled zone if the MP algorithm finds that the upwinded boundary of any sub-cell lies outside the monotonicity preserving limits given by the MP algorithm at that boundary. The MP algorithm also requires two more sub-cells on either side of the sub-cell to which it is applied. As a result, for the second order schemes the MP algorithm requires us to project the zone averages from the two neighboring zones to the left and right into the characteristic space of the zone being considered. For third and higher order schemes, the MP algorithm only requires us to project all the moments from each of the neighboring zones to the immediate left and right of the zone being considered into the characteristic space of that zone. Using all these moments we can build sub-cell averaged characteristic variables for all the sub-cells within each zone as well as in the two sub-cells to the right and left of the zone being considered. Thus say that $\overline{w}_j$ is a sub-cell averaged characteristic variable in sub-cell “j” that is being considered. To apply the MP algorithm we need $\overline{w}_{j-2}$ , $\overline{w}_{j-1}$ , $\overline{w}_j$ , $\overline{w}_{j+1}$ and $\overline{w}_{j+2}$ . If the waves associated with this characteristic variable are flowing to the right within the zone being considered then we also need $w_{j+1/2}$ which is easily available because in RKDG schemes the solution is available at any point within the zone.

Our first step is to apply a very coarse test. Thus we form a TVD bound using the interval $I[\,\overline{w}_j\,, w_{j+1/2}^{TVD}]$ where $w_{j+1/2}^{TVD}$ is a TVD limiter-based representation of the characteristic variable “w” at the sub-cell boundary “j+1/2” . If $w_{j+1/2}$ lies within the interval $I[\bar{w}_j, w^{TVD}_{j+1/2}]$ we forgo any further MP construction and say that the sub-cell “j” is trouble-free. $w^{TVD}_{j+1/2}$ is constructed using a modification of van Leer’s MC limiter as follows:

$$w^{TVD}_{j+1/2} = \bar{w}_j + 0.5\, MC\beta_Limiter\,[\, \bar{w}_{j+1} - \bar{w}_j,\ \bar{w}_j - \bar{w}_{j-1}] \qquad (3.1)$$

where

$$\begin{aligned} MC\beta_Limiter\,[a, b] &= sgn(a)\, \min(0.5\,|a+b|, \beta\,|a|, \beta\,|b|) && \text{if } a\,b > 0 \\ &= 0 && \text{otherwise} \end{aligned} \qquad (3.2)$$

Setting $\beta \in [1, 2]$ in the MCβ_Limiter produces a TVD limiter, with β=1 being the non-compressive minmod limiter and β=2 being the maximally compressive MC limiter. The inclusion of β in eqn. (3.2) allows us to pick all limiters between these two extremes.

If the sub-zone does not pass the above coarse test then we go through the MP limiter. The MP algorithm consists of realizing that the TVD condition only requires $w_{j+1/2}$ to lie within the intersection of the intervals $I[\bar{w}_j, \bar{w}_{j+1}]$ and $I[\bar{w}_j, w^{UL}_{j+1/2}]$ where:

$$w^{UL}_{j+1/2} = \bar{w}_j + \alpha\,(\bar{w}_j - \bar{w}_{j-1}) \qquad (3.3)$$

Here all values $\alpha \in [0.5, 1.0]$ will yield a TVD scheme that will function well with a CFL condition that is less than or equal to 0.5 . To make allowance for local extrema we need to include curvatures into the above equations and we do that next. We first construct curvatures $d_j$ and $d_{j+1}$ as follows:

$$d_j = \bar{w}_{j+1} - 2\,\bar{w}_j + \bar{w}_{j-1} \quad ; d_{j+1} = \bar{w}_{j+2} - 2\,\bar{w}_{j+1} + \bar{w}_j \qquad (3.4)$$

We then use these curvatures to approximate a curvature at the "j+1/2" sub-cell boundary as follows:

$$d_{j+1/2} = \tau \ \text{minmod}\left( \kappa \, d_j - d_{j+1} \, , \, \kappa \, d_{j+1} - d_j \, , d_j \, , d_{j+1} \right) \tag{3.5}$$

The ratio κ in eqn. (3.5) ensures that the curvature $d_{j+1/2}$ is zeroed out when the ratio of curvatures $d_{j+1}/d_j$ lies outside the range $[1/\kappa, \ \kappa]$ , i.e. when the curvature varies rapidly from one sub-cell to the next. The variable τ in eqn. (3.5) should be set greater than or equal to unity, where values larger than unity provide space for additional curvature to develop provided the curvatures do not fluctuate too rapidly from one sub-cell to the next. Using an equation that is entirely analogous to eqn. (3.5) we also construct the curvature $d_{j-1/2}$ . Having constructed the curvature at the "j+1/2" sub-cell boundary we wish to find a parabola with the following three conditions: a) its sub-cell averaged value in sub-cell "j" is $\overline{w}_j$ , b) its sub-cell averaged value in sub-cell "j+1" is $\overline{w}_{j+1}$ and c) its curvature is $d_{j+1/2}$ . The value of such a parabola at the "j+1/2" sub-cell boundary is given by:

$$w^{MD}_{j+1/2} = \frac{1}{2} \left( \overline{w}_j + \overline{w}_{j+1} \right) - \frac{1}{3} d_{j+1/2} \tag{3.6}$$

We also want to find a parabola with the following three conditions: a) its sub-cell averaged value in sub-cell "j" is $\overline{w}_j$ , b) its sub-cell averaged value in sub-cell "j–1" is $\overline{w}_{j-1}$ and c) its curvature is $d_{j-1/2}$ . The value of such a parabola at the "j+1/2" sub-cell boundary is given by:

$$w^{LC}_{j+1/2} = \overline{w}_j + \frac{1}{2} \left( \overline{w}_j - \overline{w}_{j-1} \right) + \frac{2}{3} d_{j-1/2} \tag{3.7}$$

The parabolic values built up in eqns. (3.6) and (3.7) are used to enhance the intervals mentioned above so that we now seek the intersection of the intervals $I[\,\overline{w}_j\,,\,\overline{w}_{j+1}\,,\,w^{MD}_{j+1/2}\,]$ and $I[\,\overline{w}_j\,,\,w^{UL}_{j+1/2}\,,\,w^{LC}_{j+1/2}\,]$. By including the parabolic profiles we make allowance for well-formed extrema. The intersection is given by the interval $I[\,w^{min}_{j+1/2}\,,\,w^{max}_{j+1/2}\,]$ where:

$$w^{min}_{j+1/2} = \max\left[\,\min\left(\overline{w}_j\,,\,\overline{w}_{j+1}\,,\,w^{MD}_{j+1/2}\right),\ \min\left(\overline{w}_j\,,\,w^{UL}_{j+1/2}\,,\,w^{LC}_{j+1/2}\right)\right] \qquad (3.8)$$

and

$$w^{min}_{j+1/2} = \min\left[\,\max\left(\overline{w}_j\,,\,\overline{w}_{j+1}\,,\,w^{MD}_{j+1/2}\right),\ \max\left(\overline{w}_j\,,\,w^{UL}_{j+1/2}\,,\,w^{LC}_{j+1/2}\right)\right] \qquad (3.9)$$

A sub-cell is said to satisfy the MP constraint if $w_{j+1/2}$ lies within the interval $I[\,w^{min}_{j+1/2}\,,\,w^{max}_{j+1/2}\,]$ . A zone is said to be free of trouble if all its sub-cells "j" satisfy the MP constraint. Zones that are not free of trouble are said to be troubled zones and the appropriate WENO scheme is used to construct the slope, as shown in Qiu and Shu [21]. The second and higher moments can be constructed using the formulation in sub-sections 2.b.i and 2.b.ii.

For typical problems associated with the Euler equations we use the following values for the free parameters defined in this section. The values change for differing RKDG schemes and the preferred values for Euler flows, obtained after extensive testing, are given below:

p=1 RKDG : $\beta = 1.3$; $\alpha = 0.7$; $\kappa = 4.0$; $\tau = 1.3$.
p=2 RKDG : $\beta = 1.3$; $\alpha = 0.7$; $\kappa = 4.0$; $\tau = 1.3$. (3.10)
p=3 RKDG : $\beta = 1.1$; $\alpha = 0.6$; $\kappa = 3.0$; $\tau = 1.1$.
Hybrid RKDG+WENO : $\beta = 1.3$; $\alpha = 0.8$; $\kappa = 4.0$; $\tau = 1.3$.

For Hybrid RKDG+WENO schemes we take each zone to be a sub-cell, just as we do for p=1 RKDG. This is so because we only have the variable and its first moment in that case. The parameters given above were optimized for use with the (local) Lax-Friedrichs flux, which we use all through this work. The above choice of parameters represents conservatively defined sets of choices. For many Euler flow problems the parameters can assume much larger values. Even with the conservatively defined choices for the parameters given above, we find the flagging to be minimal on most applications including ones with very strong shocks, as will be shown in the subsequent section. The conservatively defined choices for the parameters given above still provide enough space to ensure that most of the accuracy tests with smooth solutions run through without triggering any flagging at all, as will be shown in the next section. This completes our description of the sub-cell based algorithm for flagging troubled zones.

We make several observations below:

1) Eqns. (3.5) to (3.7) are different from the corresponding ones in Suresh and Huynh [29]. They are more consistent with the volume-averaged approach that we use in the sub-cells.

2) Notice that the flagging algorithm described here does not increase the stencil by too many zones on either side. For second order schemes it only increases the stencil by two zones on either side. For RKDG schemes that are of third and higher orders the present flagging algorithm only increases the stencil by one zone on either side.

3) The major cost of the current algorithm is the cost of producing the eigenvectors and projecting all the moments of the current zone and its neighbors into characteristic variables. This is a fixed cost that cannot be avoided for all higher order schemes because limiting strategies don't seem to work well when applied to primitive or conserved variables. The cost of the sub-cell flagging algorithm as well as the WENO reconstruction in flagged zones is modest. Likewise, the cost of transforming the moments in the flagged zones back to characteristic variables is indeed modest by comparison. Thus the present algorithm is cost-competitive with any other that carries out limiting in characteristic variables.

4) The present algorithm is capable of identifying trouble in individual characteristic fields within a zone. Unlike Qiu and Shu [21], [22] we did not see the worth of using WENO or HWENO to reconstruct all the characteristic fields in zones that are flagged. Instead we only used WENO or HWENO to reconstruct the troubled characteristic fields. This permits us to preserve more information about the solution.

5) Balsara [4] has found that the MHD system seems to need a somewhat more restrictive limiting than the Euler system. The present MP algorithm, with its adjustable parameters, provides that flexibility.

6) When formulating RKDG schemes for unstructured meshes, it is more natural to use a nodal basis set. Since nodal basis sets make it easier to motivate a sub-cell based algorithm for detecting troubled zones, it may prove easy to extend this algorithm to unstructured meshes.

7) The present strategy might also prove useful for residual distribution schemes that have sub-structure within each zone.

## 4 Accuracy Analysis

We present an accuracy analysis for the RKDG and hybrid RKDG+HWENO schemes described in this paper. In all instances the local Lax-Friedrichs flux was used. The Courant numbers were set to be 0.9 times the maximum permissible values from Cockburn, Karniadakis and Shu [11]. For each test problem the spatial and temporal accuracy were kept the same. For all tests the MP algorithm used the settings that we have found to be beneficial for the Euler equations. All the problems used in this section involve smooth solutions. Thus we present the accuracy analysis with the limiter and the flagging strategy described in the previous section and also the same accuracy analysis without any WENO limiting. If the detector for troubled zones operated optimally, it should not flag any zones as troubled. In that case the accuracy should be the same with and without the flagging algorithm. In practice, some of the more stringent test problems do trigger flagging in some zones. In that case, the WENO algorithm is invoked to limit the first and higher moments in that zone. As a result the solutions with limiting will be less accurate than the ones without. We should, nevertheless, expect to see the same formal order of accuracy. We have also advanced the viewpoint that higher moments can be reconstructed without substantial loss of accuracy for the hybrid RKDG+HWENO schemes. Thus we expect the hybrid RKDG+HWENO schemes to meet their design orders of accuracy. Moreover, on large enough meshes we might expect the hybrid RKDG+HWENO schemes to have intrinsic accuracies that are in the same range as the intrinsic accuracies of the corresponding RKDG schemes.

We also point out that the first three accuracy analyses have also been done for WENO schemes in Balsara and Shu [2]. As a general observation we find that for most of the cases considered here the RKDG schemes start out by having intrinsically better accuracy on smaller meshes. Thus while WENO schemes have higher formal order of accuracy, the RKDG schemes have better resolution on smaller meshes. The present observation led Torrilhon and Balsara [32] to conclude that RKDG schemes might have some advantages over WENO schemes in overcoming the pseudo-convergence that is observed in the MHD system.

The problems in this and the next section were all run with the (local) Lax-Friedrichs flux function. Flagging of troubled zones was done with the sub-cell based MP algorithm using the parameters given in eqn. (3.10) in all cases, except when noted otherwise. In all instances we were able to run all the test problems with Courant numbers that were set to be 0.9 times the maximum permissible values from Cockburn, Karniadakis and Shu [11]. The temporal accuracy in the Runge-Kutta time-update was set to have the same order as the spatial order of accuracy for the scheme. Here we take ( $\rho$, v, p) to be the density, velocity and pressure variables in the Euler equations.

## 4.a First Test with Advection Equation

We solve the advection equation, $u_t + u_x = 0$ , with initial condition u ( x, 0) = sin (2 $\pi$ x) on the domain [-0.5, 0.5]. Periodic boundary conditions were used and the simulation was stopped at a time of 0.5. The accuracy analysis is given in Table 1. The sinusoidal function is very smooth. As a result, we see that the solution with the limiter has the same value as the solution without the limiter in all instances, showing us that the zones were never flagged as troubled. This is the desired result, indicating that our flagging algorithm is sophisticated enough to realize that for this problem there are no troubled zones. We also notice that the hybrid RKDG+HWENO schemes have intrinsic accuracies that are in the same range as the intrinsic accuracies of the corresponding RKDG schemes, indicating that our strategy for reconstructing the higher moments is indeed an effective one for very smooth flows.

## 4.b Second Test with Advection Equation

We solve the advection equation, $u_t + u_x = 0$ , with initial condition $u(x, 0) = \sin^4(2\pi x)$ on the domain [-0.5, 0.5]. Periodic boundary conditions were used and the simulation was stopped at a time of 0.5. The accuracy analysis is given in Table 2. The present initial conditions have a very rapidly varying curvature. As a result, we expect that there will be instances where the flagging algorithm will falsely flag zones as being troubled. Thus we see that on some of the smaller meshes the solution with the limiter has accuracy that differs from the accuracy of the solution without the limiter. We do, however, expect that as the mesh is refined fewer zones will be flagged as troubled. Table 2 shows us that this is indeed the case. Thus on larger meshes we find that the accuracy of the solution with the limiter becomes comparable to the accuracy of the solution without the limiter. In all instances we see that the schemes meet their design orders of accuracy. On the larger meshes we also observe that the p=2 RKDG+HWENO scheme has accuracy that is almost the same as the corresponding p=2 RKDG scheme. This shows that our strategy of reconstructing the second moments was an effective one. The p=3 RKDG+HWENO scheme reconstructs the second and third moments. On large meshes we see that it is less accurate than the p=3 RKDG scheme by a couple of orders of magnitude. Thus, because a larger number of moments are reconstructed in the p=3 RKDG+HWENO scheme, the p=3 RKDG scheme retains an advantage over the p=3 RKDG+HWENO scheme. Both schemes do, however, meet their designed order of accuracy.

## 4.c Burgers Equation Test

We solve the nonlinear scalar Burgers equation, $u_t + u\, u_x = 0$, with initial condition $u(x, 0) = 0.25 + 0.5 \sin(2 \pi x)$ on the domain [-0.5, 0.5]. Periodic boundary conditions were used and the simulation was stopped at a time of $0.5/\pi$ which corresponds to a time before any shocks form. The accuracy analysis is given in Table 3. The initial condition is very smooth. As a result, we see that the solution with the limiter has the same value as the solution without the limiter in all instances, showing us that the zones were never flagged as troubled. We also notice that the hybrid RKDG+HWENO schemes have intrinsic accuracies that are in the same range as the intrinsic accuracies of the corresponding RKDG schemes, indicating that our strategy for reconstructing the higher moments is indeed an effective one for very smooth flows.

## 4.d One Dimensional Test with Euler Equations

We solve the Euler equations with density profile $\rho(x, 0) = 1.0 + 0.25 \sin(2 \pi x)$ and pressure and velocity set to unity. The problem was run on the domain [-0.5, 0.5] with periodic boundaries and stopped at a time of 1.0. The results are shown in Table 4. Since the initial condition is very smooth, we see that the solution with the limiter has the same value as the solution without the limiter in all instances. This shows us that the zones were never flagged as troubled, indicating that our flagging algorithm is sophisticated enough to realize that for this problem there are no troubled zones. On the larger meshes we also observe that the p=2 RKDG+HWENO scheme has accuracy that is almost the same as the corresponding p=2 RKDG scheme. On large meshes we see that the p=3 RKDG+HWENO scheme is less accurate than the p=3 RKDG scheme by a couple of orders of magnitude. Thus as larger numbers of moments are reconstructed the p=3 RKDG scheme retains an advantage over the p=3 RKDG+HWENO scheme for some test problems. Both schemes do, however, meet their designed order of accuracy.

## 4.e Multi-Dimensional Vortex Test with Euler Equations

The previous tests have shown that the methods developed in Section 3 are good at flagging troubled zones and that the methods presented in Sub-Section 2.b successfully reconstruct the moments when the zones are flagged. We are now interested in examining the performance of the cross-terms developed in Sub-Section 2.c on multi-dimensional problems. we analyze the propagation of a strong vortex at a supersonic Mach number. The vortex propagates at $45^0$ to the grid lines which gives ample opportunity for the effects of multidimensional propagation to manifest themselves in this test problem. The problem is initialized on the two dimensional domain given by [-5,5]X[-5,5]. An unperturbed flow of the Euler equations with $\left( \rho , P , v_x , v_y \right) = \left( 1, 1, 1, 1 \right)$ and a ratio of specific heats given by $\gamma = 1.4$ is initialized on the computational domain. The temperature and entropy are defined as $T = P/\rho$ and $S = P/\rho^{\gamma}$ . The vortex is defined as a fluctuation to this mean flow given by

$$\left( \delta v_x , \delta v_y \right) = \frac{\varepsilon}{2 \pi} e^{0.5 \left( 1 - r^2 \right)} \left( - y , x \right)$$
$$\delta T = - \frac{\left( \gamma - 1 \right) \varepsilon^2}{8 \gamma \pi^2} e^{\left( 1 - r^2 \right)} ; \quad \delta S = 0 \tag{4.1}$$

where $r^2 = x^2 + y^2$ and the vortex strength $\varepsilon = 5$ . We utilize periodic boundary conditions. The accuracy was evaluated at a time of 10.0 code units. Table 5 shows the results. The problem is run on mesh sizes ranging from 10X10 to 160X160. On the smallest meshes a significant amount of limiting is triggered but we see that asymptotically, the error with a limiter becomes comparable to the error without the limiter. This includes situations where limiting of the cross terms is triggered, showing that the limited cross terms do not cause any degradation in the formal order of accuracy of the scheme. In Sub-section 2.3 we showed that the limiting could be accomplished either with a MinMod limiter or with the use of smoothness measures in a WENO-like fashion. Table 5 shows the order of accuracy with either style of limiting applied to the cross-terms. We see that use of the WENO-like, smoothness measure-based limiter produces slightly better accuracy, albeit at a substantially increased computational cost. It is, however, satisfying to notice that both strategies for limiting the cross-terms meet their design accuracy. We also see that the accuracy of the solution with a limiter begins to approach the accuracy of the solution without the limiter on 80X80 meshes. This is expected because as the resolution increases, the limiter views the solution as more smoothly represented on the mesh and, therefore, flags fewer zones for reconstruction.

## 5 Test Problems

In these tests we emphasize the third and fourth order schemes. The p=1 RKDG scheme also works well on these test problems but it produces results that are not so much better than a very, very good TVD scheme. As a result, it is not shown here.

For each of the last three tests in this section we also provide a table that gives us the percentage of flagged zones for each of the schemes tested as a function of increasing number of zones. We should expect that for the same physical problem the percentage of flagged zones decreases with increasing number of zones. We will see that the expectation is borne out in practically all cases. For RKDG schemes this implies that as the mesh is refined, most of the zones will have moments from the more accurate RKDG formulation rather than the less accurate WENO reconstruction. For hybrid RKDG+HWENO schemes this implies that the higher moments will be reconstructed using the more accurate first moments from the more accurate RKDG formulation rather than the less accurate WENO reconstruction. Thus we see that as the mesh is refined the intrinsic accuracy of the schemes presented here is closer to the intrinsic accuracy of the RKDG scheme. In some scientific problems, such as the pseudo-convergence of MHD Riemann problems that was explored in Torrilhon and Balsara [32], the intrinsic accuracy of the scheme is more important than its formal order of accuracy. For this reason, Torrilhon and Balsara [32] concluded that RKDG schemes that respect the divergence-free evolution of the magnetic field might have some special advantages for numerical MHD.

## 5.a Advection Test

Our first test problem consists of testing the behavior of the scheme on a rather stringent scalar advection test problem. It is the same test problem that was catalogued in Jiang and Shu [18]. Thus we solve the problem

$$\begin{aligned} & u_t + u_x = 0 && -1 < x < 1 \\ & u(x, 0) = u_0(x) && \text{periodic} \end{aligned} \tag{5.1}$$

with

$$\begin{aligned} u_0(x) &= \frac{1}{6}\left[ G(x, \phi, z-\delta) + G(x, \phi, z+\delta) + 4\,G(x, \phi, z) \right] && -0.8 \le x \le -0.6 \\ &= 1 && -0.4 \le x \le -0.2 \\ &= 1 - \left| 10\,(x - 0.1) \right| && 0.0 \le x \le 0.2 \\ &= \frac{1}{6}\left[ F(x, \psi, a-\delta) + F(x, \psi, a+\delta) + 4\,F(x, \psi, a) \right] && 0.4 \le x \le 0.6 \\ &= 0 && \text{otherwise} \end{aligned} \tag{5.2}$$

$$\begin{aligned} G(x, \phi, z) &= e^{-\phi (x-z)^2} \\ F(x, \psi, a) &= \sqrt{\max\left(1 - \psi^2 (x-a)^2 , 0\right)} \end{aligned} \tag{5.3}$$

The constants in eqs (5.2) and (5.3) are given by

$$a = 0.5 \;;\; z = -0.7 \;;\; \delta = 0.005 \;;\; \psi = 10 \;;\; \phi = \frac{\log 2}{36\,\delta^2} \tag{5.4}$$

The problem has several shapes that are difficult to advect with fidelity. The shapes consist of : 1) a combination of Gaussians, 2) a square wave, 3) a sharply peaked triangle and 4) a half ellipse arranged initially from left to right. The reasons that make it a stringent test problem are catalogued in Balsara and Shu [2]. The problem was initialized on a mesh of 200 zones. It was run for a simulation time of 20 which corresponds to ten traversals around the mesh. In doing so, the features catalogued in eqs (5.2) and (5.3) were advected over 2000 mesh points.

Figs 3a to 3d show the solutions obtained from the p=2 RKDG, p=2 RKDG+HWENO, p=3 RKDG and p=3 RKDG+HWENO schemes respectively. The reference solution is also shown as an overlaid solid line. We see that the p=2 RKDG and p=2 RKDG+HWENO have done equally well in advecting the profile without much distortion of the shape or clipping of extrema. Also notice that both these schemes are just third order schemes. Nevertheless, on comparing the results from Figs. 3a and 3b to the result from the ninth order accurate r=5 WENO scheme from Balsara and Shu [2] we see that the results from p=2 RKDG and p=2 RKDG+HWENO are entirely competitive. Thus the third order accurate RKDG family of schemes have performed just as well as a ninth order accurate WENO scheme. The advantage of the RKDG family of schemes stems from their smaller stencil and their substantially lower dissipation on smaller meshes. Fig 1c shows that the fourth order accurate p=3 RKDG and p=3 RKDG+HWENO schemes perform even better than their third order counterparts. Moreover, the p=3 RKDG scheme clearly outperforms the ninth order accurate r=5 WENO scheme. We also see that the p=3 RKDG scheme has a slight edge over the p=3 RKDG+HWENO scheme, as expected. Note though that the r=5 WENO scheme lends itself to easy modification so that we can include the artificial compression method (ACM) from Yang [34] to steepen the profile of the square pulse. The RKDG algorithm has not been similarly modified in the currently available literature.

## 5.b The Lax Problem

The Lax Riemann problem consists of the following specification:

$$\begin{aligned} (\rho, v, p) &= (0.445, 0.698, 3.528) && \text{for } -0.5 \le x \le 0.0 \\ &= (0.5, 0, 0.571) && \text{for } 0.0 \le x \le 0.5 \end{aligned} \qquad (5.5)$$

The problem was run on a 200 zone mesh to a time of 1.3. Figs 4a, 4b, 4c and 4d show the resulting density for the p=2 RKDG, p=2 RKDG+HWENO, p=3 RKDG and p=3 RKDG+HWENO schemes respectively. The reference solution is also shown as an overlaid solid line. Figs. 4e, 4f, 4g and 4h show the history of flagged points in space-time for the above four schemes. All schemes show exceptional treatment of the contact discontinuity, owing to the fact that they are all better than second order and have small stencils. We also see that the sub-cell based flagging algorithm has always been effective at detecting the shock and both ends of the rarefaction fans. In some instances the contact discontinuity is also flagged. Figs. 4e to 4h also show us that the maximal flagging occurs towards the end of the simulation. Table 6 provides the percentage of zones that were flagged in the last timestep for each of the schemes tested as a function of increasing number of zones. We see that, in keeping with our expectations, the percentage of flagged zones decreases as a function of increasing number of zones.

## 5.c The Shock-Entropy Wave Interaction Problem

This problem was first presented in Shu and Osher [27]. A moving Mach 3 shock is made to interact with a sinusoidal density fluctuation. The initial conditions are given by:

$$\begin{aligned}(\rho, \mathrm{v}, \mathrm{p}) &= (3.857143, 2.629369, 10.333333) && \text{for } -5 \le x \le -4 \\ &= (1 + 0.2 \sin(5x), 0, 1) && \text{for } -4 \le x \le 5\end{aligned} \qquad (5.6)$$

The problem provides an example of the interaction of a shock with a smooth flow having oscillations. The simulation was run on a 200 zone mesh and stopped at a time of 1.8. Figs 5a, 5b, 5c and 5d show the resulting density for the p=2 RKDG, p=2 RKDG+HWENO, p=3 RKDG and p=3 RKDG+HWENO schemes respectively. The reference solution is also shown as an overlaid solid line. Figs 5e, 5f, 5g and 5h show the history of flagged points in space-time for the above four schemes. We see that the density profile from the p=2 RKDG scheme is very marginally better than the p=2 RKDG+HWENO, but that is strongly dependent on the choice of parameters in the sub-cell based algorithm for detecting troubled zones. The p=3 RKDG and p=3 RKDG+HWENO schemes perform comparably well. In all cases the post-shock oscillations are well resolved, showing that our algorithm does not unduly destroy structures. Figs. 5e through 5h show that the flagging algorithm has accurately detected the Mach 3 shock and also the smaller shocks that form behind it. Figs. 5e to 5h also show us that the maximal flagging occurs towards the end of the simulation. Table 7 provides the percentage of zones that were flagged in the last timestep for each of the schemes tested as a function of increasing number of zones. We see that, in keeping with our expectations, the percentage of flagged zones decreases as a function of increasing number of zones.

## 5.d Interaction of Blast Waves

The present problem was suggested by Woodward and Colella [33] and considers the interaction of blast waves. The initial conditions are :

$$
\begin{aligned}
(\rho, \mathrm{v}, \mathrm{p}) &= (1, 0, 1000) \qquad \text{for } -0.5 \le x \le -0.4 \\
&= (1, 0, 0.01) \qquad \text{for } -0.4 \le x \le 0.4 \\
&= (1, 0, 100) \qquad \text{for } \ \ 0.4 \le x \le 0.5
\end{aligned}
\tag{5.7}
$$

Reflecting boundary conditions are used at both ends of the computational domain with 400 zones and the problem was run to a time of 0.038. Figs 6a, 6b, 6c and 6d show the resulting density for the p=2 RKDG, p=2 RKDG+HWENO, p=3 RKDG and p=3 RKDG+HWENO schemes respectively. Figs 6e, 6f, 6g and 6h show the history of flagged points in space-time for the above four schemes. All the schemes were run with the parameters from eqn. (3.10) in the sub-cell based algorithm for detecting troubled zones with the exception of the p=3 RKDG+HWENO scheme. The latter was run with $\beta = 1.1$; $\alpha = 0.7$; $\kappa = 3.0$; $\tau = 1.15$ . For just this problem we preferred to use the third-order accurate Runge-Kutta time-stepping strategy for all the runs because of its good TVD preserving properties. We see that in all instances the density profile compares well with the reference solution, which is shown with a solid line. The p=2 RKDG and p=2 RKDG+HWENO schemes seem to require minimal amount of flagging and closely track the strong shocks. The p=3 RKDG scheme uses only slightly more flagging. Because of the reduced parameters in the flagging algorithm, we see that the p=3 RKDG+HWENO scheme produces a somewhat larger number of flagged points. Figs. 6e to 6h also show us that the flagging that occurs towards the end of the simulation is representative of the amount of flagging that occurs all through the run. Table 8 provides the percentage of zones that were flagged in the last timestep for each of the schemes tested as a function of increasing number of zones. We see that, in keeping with our expectations, the percentage of flagged zones decreases as a function of increasing number of zones.

### 5.e) Mach 3 Wind Tunnel with a Forward Facing Step

This problem has been proposed by Woodward and Colella [33]. It has been simulated by Cockburn and Shu [10] with RKDG schemes showing the vortex sheet roll up that sets in with increasing resolution. Our purpose is not to make such a resolution study but rather to validate the robust and accurate behavior of the schemes proposed here. For this reason we have simulated this test problem at the same resolution and twice the resolution as Woodward and Colella [33]. The problem consists of a wind tunnel that is initialized on a two dimensional grid with rectangular zones that span the region [0,3]X[0,1]. A forward facing step is set up with the corner of the step at (0.6,0.2). The left boundary is initialized as an inflow boundary that has a Mach 3 gas with density of 1.4 and unit pressure flowing in. The gas has a ratio of specific heats given by 1.4 . The right boundary is taken to be an outflow boundary. Reflective boundary conditions are applied to the walls of the tunnel. We treated the singularity at the corner with the same technique that was suggested in Woodward and Colella [33]. The problem was run with p=2 and p=3 RKDG schemes using the limiter given here on grids of 240X80 till a simulation time of 4.0 .

Fig. 7 shows the density from the simulation of the Mach 3 forward-facing step problem with 240X80 zone resolution at a time of 4.0 for the p=2 RKDG scheme. The problem was run with the troubled zone detection strategy described in Section 3 and methods described in Section 2 were used to reconstruct the variables in those zones. We see clearly that all shocks have sharp profiles and are well-captured on the computing grids that have been used. The vortex sheet that emanates from the Mach stem is properly resolved with just a few zones across the vortex sheet.

## 6 Conclusions

We arrive at the following conclusions:

1) An effective indicator of troubled zones in RKDG schemes should be based on a sub-cell algorithm for detecting troubled zones. This is so because RKDG schemes can retain meaningful sub-cell structure that does not need to be limited.

2) We have recast the MP algorithm of Suresh and Huynh [29] so that it detects troubled zones by examining the sub-cells of a zone. This makes it an effective, scale-free, problem-independent detector of troubled zones. Our algorithm for detecting troubled zones has been applied successfully to several stringent test problems.

3) We have also realized that in most situations, the variable and its first moment carry a majority of the information in the flow. Building on that, we have designed hybrid RKDG+HWENO schemes that reconstruct the second and third moments by using the information contained in the variable and its first moment. Explicit formulae have been provided for one dimensional and two dimensional cases and the formulae are suitable for implementation in practical numerical codes. The resulting schemes have the same order of accuracy as the corresponding RKDG schemes.

4) The hybrid RKDG+HWENO schemes are low-storage alternatives to the RKDG schemes that usually perform almost as well as the RKDG schemes. This has been shown via several accuracy analyses and stringent test problems in one and two dimensions.

## Acknowledgements

DSB acknowledges support via NSF grants AST-005569-001, AST-0607731 and NSF-PFC grant PHY02-16783 and NASA grant HST-AR-10934.01-A. Computer support from NCSA and Notre Dame’s HPCC cluster is also acknowledged.

## Appendix

The coefficients in eqn. (2.3) can be obtained in any zone by using the mass matrix and the orthogonality of the modal bases. We assume that a zone has local coordinates that span [-0.5,0.5]. The zeroth moment is given by:

$$u_0\ (t) = \int_{-1/2}^{1/2} u\ (x,\ t)\ dx$$

The first moment is given by:

$$u_1\ (t) = 12 \int_{-1/2}^{1/2} u\ (x,\ t)\ x\ dx$$

The second moment is given by:

$$u_2\ (t) = 180 \int_{-1/2}^{1/2} u\ (x,\ t) \left( x^2 - \frac{1}{12} \right) dx$$

The third moment is given by:

$$u_3\ (t) = 2800 \int_{-1/2}^{1/2} u\ (x,\ t) \left( x^3 - x \frac{3}{20} \right) dx$$

Gauss or Gauss-Lobatto quadrature points can be used from Stroud and Secrest [28] to evaluate these integrals numerically. This can be used to initialize the solution on the mesh with the desired level of accuracy and it can also be used for reconstructing the moments for the RKDG and hybrid RKDG+HWENO schemes when that is needed.

## Tables

Table 1: Advection equation $u_t + u_x = 0$ with $u(x, 0) = \sin(2 \pi x)$ profile. The problem was run on the domain [-0.5, 0.5] with periodic boundaries and stopped at a time of 0.5. Comparing DG with and without limiter. The MP detection algorithm with the settings for Euler equations was used.

| | | DG or DG+HWENO With WENO Limiter | | | | DG or DG+HWENO Without Limiter | | | |
|---|---|---|---|---|---|---|---|---|---|
| | N | $L_1$ err | order | $L_\infty$ err | order | $L_1$ err | order | $L_\infty$ err | order |
| | 10 | 1.51e-2 | | 2.34e-2 | | 1.51e-2 | | 2.34e-2 | |
| $P^1$- | 20 | 3.29e-3 | 2.20 | 5.19e-3 | 2.17 | 3.29e-3 | 2.20 | 5.19e-3 | 2.17 |
| RKDG | 40 | 7.76e-4 | 2.08 | 1.22e-3 | 2.09 | 7.76e-4 | 2.08 | 1.22e-3 | 2.09 |
| | 80 | 1.89e-4 | 2.04 | 2.97e-4 | 2.04 | 1.89e-4 | 2.04 | 2.97e-4 | 2.04 |
| | 10 | 1.58e-4 | | 2.44e-4 | | 1.58e-4 | | 2.44e-4 | |
| $P^2$- | 20 | 1.79e-5 | 3.14 | 2.80e-5 | 3.13 | 1.79e-5 | 3.14 | 2.80e-5 | 3.13 |
| RKDG | 40 | 2.16e-6 | 3.05 | 3.40e-6 | 3.04 | 2.16e-6 | 3.05 | 3.40e-6 | 3.04 |
| | 80 | 2.68e-7 | 3.01 | 4.22e-7 | 3.01 | 2.68e-6 | 3.01 | 4.22e-7 | 3.01 |
| | 10 | 8.97e-7 | | 1.38e-6 | | 8.97e-7 | | 1.38e-6 | |
| | 20 | 4.58e-8 | 4.29 | 7.16e-8 | 4.27 | 4.58e-8 | 4.29 | 7.16e-8 | 4.27 |
| $P^3$- | 40 | 2.93e-9 | 3.97 | 4.59e-9 | 3.96 | 2.93e-9 | 3.97 | 4.59e-9 | 3.96 |
| RKDG | 80 | 1.83e-10 | 3.99 | 2.88e-10 | 3.99 | 1.83e-10 | 3.99 | 2.88e-10 | 3.99 |
| $P^2$- | 10 | 1.06e-3 | | 2.07e-3 | | 1.06e-3 | | 2.07e-3 | |
| RKDG | 20 | 3.51e-5 | 4.92 | 6.54e-5 | 4.98 | 3.51e-5 | 4.92 | 6.54e-5 | 4.98 |
| HWEN | 40 | 2.58e-6 | 3.76 | 4.09e-6 | 4.00 | 2.58e-6 | 3.76 | 4.09e-6 | 4.00 |
| O | 80 | 2.65e-7 | 3.29 | 4.17e-7 | 3.29 | 2.65e-7 | 3.29 | 4.17e-7 | 3.29 |
| $P^3$- | 10 | 2.10e-4 | | 4.37e-4 | | 2.10e-4 | | 4.37e-4 | |
| RKDG | 20 | 2.61e-6 | 6.33 | 6.09e-6 | 6.17 | 2.61e-6 | 6.33 | 6.09e-6 | 6.17 |
| HWEN | 40 | 4.85e-8 | 5.75 | 1.48e-7 | 5.35 | 4.85e-8 | 5.75 | 1.48e-7 | 5.35 |
| O | 80 | 1.13e-9 | 5.41 | 4.04e-9 | 5.20 | 1.13e-9 | 5.41 | 4.04e-9 | 5.20 |

Table 2: Advection equation $u_t + u_x = 0$ with $u(x, 0) = \sin^4(2\pi x)$ profile. The problem was run on the domain [-0.5, 0.5] with periodic boundaries and stopped at a time of 0.5. Comparing DG with and without limiter. The MP detection algorithm with the settings for Euler equations was used.

| | | DG or DG+HWENO With WENO Limiter | | | | DG or DG+HWENO Without Limiter | | | |
|---|---|---|---|---|---|---|---|---|---|
| | N | $L_1$ err | order | $L_\infty$ err | order | $L_1$ err | order | $L_\infty$ err | order |
| | 40 | 9.39e-3 | | 1.73e-2 | | 7.17e-3 | | 1.43e-2 | |
| $P^1$- | 80 | 1.67e-3 | 2.48 | 3.49e-3 | 2.31 | 1.67e-3 | 2.48 | 3.49e-3 | 2.31 |
| RKDG | 160 | 4.02e-4 | 2.06 | 8.09e-4 | 2.11 | 4.02e-4 | 2.06 | 8.09e-4 | 2.11 |
| | 320 | 9.91e-5 | 2.02 | 2.00e-4 | 2.01 | 9.91e-5 | 2.02 | 2.00e-4 | 2.01 |
| | 40 | 5.05e-4 | | 2.44e-3 | | 8.11e-5 | | 1.52e-4 | |
| $P^2$- | 80 | 3.45e-5 | 3.87 | 2.35e-4 | 3.38 | 9.08e-6 | 3.16 | 1.74e-5 | 3.13 |
| RKDG | 160 | 1.88e-6 | 4.19 | 1.71e-5 | 3.78 | 1.10e-6 | 3.04 | 2.13e-6 | 3.03 |
| | 320 | 1.51e-7 | 3.65 | 1.02e-6 | 4.06 | 1.37e-7 | 3.01 | 2.64e-7 | 3.01 |
| | 40 | 2.14e-4 | | 1.44e-3 | | 3.65e-7 | | 6.19e-7 | |
| | 80 | 1.07e-5 | 4.32 | 1.56e-4 | 3.21 | 2.34e-8 | 3.96 | 3.99e-8 | 3.96 |
| $P^3$- | 160 | 4.76e-7 | 4.49 | 1.09e-5 | 3.84 | 1.47e-9 | 3.99 | 2.51e-9 | 3.99 |
| RKDG | 320 | 9.23e-11 | 12.33 | 1.57e-10 | 16.07 | 9.23e-11 | 4.00 | 1.57e-10 | 4.00 |
| $P^2$- | 40 | 5.14e-4 | | 1.22e-3 | | 4.89e-4 | | 1.05e-3 | |
| RKDG | 80 | 3.41e-5 | 3.91 | 1.21e-4 | 3.34 | 3.32e-5 | 3.89 | 1.34e-4 | 2.97 |
| HWENO | 160 | 2.42e-6 | 3.82 | 1.36e-5 | 3.15 | 2.39e-6 | 3.79 | 1.49e-5 | 3.16 |
| | 320 | 1.96e-7 | 3.62 | 1.46e-6 | 3.22 | 1.97e-7 | 3.60 | 1.57e-6 | 3.24 |
| $P^3$- | 40 | 3.04e-4 | | 1.47e-3 | | 1.84e | | 7.35e-4 | |
| RKDG | 80 | 1.45e-5 | 4.39 | 1.41e-4 | 3.38 | 1.00e-5 | 4.19 | 6.81e-5 | 3.43 |
| HWENO | 160 | 6.99e-7 | 4.38 | 9.12e-6 | 3.95 | 5.32e-7 | 4.24 | 4.99e-6 | 3.77 |
| | 320 | 2.79e-8 | 4.65 | 5.71e-7 | 3.99 | 2.36e-8 | 4.49 | 4.95e-7 | 3.33 |

Table 3: Burgers equation $u_t + u\, u_x = 0$ with $u(x, 0) = 0.25 + 0.5 \sin(2 \pi x)$ profile. The problem was run on the domain [-0.5, 0.5] with periodic boundaries and stopped at a time of $0.5/\pi$ . Comparing DG with and without limiter. The MP detection algorithm with the settings for Euler equations was used.

| | | DG or DG+HWENO With WENO Limiter | | | | DG or DG+HWENO Without Limiter | | | |
|---|---|---|---|---|---|---|---|---|---|
| | N | $L_1$ err | order | $L_\infty$ err | order | $L_1$ err | order | $L_\infty$ err | order |
| | 10 | 7.40e-4 | | 4.08e-3 | | 4.34e-4 | | 1.82e-3 | |
| $P^1$- | 20 | 9.17e-5 | 3.01 | 4.42e-4 | 3.20 | 9.17e-5 | 2.24 | 4.42e-4 | 2.04 |
| RKDG | 40 | 1.96e-5 | 2.22 | 1.12e-4 | 1.98 | 1.96e-5 | 2.22 | 1.12e-4 | 1.98 |
| | 80 | 4.45e-6 | 2.14 | 2.90e-5 | 1.95 | 4.45e-6 | 2.14 | 2.90e-5 | 1.95 |
| | 10 | 1.82e-5 | | 1.26e-4 | | 1.82e-5 | | 1.26e-4 | |
| $P^2$- | 20 | 1.44e-6 | 3.65 | 8.57e-6 | 3.88 | 1.44e-6 | 3.65 | 8.57e-6 | 3.88 |
| RKDG | 40 | 1.10e-7 | 3.72 | 6.28e-7 | 3.77 | 1.10e-7 | 3.72 | 6.28e-7 | 3.77 |
| | 80 | 1.00e-8 | 3.46 | 6.09e-8 | 3.37 | 1.00e-8 | 3.46 | 6.09e-8 | 3.37 |
| | 10 | 4.98e-7 | | 4.05e-6 | | 4.98e-7 | | 4.05e-6 | |
| | 20 | 2.83e-8 | 4.14 | 3.45e-7 | 3.56 | 2.83e-8 | 4.14 | 3.45e-7 | 3.56 |
| $P^3$- | 40 | 1.10e-9 | 4.68 | 2.96e-8 | 3.54 | 1.10e-9 | 4.68 | 2.96e-8 | 3.54 |
| RKDG | 80 | 3.35e-11 | 5.04 | 1.64e-9 | 4.17 | 3.35e-11 | 5.04 | 1.64e-9 | 4.17 |
| $P^2$- | 10 | 3.25e-5 | | 1.81e-4 | | 3.25e-5 | | 1.81e-4 | |
| RKDG | 20 | 3.23e-6 | 3.33 | 3.09e-5 | 2.55 | 3.23e-6 | 3.33 | 3.09e-5 | 2.55 |
| HWEN | 40 | 2.64e-7 | 3.61 | 2.49e-6 | 3.63 | 2.64e-7 | 3.61 | 2.49e-6 | 3.63 |
| O | 80 | 2.17e-8 | 3.61 | 1.90e-7 | 3.71 | 2.17e-8 | 3.61 | 1.90e-7 | 3.71 |
| $P^3$- | 10 | 1.08e-5 | | 9.98e-5 | | 1.08e-5 | | 9.98e-5 | |
| RKDG | 20 | 2.78e-7 | 5.29 | 3.49e-6 | 4.84 | 2.78e-7 | 5.29 | 3.49e-6 | 4.84 |
| HWEN | 40 | 4.58e-9 | 5.92 | 5.26e-8 | 6.05 | 4.58e-9 | 5.92 | 5.26e-8 | 6.05 |
| O | 80 | 2.25e-10 | 4.35 | 3.39e-9 | 3.96 | 2.25e-10 | 4.35 | 3.39e-9 | 3.96 |

Table 4: Euler equations with density profile $\rho(x, 0) = 1.0 + 0.25 \sin(2\pi x)$ and pressure and velocity set to unity. The problem was run on the domain [-0.5, 0.5] with periodic boundaries and stopped at a time of 1.0. Comparing DG with and without limiter. The MP detection algorithm with the settings for Euler equations was used.

| | | DG or DG+HWENO With WENO Limiter | | | | DG or DG+HWENO Without Limiter | | | |
|---|---|---|---|---|---|---|---|---|---|
| | N | $L_1$ err | order | $L_\infty$ err | order | $L_1$ err | order | $L_\infty$ err | order |
| | 10 | 1.84e-3 | | 3.11e-3 | | 1.84e-3 | | 3.11e-3 | |
| $P^1$- | 20 | 3.25e-4 | 2.50 | 5.23e-4 | 2.57 | 3.25e-4 | 2.50 | 5.23e-4 | 2.57 |
| RKDG | 40 | 7.17e-5 | 2.18 | 1.17e-4 | 2.16 | 7.17e-5 | 2.18 | 1.17e-4 | 2.16 |
| | 80 | 1.71e-5 | 2.07 | 2.75e-5 | 2.09 | 1.71e-5 | 2.07 | 2.75e-5 | 2.09 |
| | 10 | 2.71e-5 | | 4.72e-5 | | 2.71e-5 | | 4.72e-5 | |
| $P^2$- | 20 | 1.46e-6 | 4.21 | 2.65e-6 | 4.15 | 1.46e-6 | 4.21 | 2.65e-6 | 4.15 |
| RKDG | 40 | 1.08e-7 | 3.76 | 1.84e-7 | 3.85 | 1.08e-7 | 3.76 | 1.84e-7 | 3.85 |
| | 80 | 1.10e-8 | 3.30 | 1.78e-8 | 3.37 | 1.10e-8 | 3.30 | 1.78e-8 | 3.37 |
| | 10 | 4.97e-8 | | 9.03e-8 | | 4.97e-8 | | 9.03e-8 | |
| $P^3$- | 20 | 5.68e-10 | 6.45 | 9.57e-10 | 6.56 | 5.68e-10 | 6.45 | 9.57e-10 | 6.56 |
| RKDG | 40 | 4.66e-11 | 3.61 | 7.35e-11 | 3.70 | 4.66e-11 | 3.61 | 7.35e-11 | 3.70 |
| | 80 | 2.93e-12 | 3.99 | 4.70e-12 | 3.97 | 2.93e-12 | 3.99 | 4.70e-12 | 3.97 |
| $P^2$- | 10 | 5.43e-4 | | 8.12e-4 | | 5.43e-4 | | 8.12e-4 | |
| RKDG | 20 | 1.47e-5 | 5.20 | 2.37e-5 | 5.10 | 1.47e-5 | 5.20 | 2.37e-5 | 5.10 |
| HWEN | 40 | 8.69e-7 | 4.09 | 1.34e-6 | 4.15 | 8.69e-7 | 4.09 | 1.34e-6 | 4.15 |
| O | 80 | 5.32e-8 | 4.03 | 8.37e-8 | 4.00 | 5.32e-8 | 4.03 | 8.37e-8 | 4.00 |
| $P^3$- | 10 | 1.41e-4 | | 2.35e-4 | | 1.41e-4 | | 2.35e-4 | |
| RKDG | 20 | 1.20e-6 | 6.87 | 3.34e-6 | 6.13 | 1.20e-6 | 6.87 | 3.34e-6 | 6.13 |
| HWEN | 40 | 1.78e-8 | 6.07 | 8.10e-8 | 5.37 | 1.78e-8 | 6.07 | 8.10e-8 | 5.37 |
| O | 80 | 3.23e-10 | 5.78 | 2.09e-9 | 5.28 | 3.23e-10 | 5.78 | 2.09e-9 | 5.28 |

Table 5: Euler equations for the multi-dimensional vortex problem catalogued in Sub-section 4.e. The problem was run on the domain [-5, 5]X[-5, 5] with periodic boundaries and stopped at a time of 10.0. Comparing DG with and without limiter. The MP detection algorithm with the settings for Euler equations was used. In Sub-section 2.c we showed that the cross-terms can be limited using either a MinMod limiter or one that is based on constructing smoothness measures in a WENO-like fashion. Results from both limiting strategies are shown below.

| | | DG With WENO Limiter | | | | DG Without Limiter | | | |
|---|---|---|---|---|---|---|---|---|---|
| | N | $L_1$ err | order | $L_\infty$ err | order | $L_1$ err | order | $L_\infty$ err | order |
| | 10X10 | 1.70e-0 | | 3.47e-1 | | 2.97e-1 | | 5.43e-2 | |
| $P^2$- | 20X20 | 2.65e-1 | 2.68 | 3.69e-2 | 3.24 | 4.09e-2 | 2.86 | 1.22e-2 | 2.16 |
| RKDG | 40X40 | 2.72e-2 | 3.30 | 4.29e-3 | 3.10 | 3.19e-3 | 3.68 | 7.90e-4 | 3.95 |
| (minmod) | 80X80 | 1.29e-3 | 4.40 | 3.04e-4 | 3.82 | 3.46e-4 | 3.20 | 9.29e-5 | 3.09 |
| | 160X160 | 8.11e-5 | 3.99 | 2.77e-5 | 3.46 | 5.24e-5 | 2.72 | 1.41e-5 | 2.72 |
| | 10X10 | 1.58e-0 | | 3.44e-1 | | 1.17e-1 | | 4.55e-2 | |
| | 20X20 | 2.06e-1 | 2.94 | 6.45e-2 | 2.42 | 5.07e-3 | 4.53 | 2.32e-3 | 4.30 |
| $P^3$- | 40X40 | 1.39e-2 | 3.88 | 1.97e-3 | 5.03 | 1.89e-4 | 4.75 | 5.23e-5 | 5.47 |
| RKDG | 80X80 | 1.54e-4 | 6.49 | 6.06e-5 | 5.02 | 7.83e-6 | 4.59 | 3.27e-6 | 4.00 |
| (minmod) | 160X160 | 1.27e-6 | 6.92 | 1.40e-6 | 5.44 | 3.71e-7 | 4.40 | 1.70e-7 | 4.27 |
| $P^2$- | 10X10 | 1.60e-0 | | 3.37e-1 | | 2.97e-1 | | 5.43e-2 | |
| RKDG | 20X20 | 2.03e-1 | 2.98 | 2.62e-2 | 3.68 | 4.09e-2 | 2.86 | 1.22e-2 | 2.16 |
| (smooth) | 40X40 | 1.56e-2 | 3.70 | 2.95e-3 | 3.15 | 3.19e-3 | 3.68 | 7.90e-4 | 3.95 |
| | 80X80 | 5.88e-4 | 4.73 | 3.13e-4 | 3.23 | 3.46e-4 | 3.20 | 9.29e-5 | 3.09 |
| | 160X160 | 5.81e-5 | 3.34 | 1.88e-5 | 4.06 | 5.24e-5 | 2.72 | 1.41e-5 | 2.72 |
| $P^3$- | 10X10 | 1.48e-0 | | 3.26e-1 | | 1.17e-1 | | 4.55e-2 | |
| RKDG | 20X20 | 1.59e-1 | 3.23 | 4.69e-2 | 2.80 | 5.07e-3 | 4.53 | 2.32e-3 | 4.30 |
| (smooth) | 40X40 | 1.19e-2 | 3.73 | 1.94e-3 | 4.60 | 1.89e-4 | 4.75 | 5.23e-5 | 5.47 |
| | 80X80 | 8.65e-5 | 7.11 | 4.86e-5 | 5.32 | 7.83e-6 | 4.59 | 3.27e-6 | 4.00 |
| | 160X160 | 2.53e-6 | 5.10 | 3.28e-6 | 3.89 | 3.71e-7 | 4.40 | 1.70e-7 | 4.27 |

Table 6: Percentage of zones that were flagged in the final time step of the Lax problem for meshes with increasing number of zones:

| | P=2 RKDG | P=2 RKDG +HWENO | P=3 RKDG | P=3 RKDG +HWENO |
|---|---|---|---|---|
| 200 | 9.16 | 16.83 | 8.75 | 12.13 |
| 400 | 3.65 | 4.16 | 3.38 | 9.62 |
| 800 | 2.66 | 2.13 | 1.81 | 7.59 |

Table 7: Percentage of zones that were flagged in the final time step of the shock-entropy wave interaction problem for meshes with increasing number of zones:

| | P=2 RKDG | P=2 RKDG +HWENO | P=3 RKDG | P=3 RKDG +HWENO |
|---|---|---|---|---|
| 200 | 28.33 | 26.33 | 33.00 | 25.75 |
| 400 | 13.75 | 13.00 | 12.25 | 14.44 |
| 800 | 4.99 | 7.08 | 4.81 | 8.09 |

Table 8: Percentage of zones that were flagged in the final time step of the interacting blast waves problem for meshes with increasing number of zones:

| | P=2 RKDG | P=2 RKDG +HWENO | P=3 RKDG | P=3 RKDG +HWENO |
|---|---|---|---|---|
| 200 | 8.33 | 7.16 | 13.00 | 13.50 |
| 400 | 4.00 | 4.33 | 6.08 | 5.75 |
| 800 | 2.16 | 2.92 | 3.54 | 5.50 |

## Figure Captions

Fig. 1 shows the nine zones around zone (i,j) that are used to carry out spatially third order accurate reconstruction. The maximally compact eight stencils that yield a third order accurate interpolant are also shown. The stencils are labeled $S_1$ through $S_8$ .

Fig. 2 shows the twenty-one zones around zone (i,j) that are used to carry out spatially fourth order accurate reconstruction. The maximally compact eight stencils that yield a fourth order accurate interpolant are also shown. The stencils are labeled $S_1$ through $S_8$ .

Figs 3a, 3b, 3c and 3d show the results for the advection test problem for the p=2 RKDG, p=2 RKDG+HWENO, p=3 RKDG and p=3 RKDG+HWENO schemes respectively. The reference solution is shown as a solid line. The problem was solved on a 200 zone mesh and the diamonds show the solution from that mesh. The reference solution is shown as a solid line.

Figs 4a, 4b, 4c and 4d show the results for the Lax test problem for the p=2 RKDG, p=2 RKDG+HWENO, p=3 RKDG and p=3 RKDG+HWENO schemes respectively. The reference solution is shown as a solid line. The problem was solved on a 200 zone mesh and the diamonds show the solution from that mesh. Figs 4e, 4f, 4g and 4h show the history of flagged points in space-time for the above four schemes.

Figs 5a, 5b, 5c and 5d show the results for the shock-entropy wave test problem for the p=2 RKDG, p=2 RKDG+HWENO, p=3 RKDG and p=3 RKDG+HWENO schemes respectively. The reference solution is shown as a solid line. The problem was solved on a 200 zone mesh and the diamonds show the solution from that mesh. Figs 5e, 5f, 5g and 5h show the history of flagged points in space-time for the above four schemes.

Figs 6a, 6b, 6c and 6d show the results for the blast wave interaction test problem for the p=2 RKDG, p=2 RKDG+HWENO, p=3 RKDG and p=3 RKDG+HWENO schemes respectively. The reference solution is shown as a solid line. The problem was solved on a 400 zone mesh and the diamonds show the solution from that mesh. Figs 6e, 6f, 6g and 6h show the history of flagged points in space-time for the above four schemes.

Figs. 7 shows the density from the simulation of the Mach 3 forward-facing step problem with 240X80 zone resolution at a time of 4.0 for the p=2 RKDG scheme.

Fig. 1

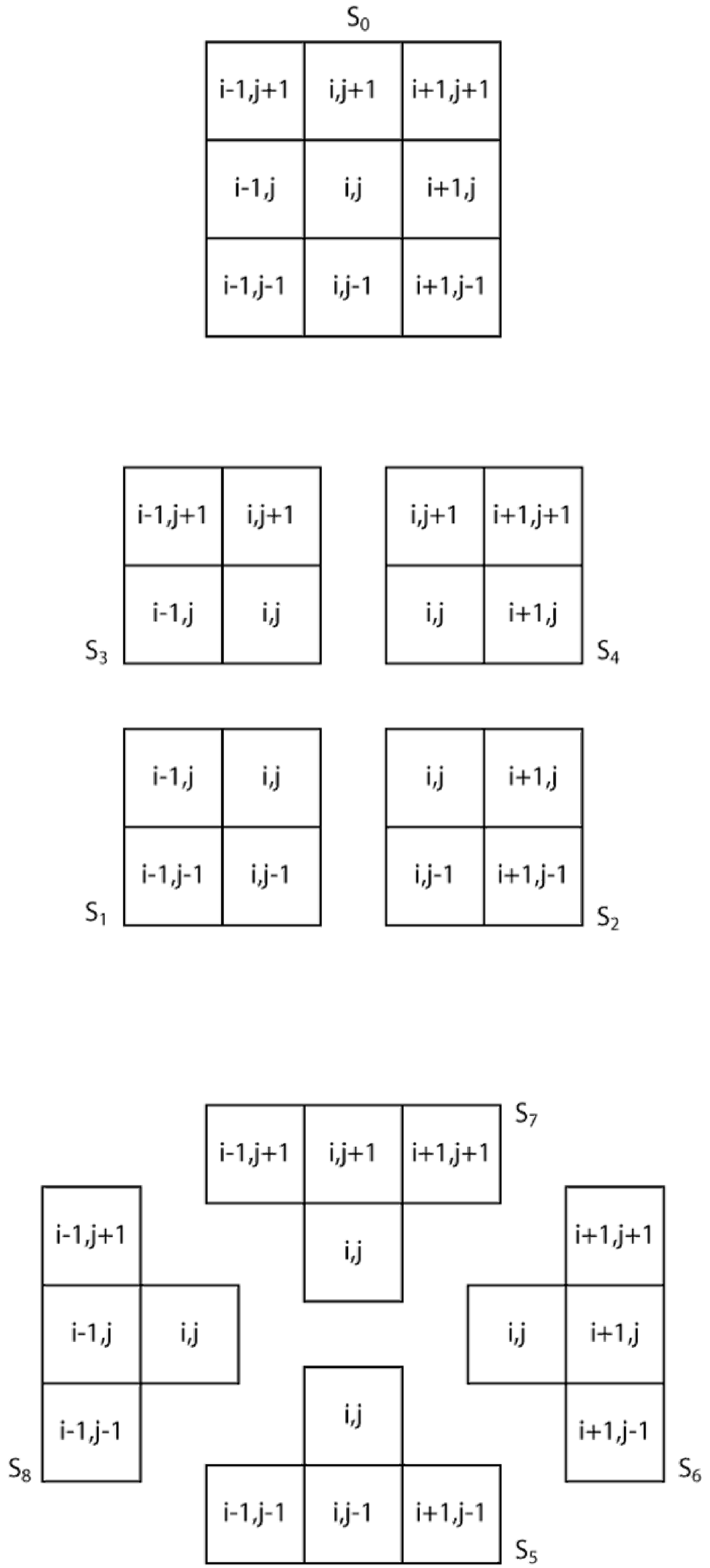

Fig. 2

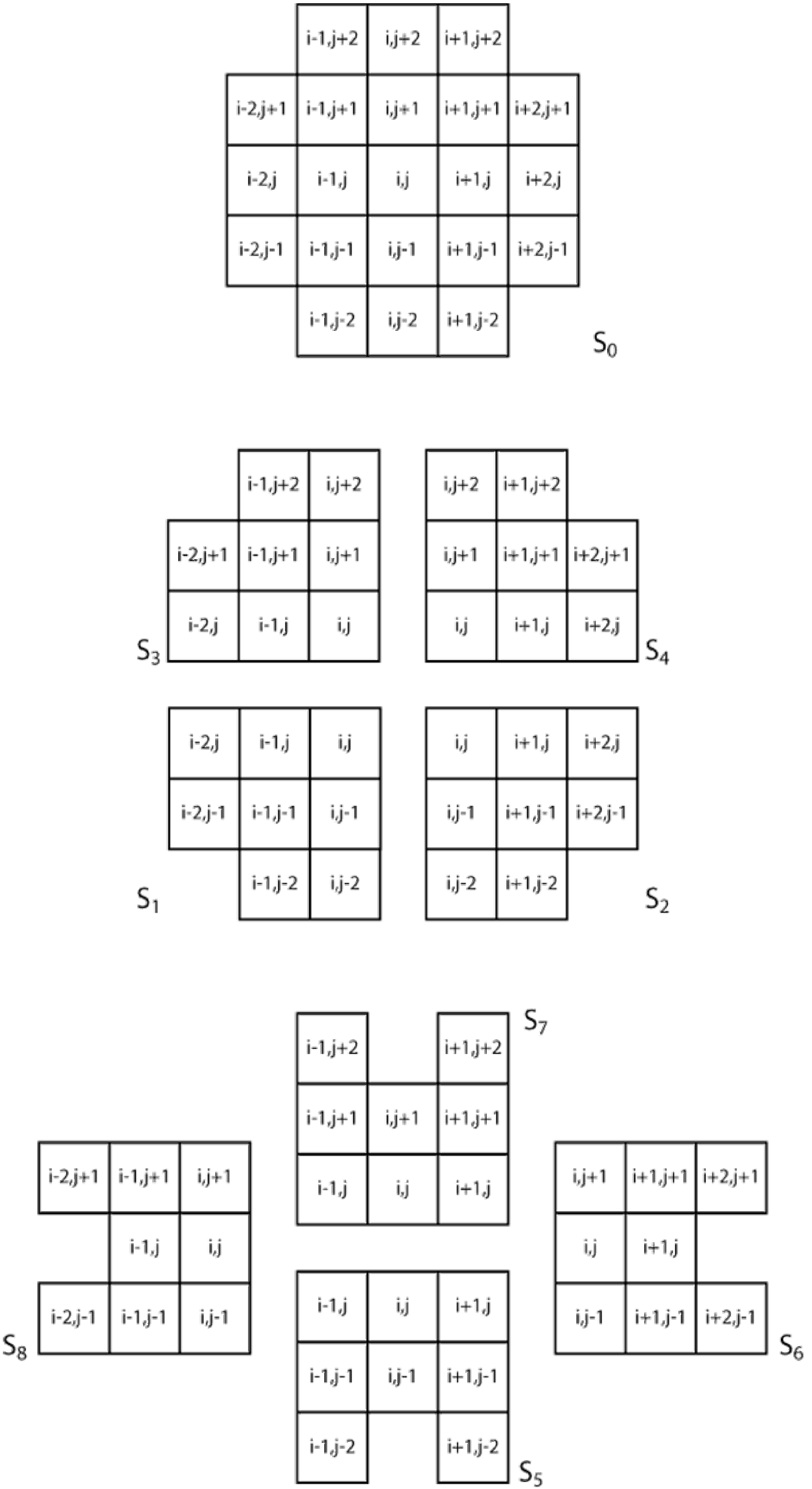

Fig. 3a

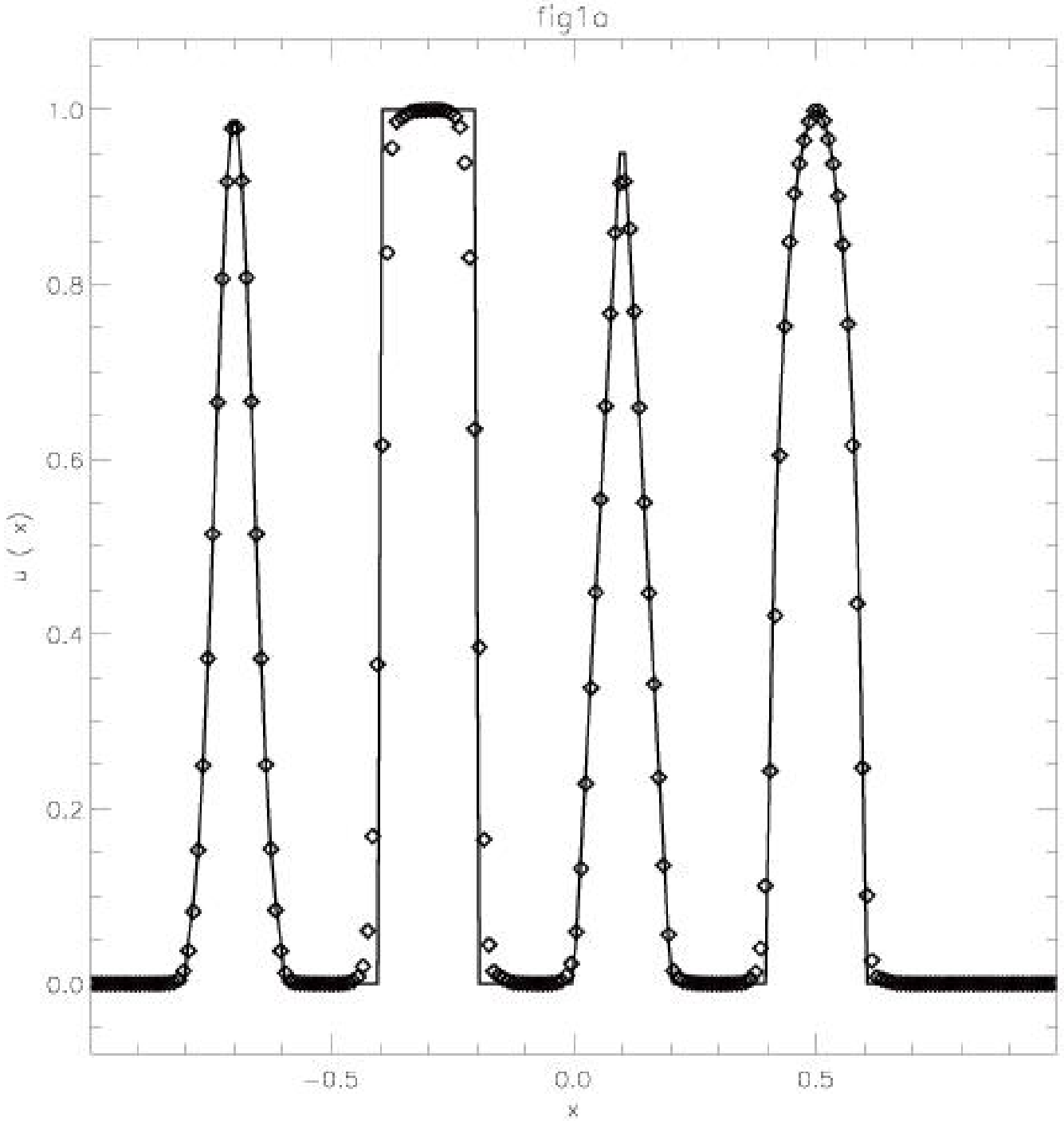

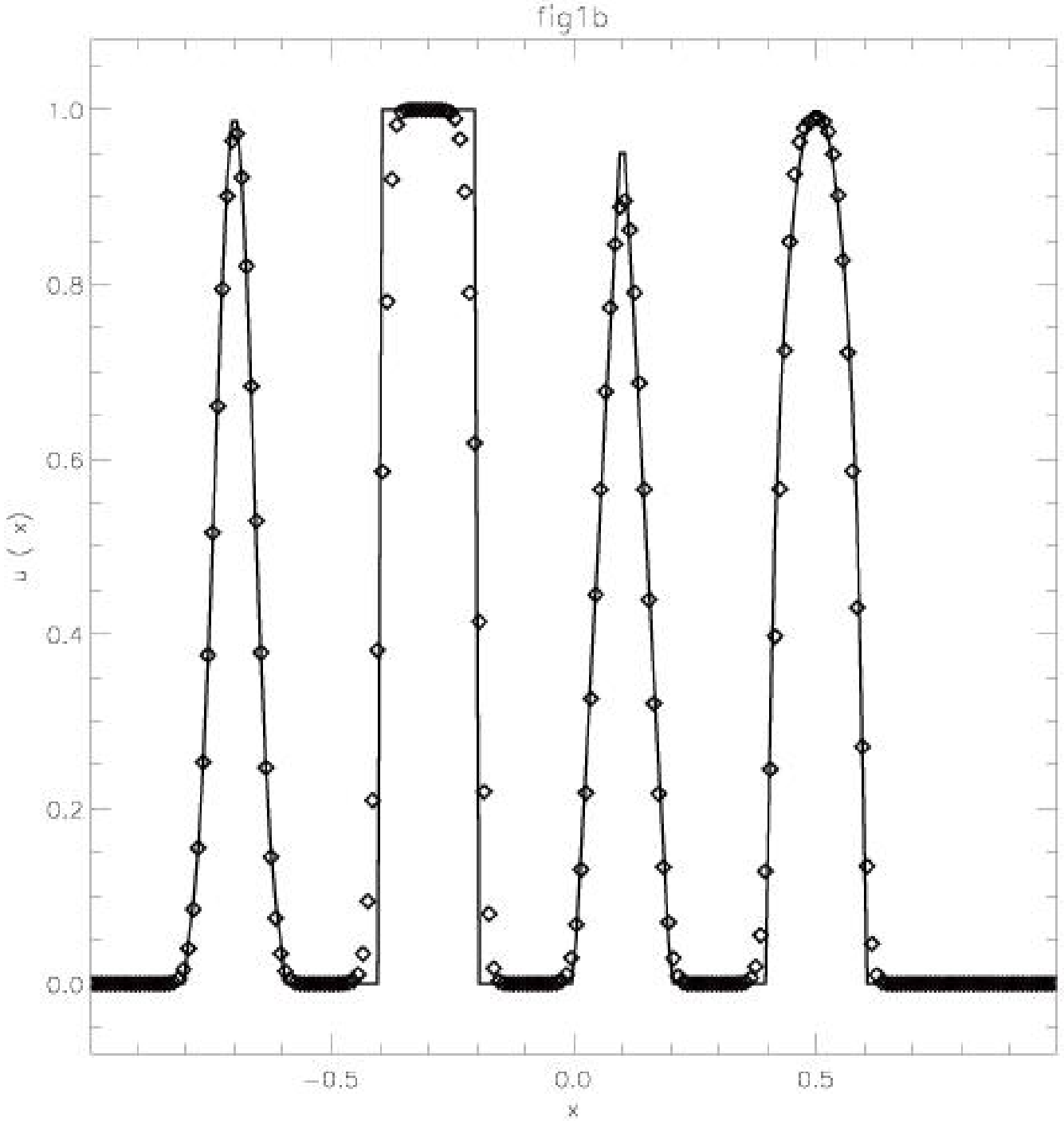

Fig. 3b

Fig. 3c

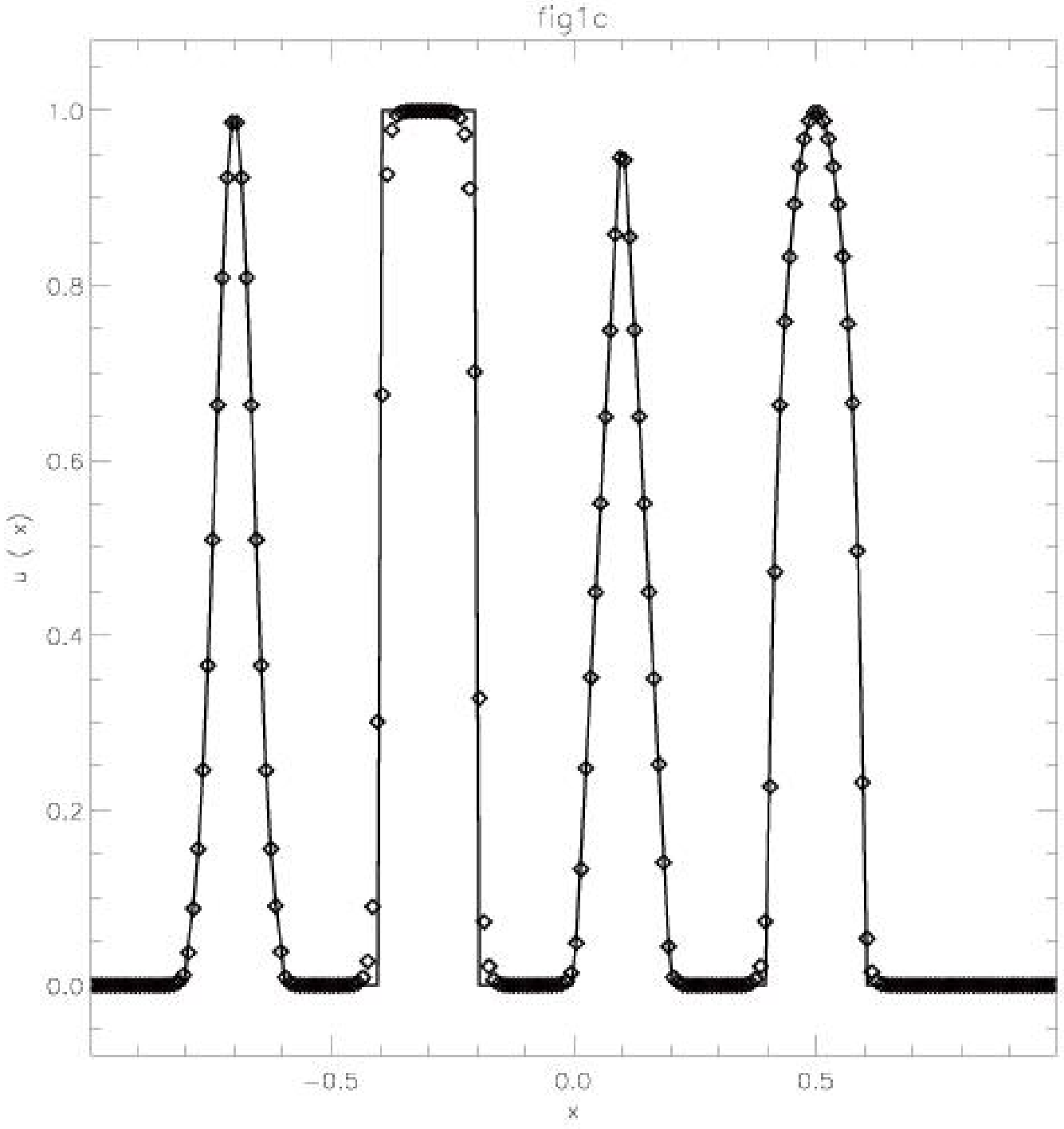

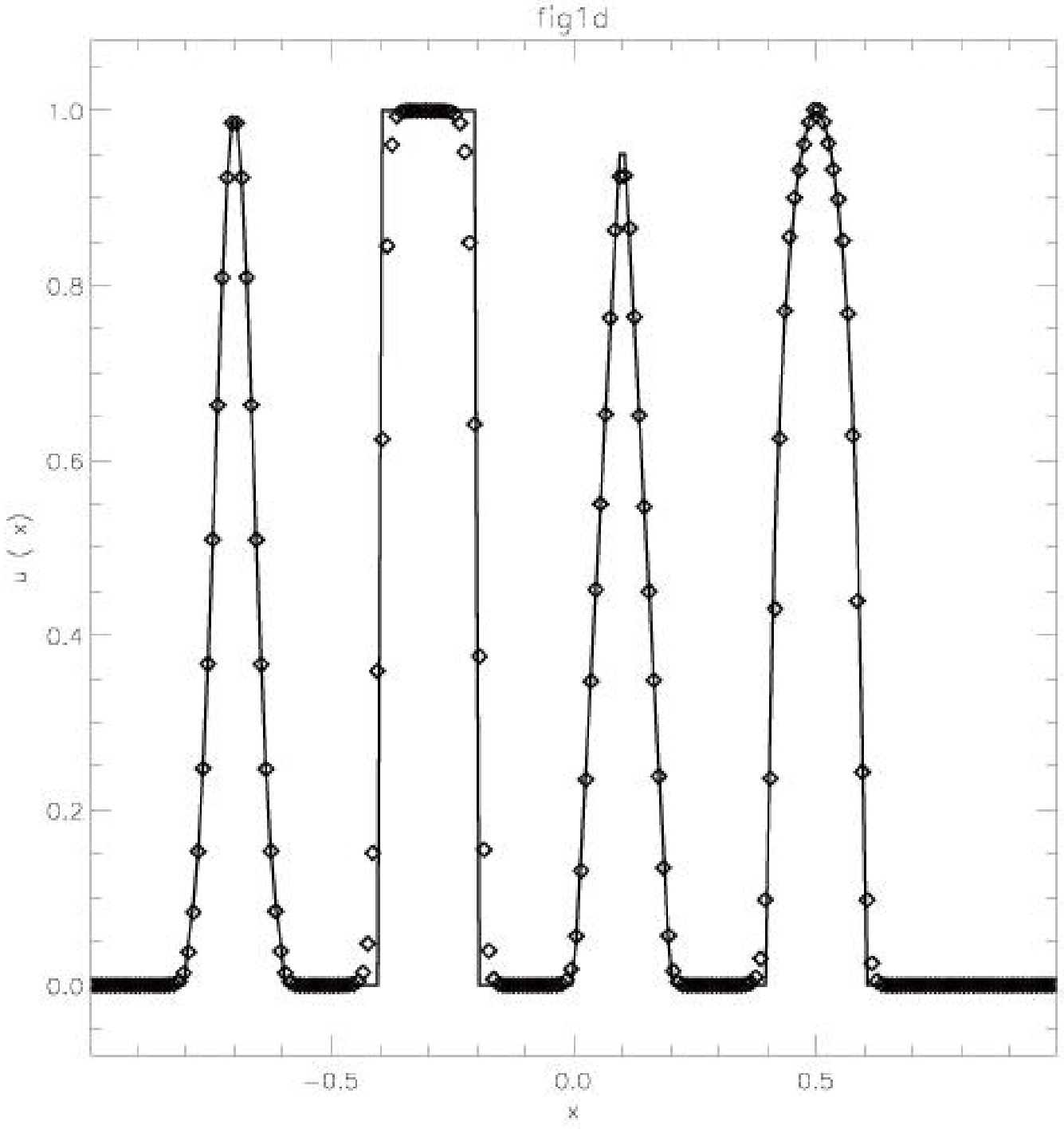

Fig. 3d

Fig. 4a

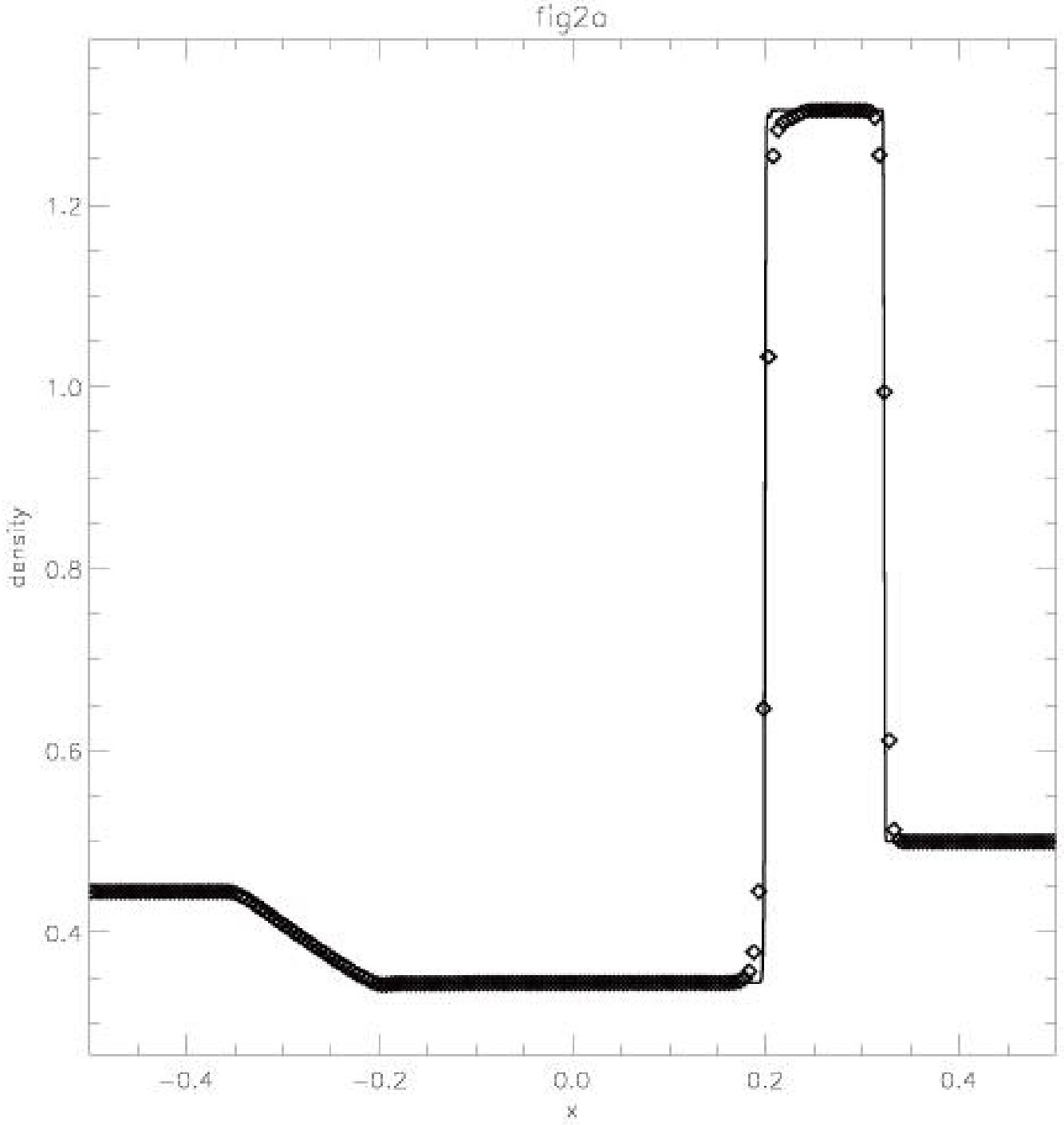

Fig. 4b

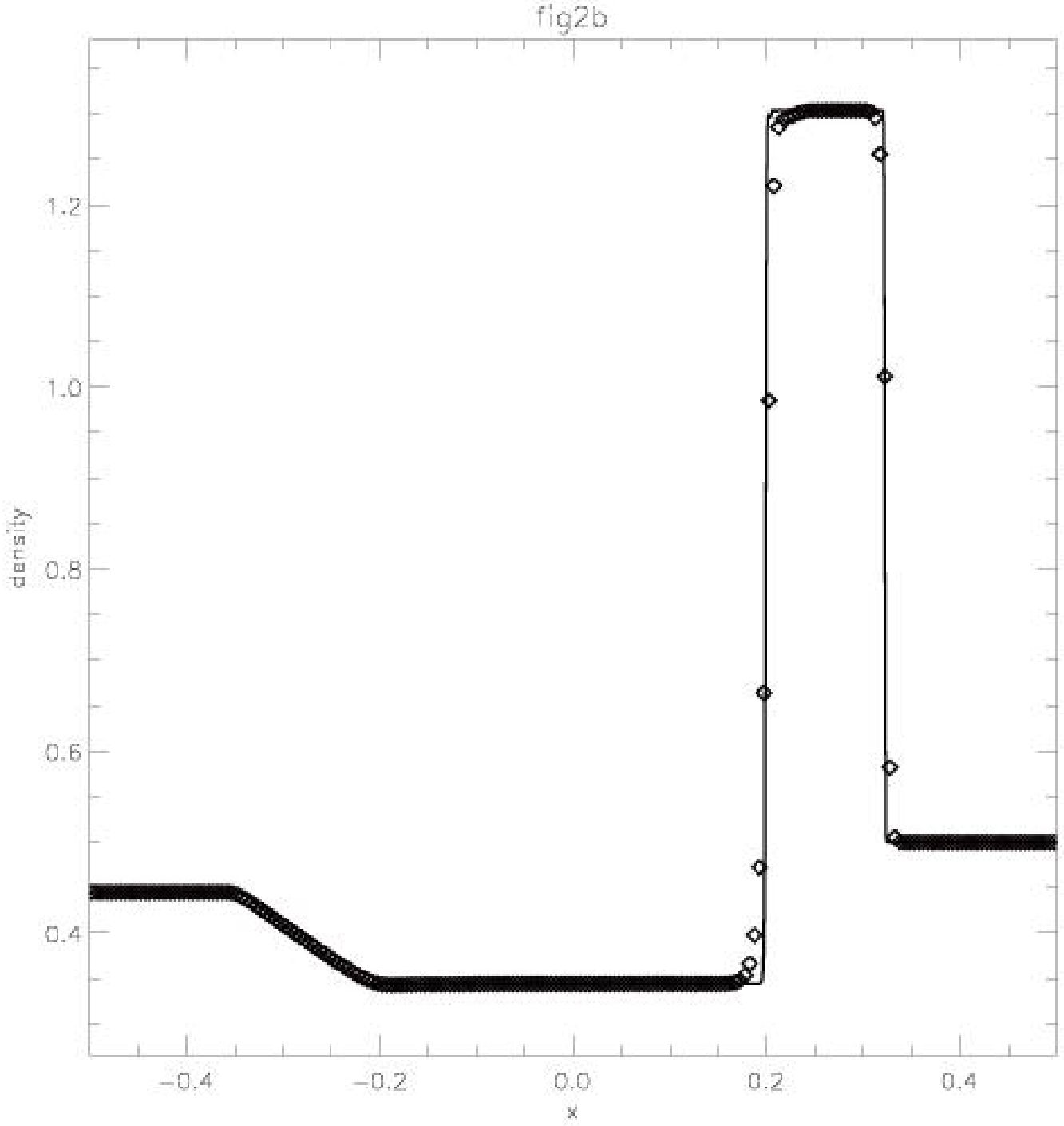

Fig. 4c

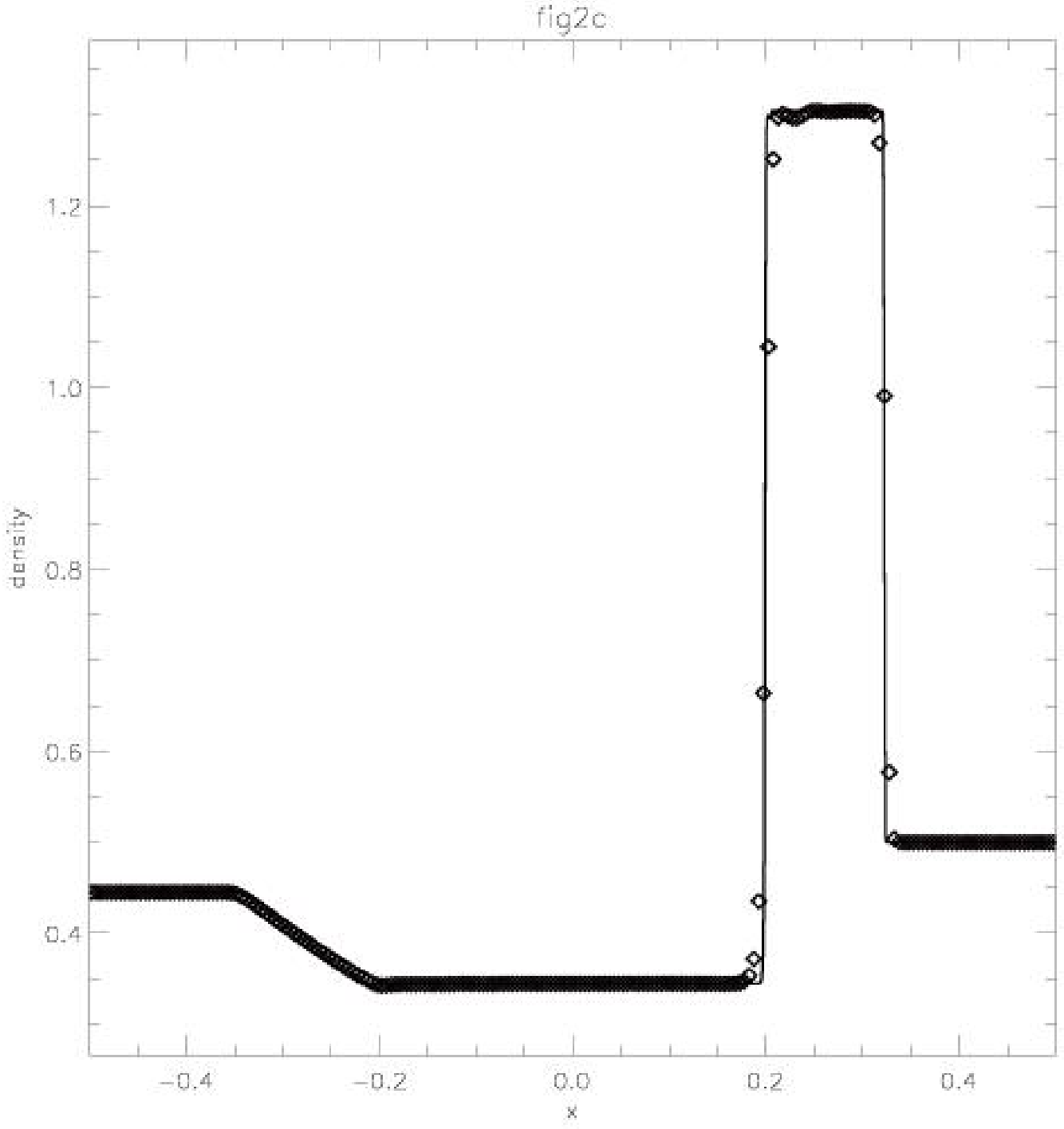

Fig. 4d

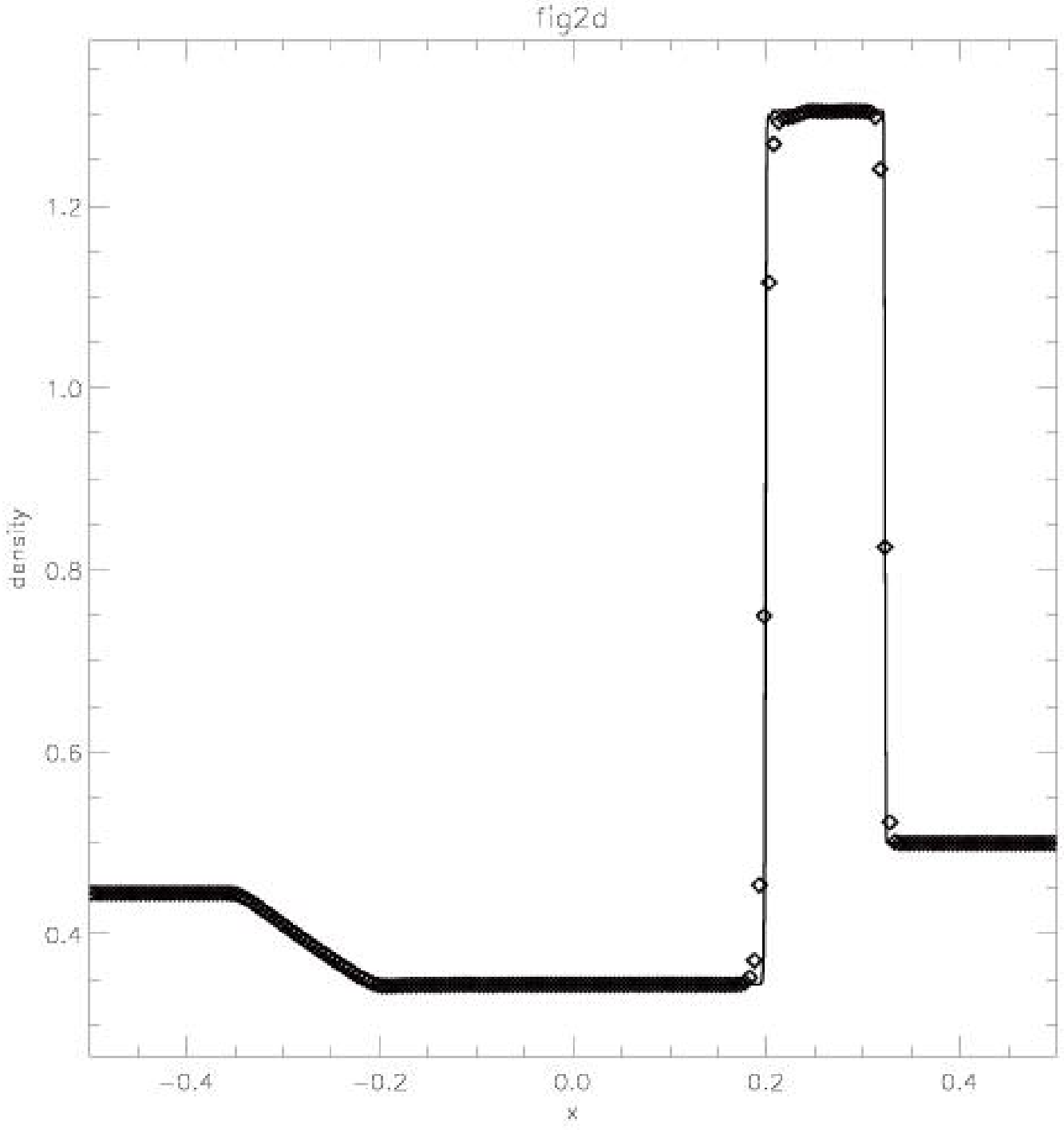

Fig. 4e

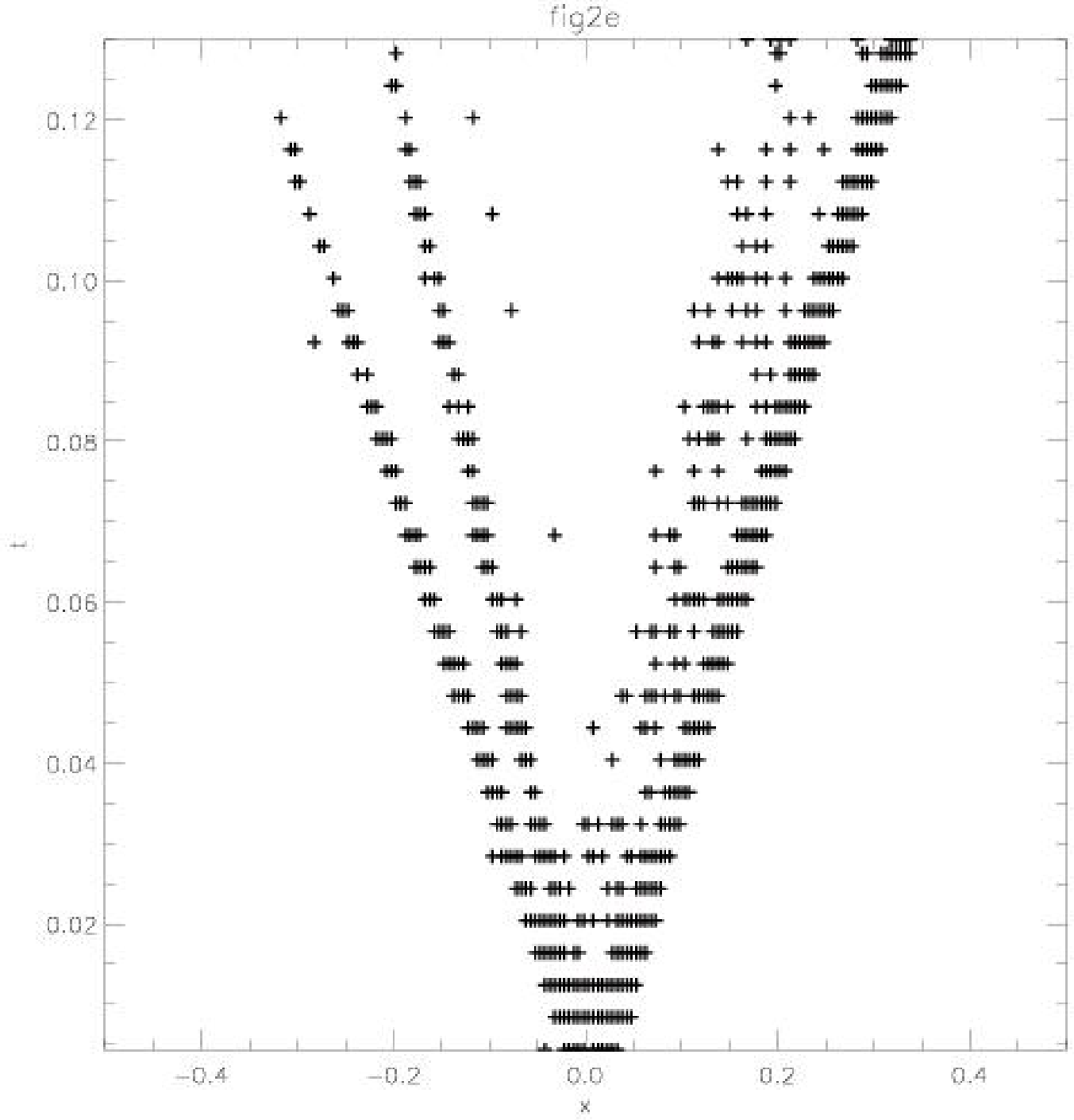

Fig. 4f

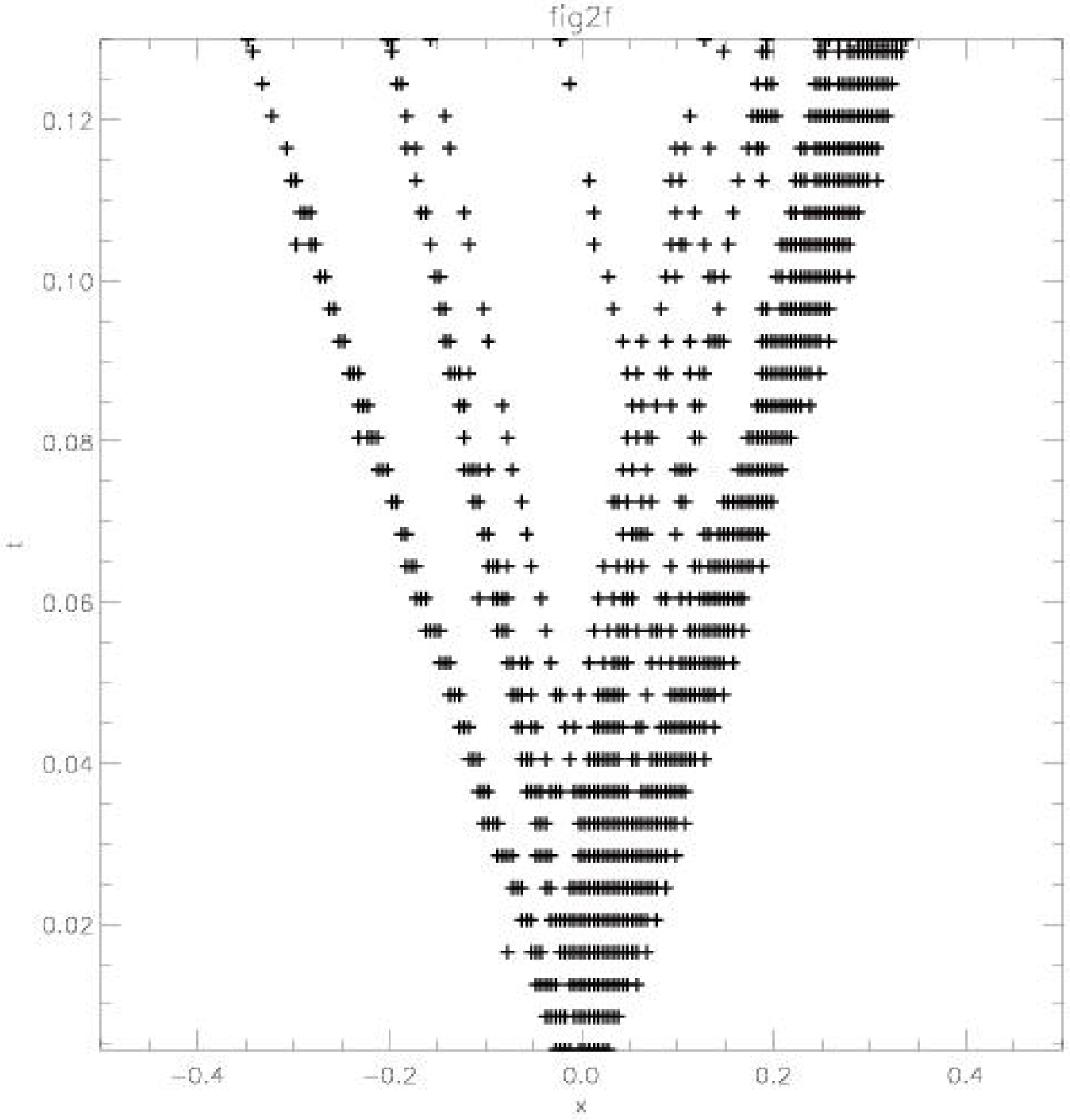

Fig. 4g

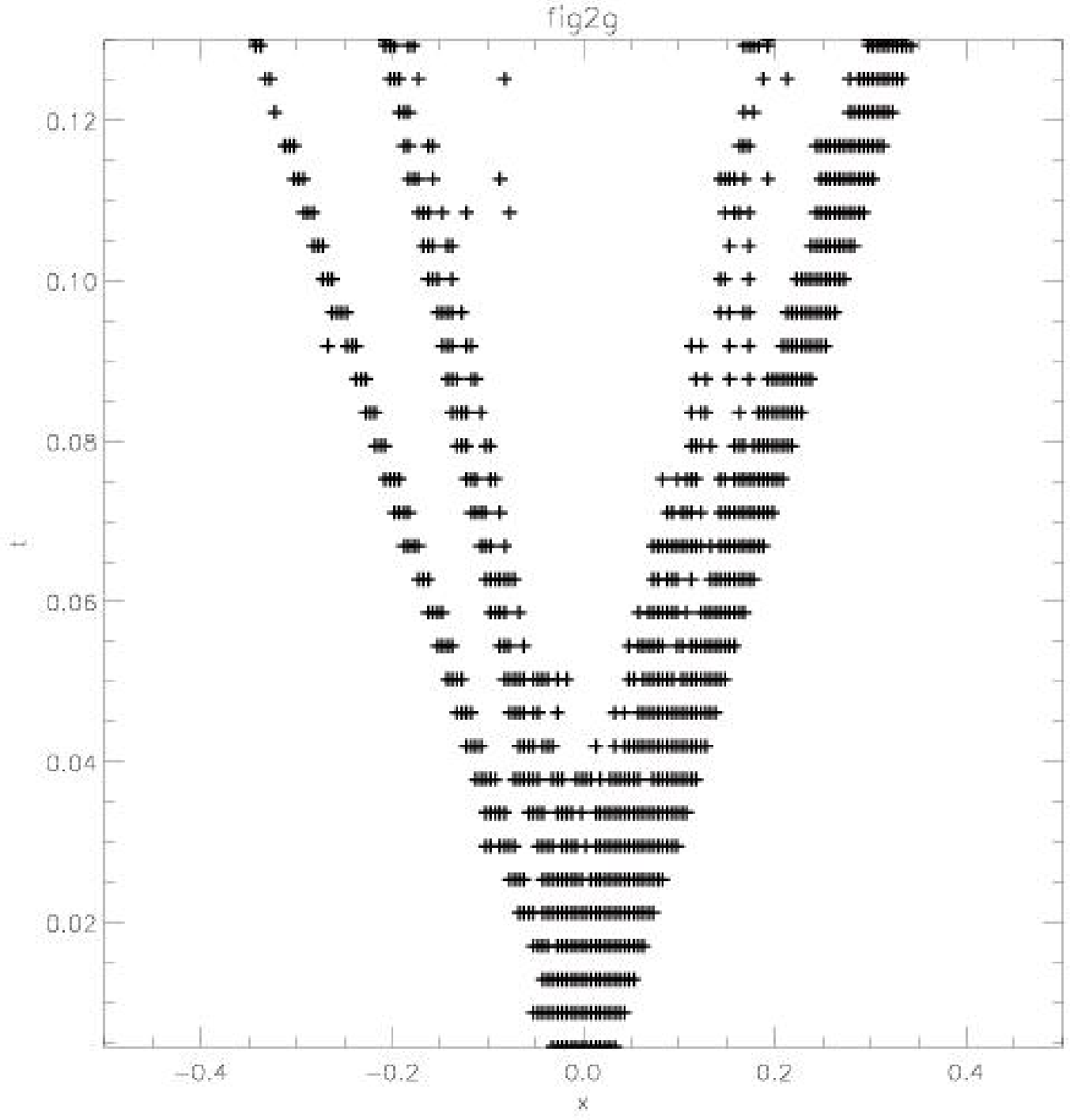

Fig. 4h

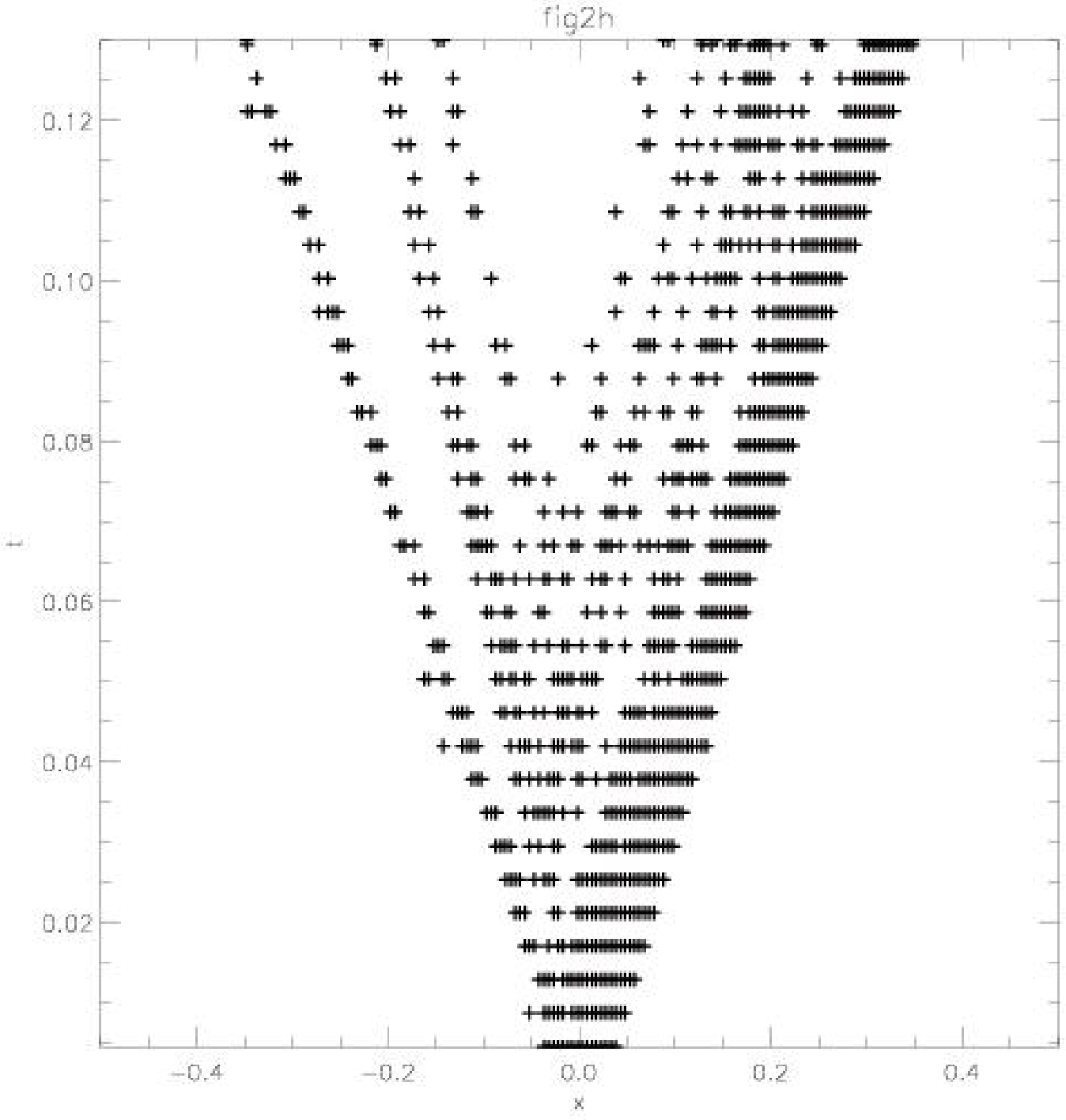

Fig. 5a

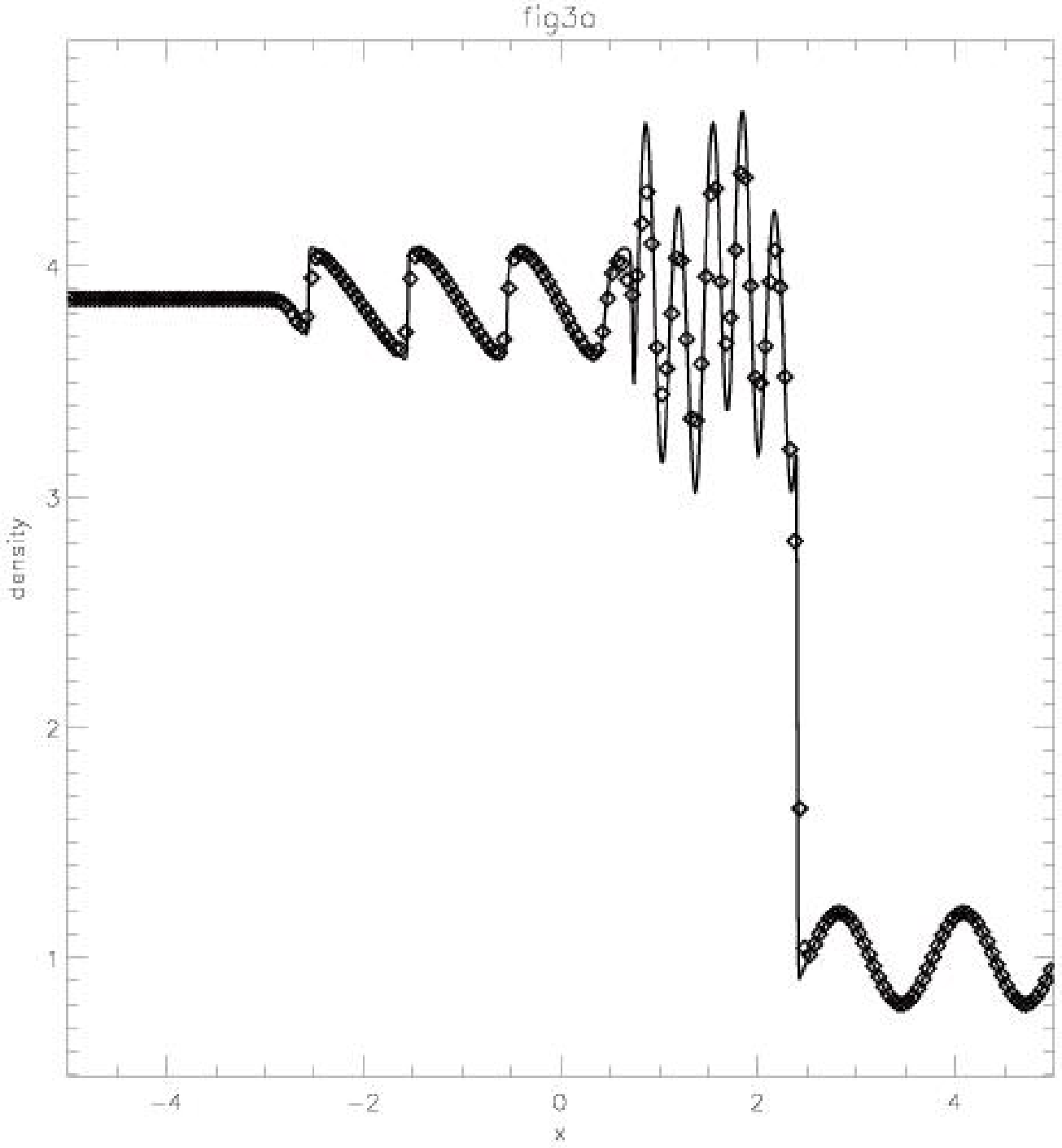

Fig. 5b

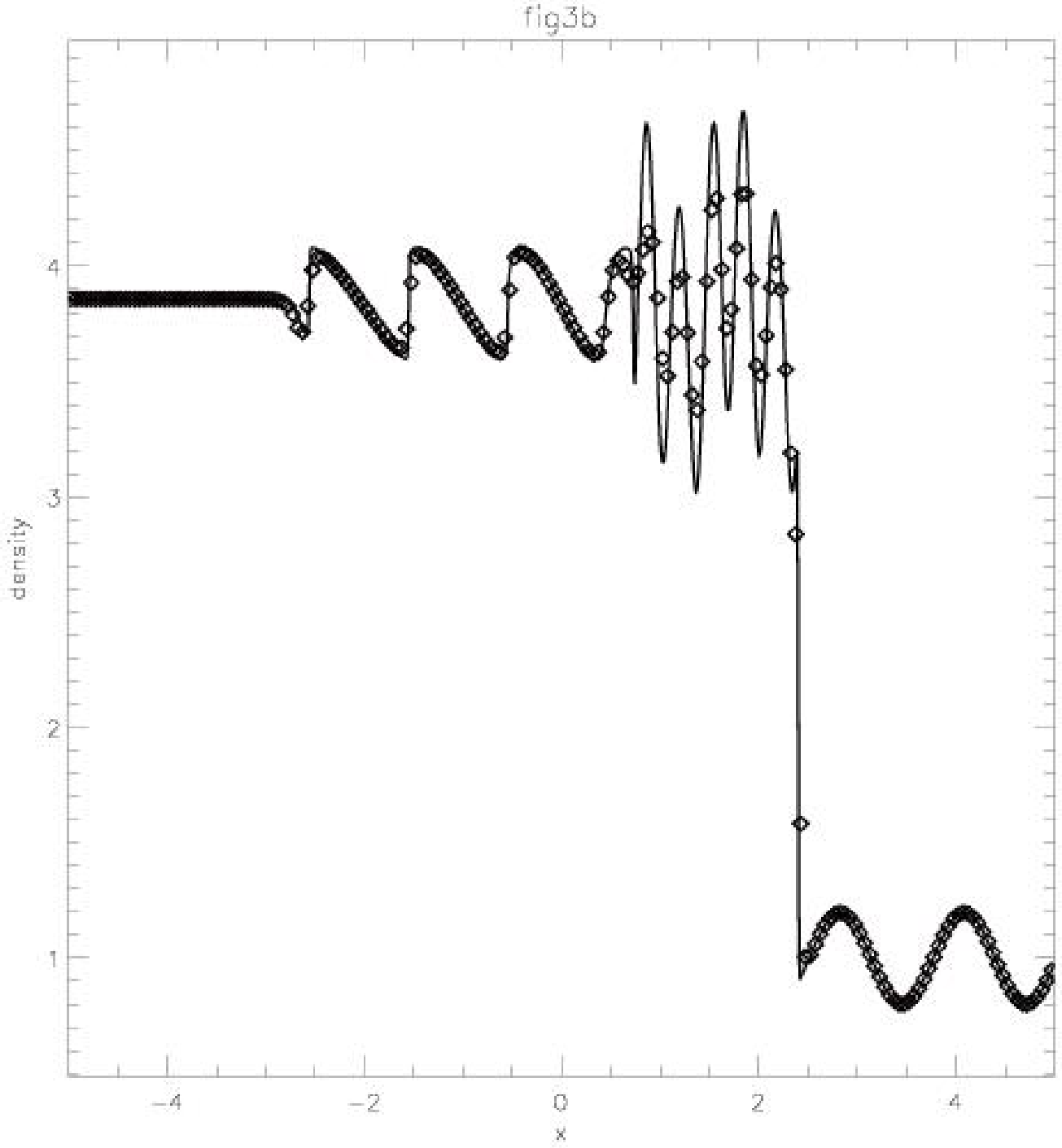

Fig. 5c

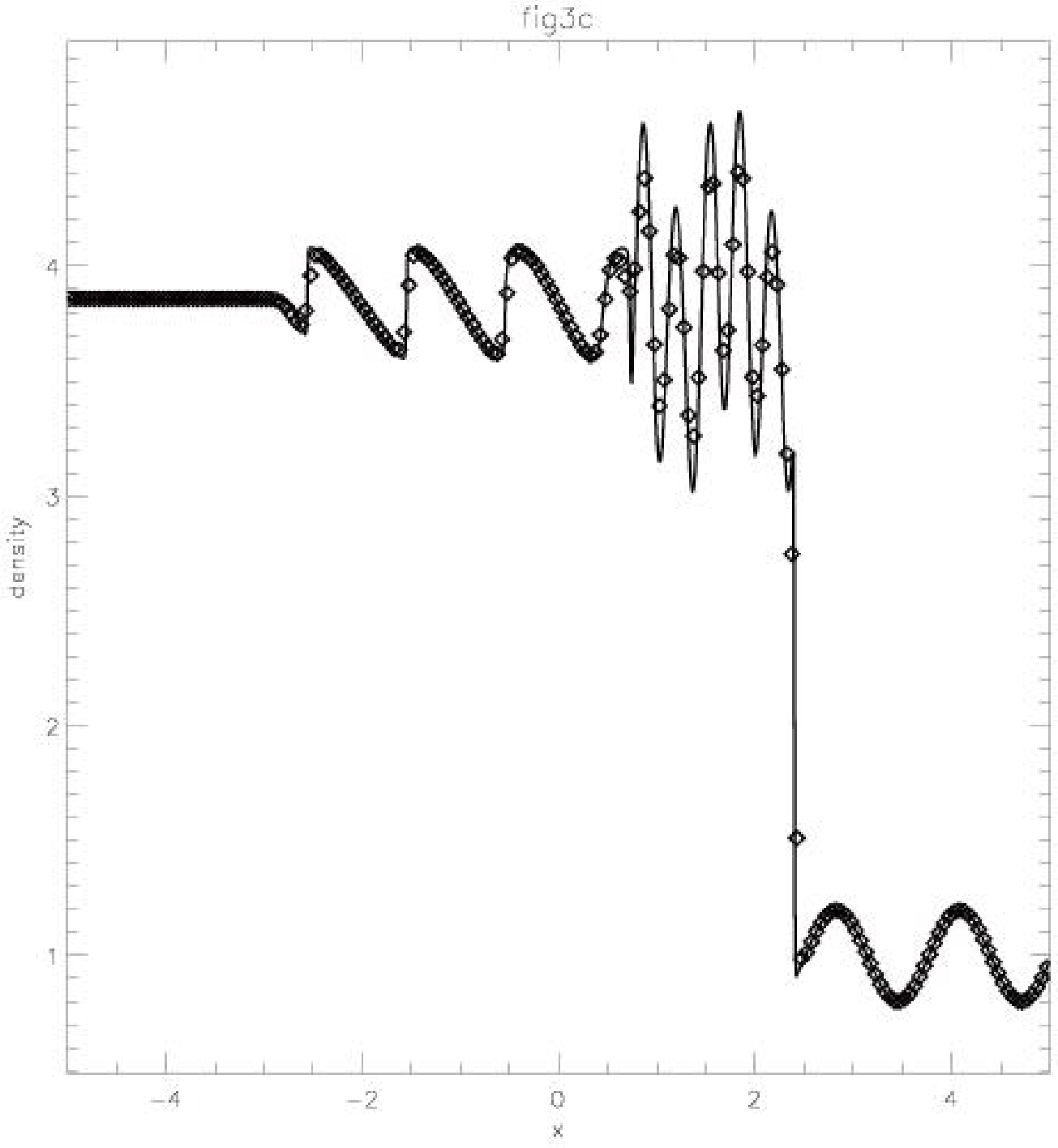

Fig. 5d

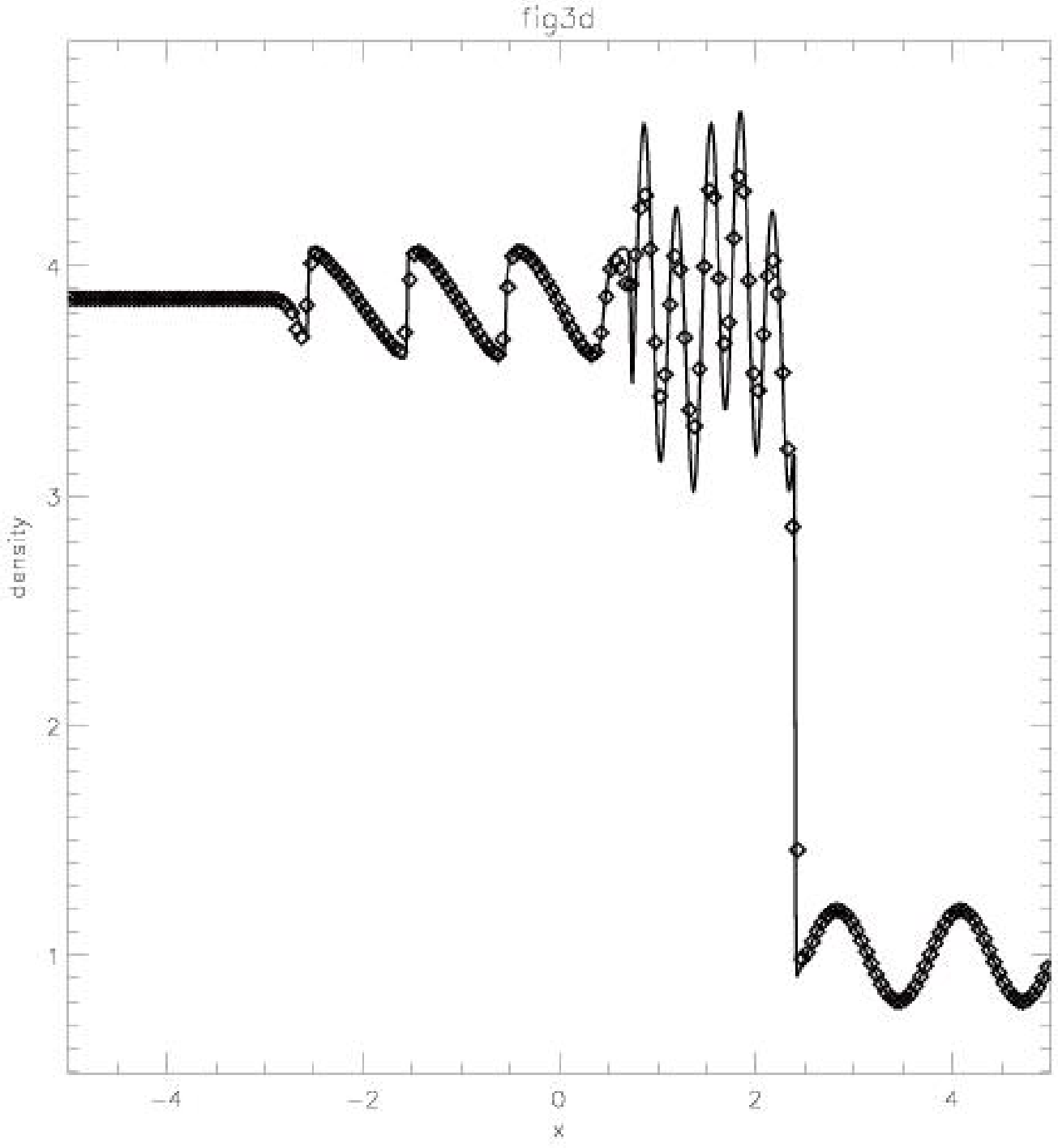

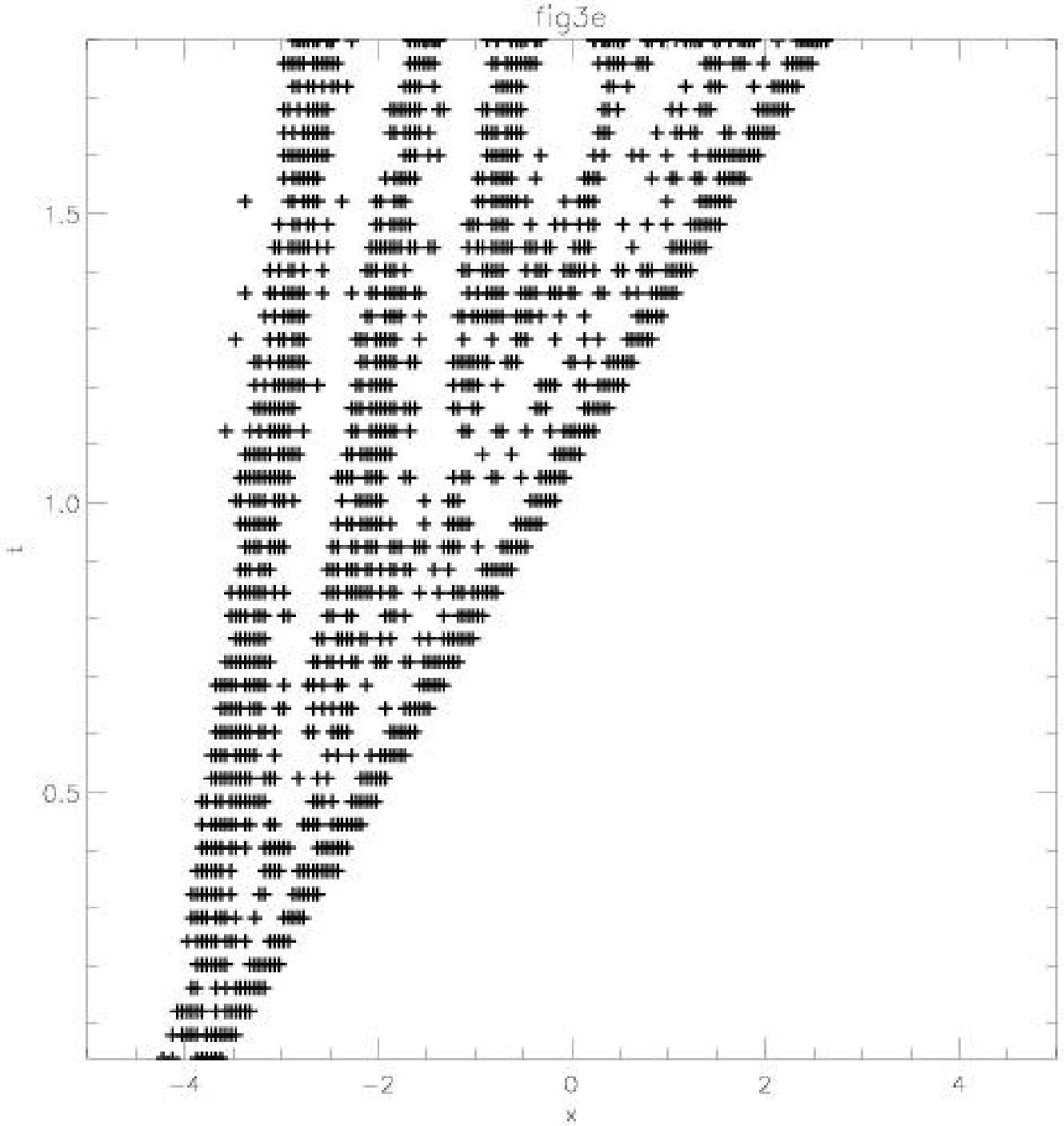

Fig. 5e

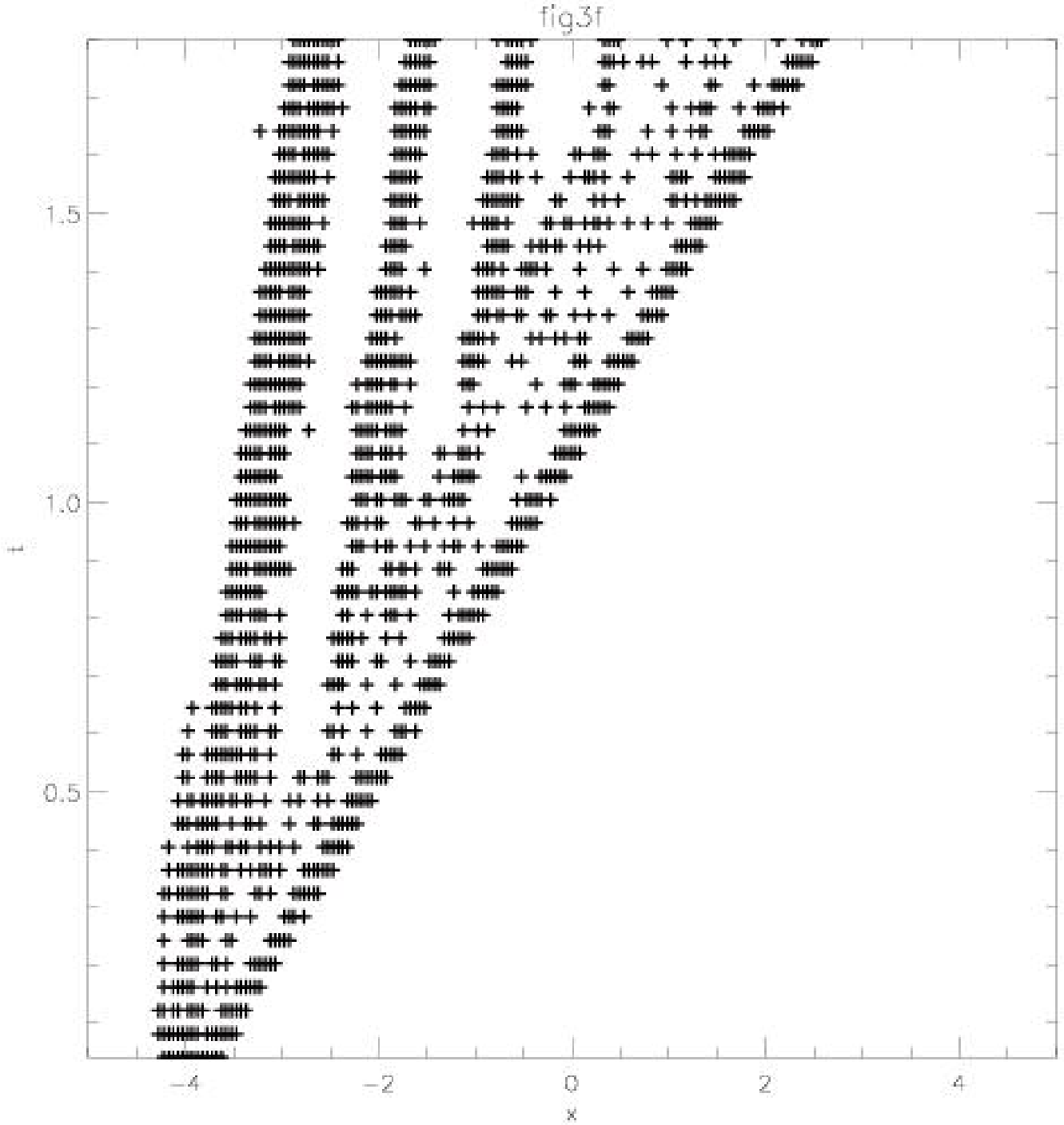

Fig. 5f

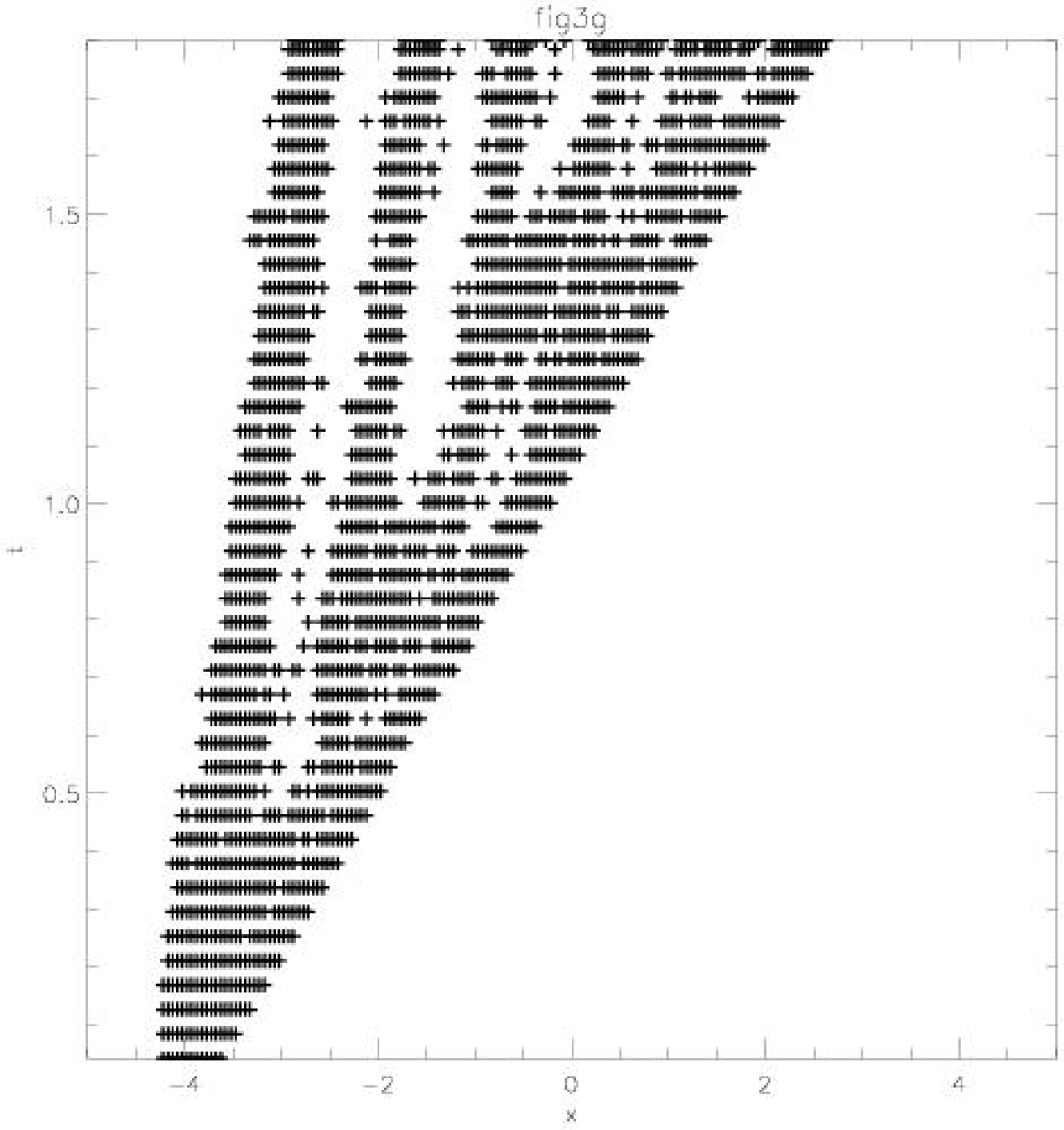

Fig. 5g

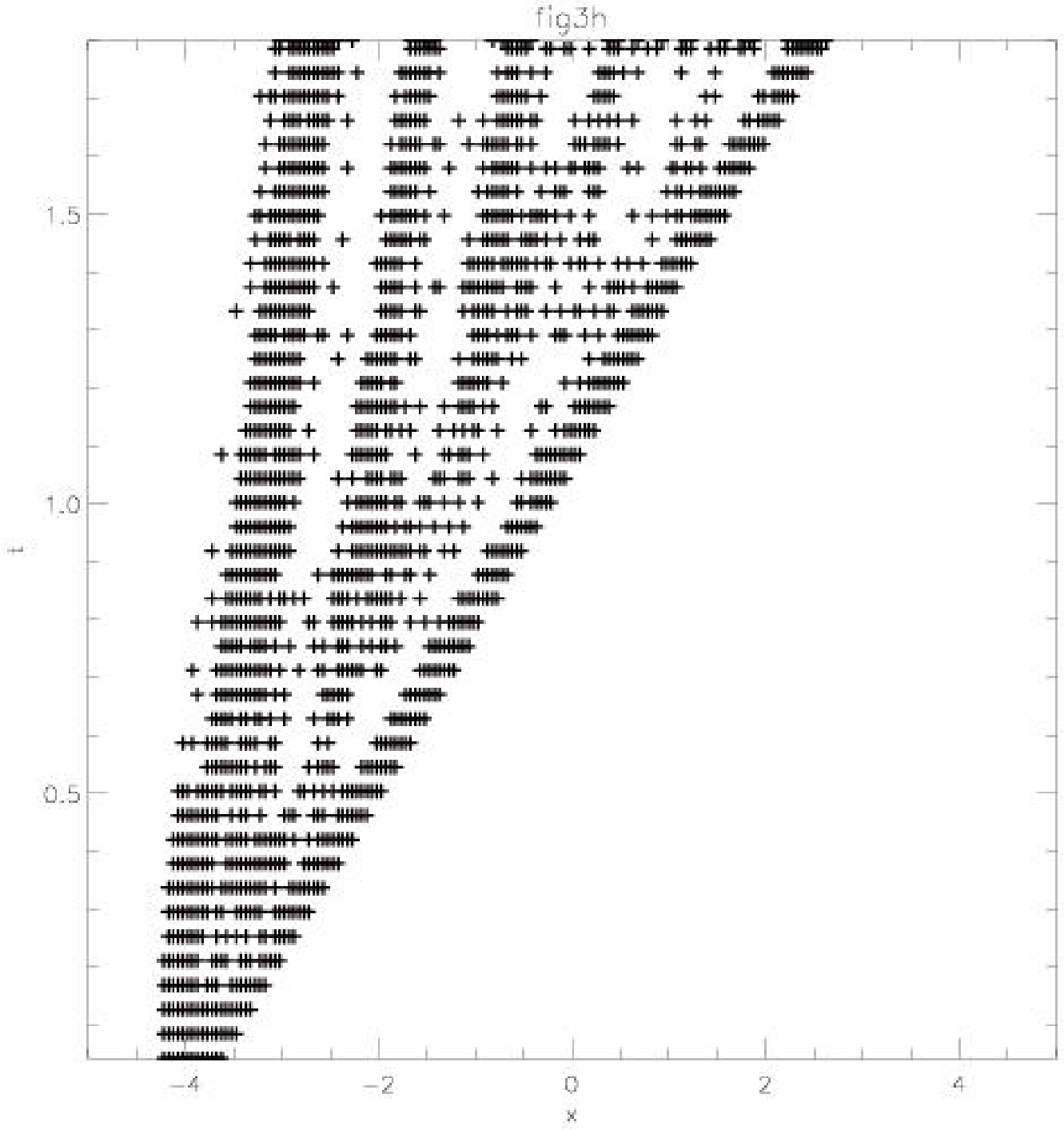

Fig. 5h

Fig. 6a

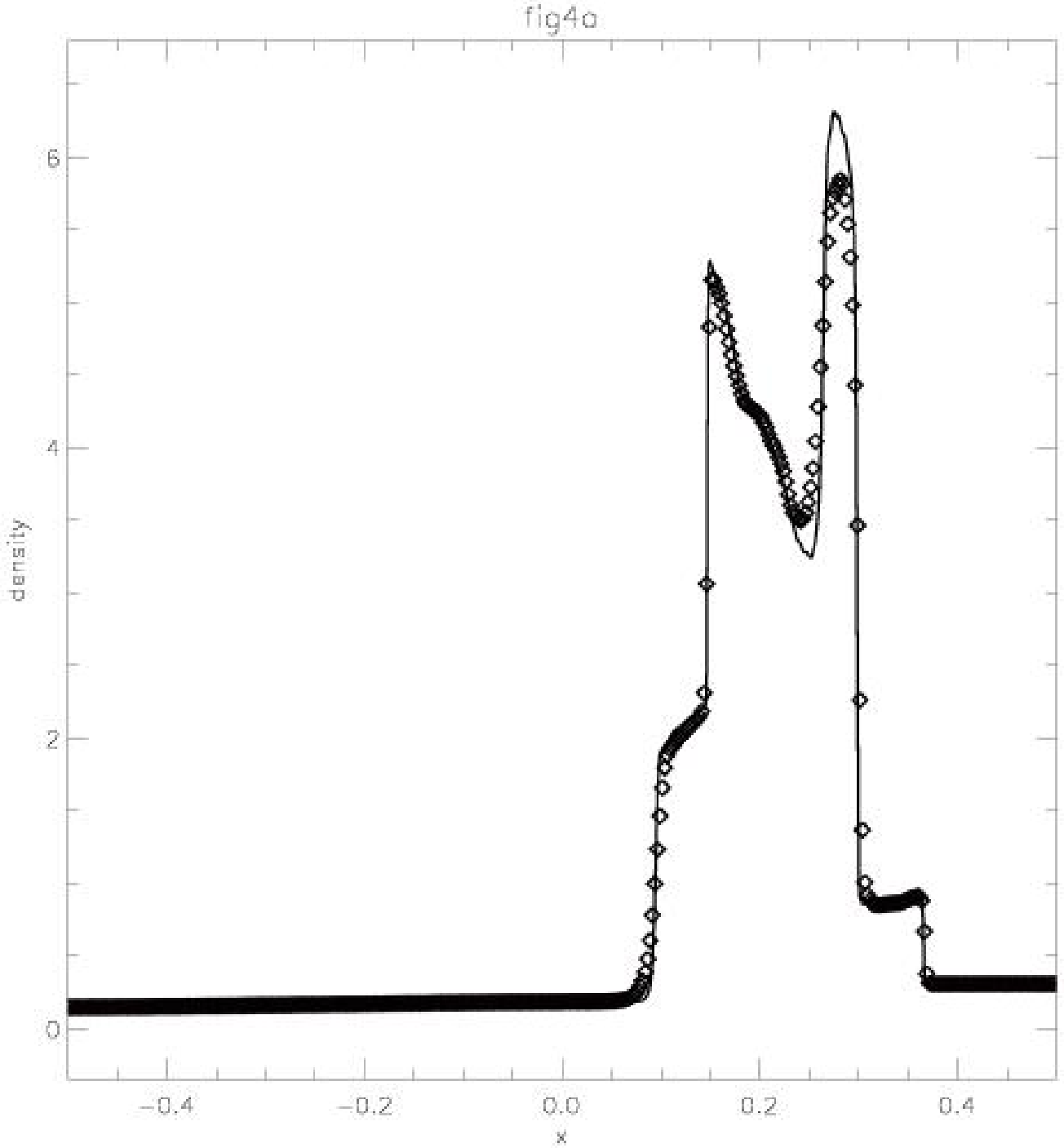

Fig. 6b

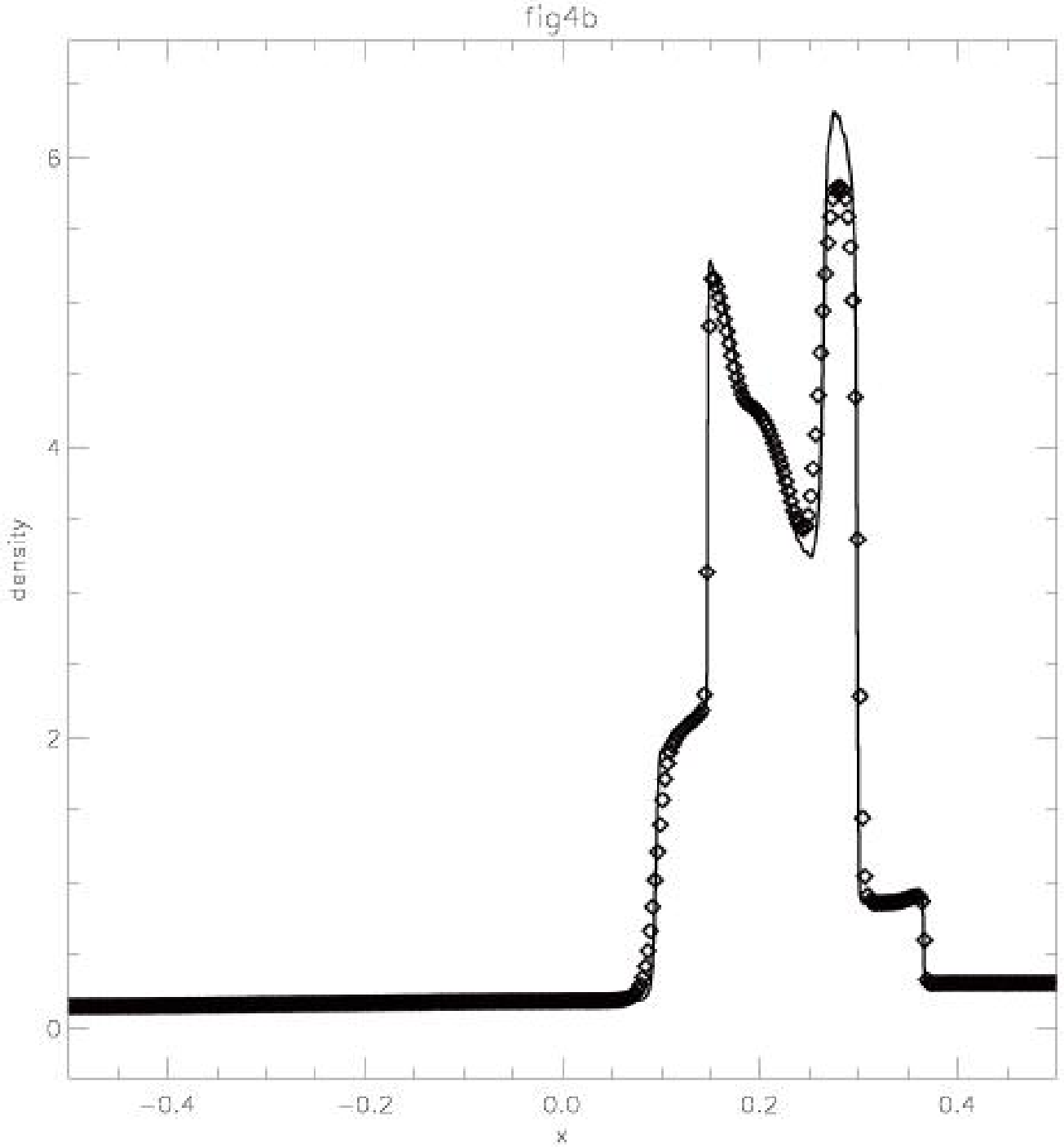

Fig. 6c

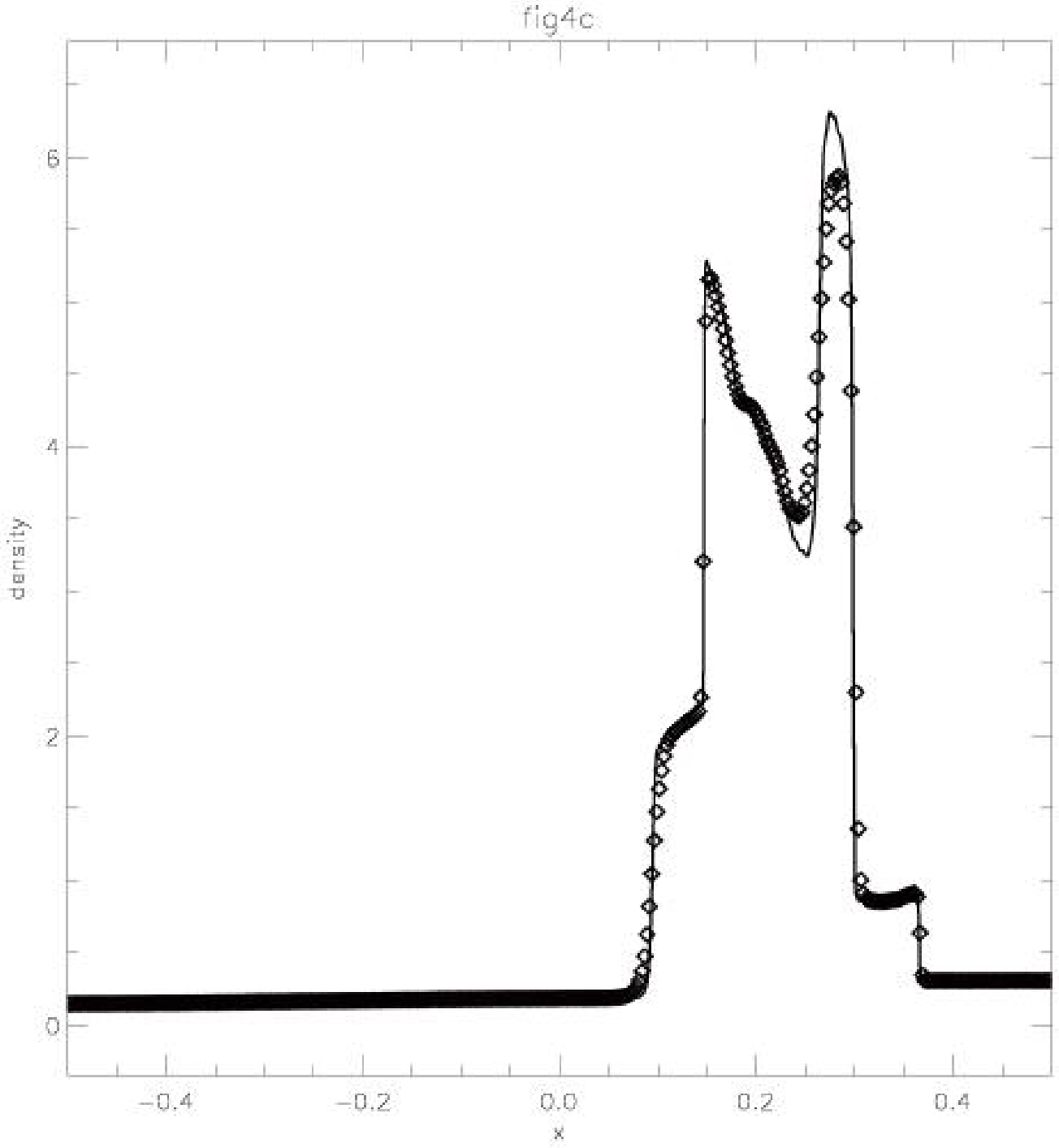

Fig. 6d

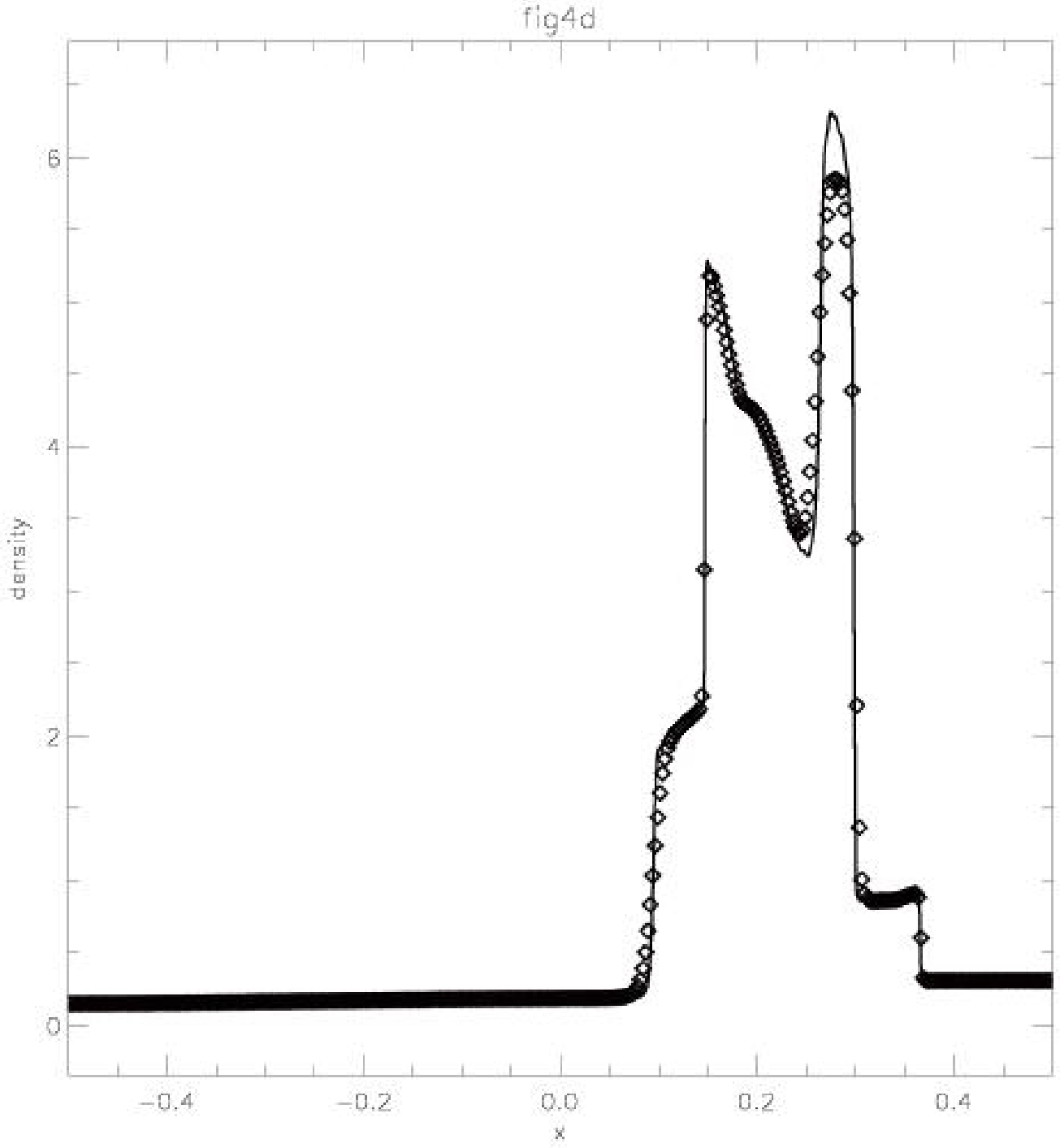

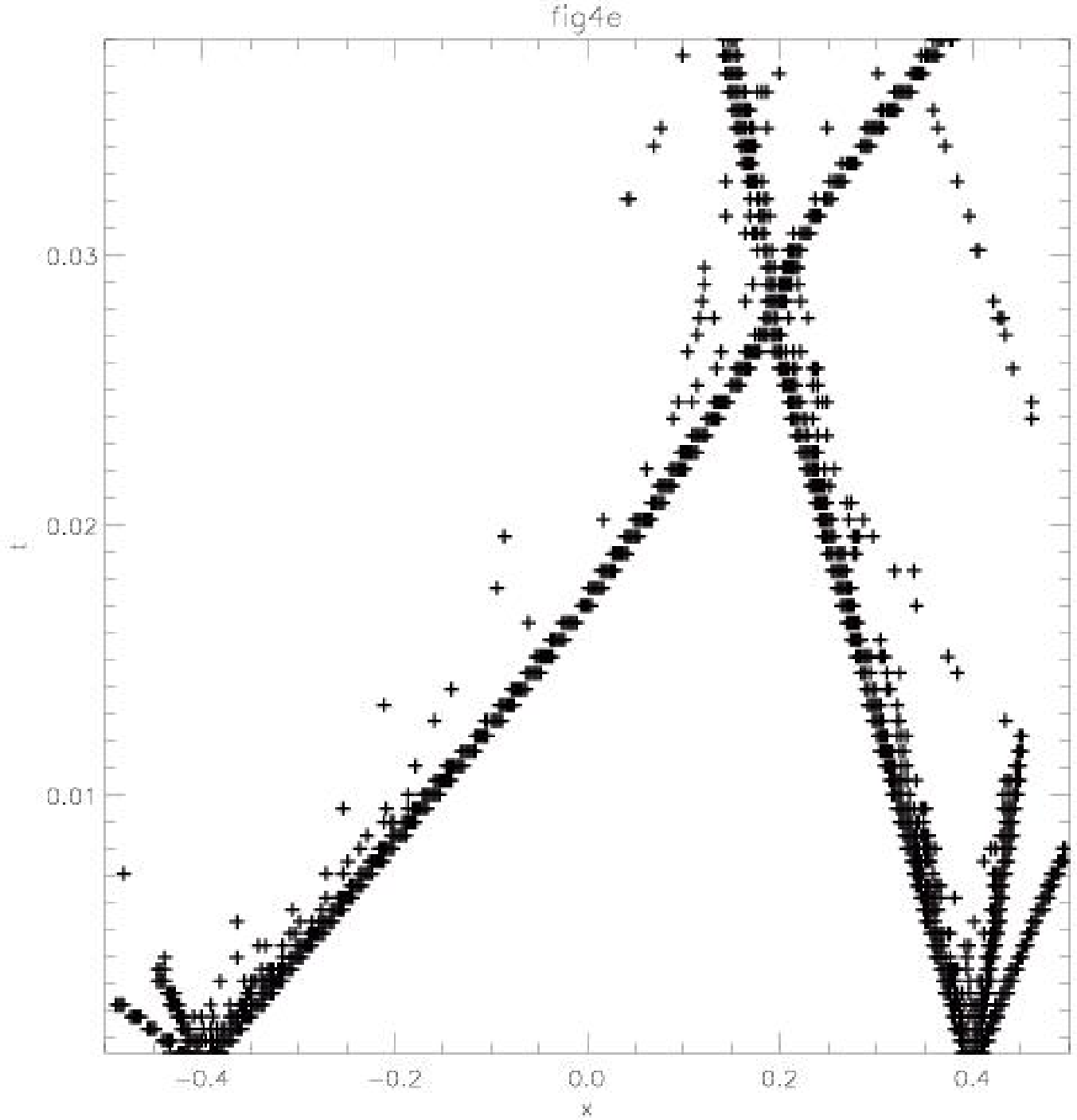

Fig. 6e

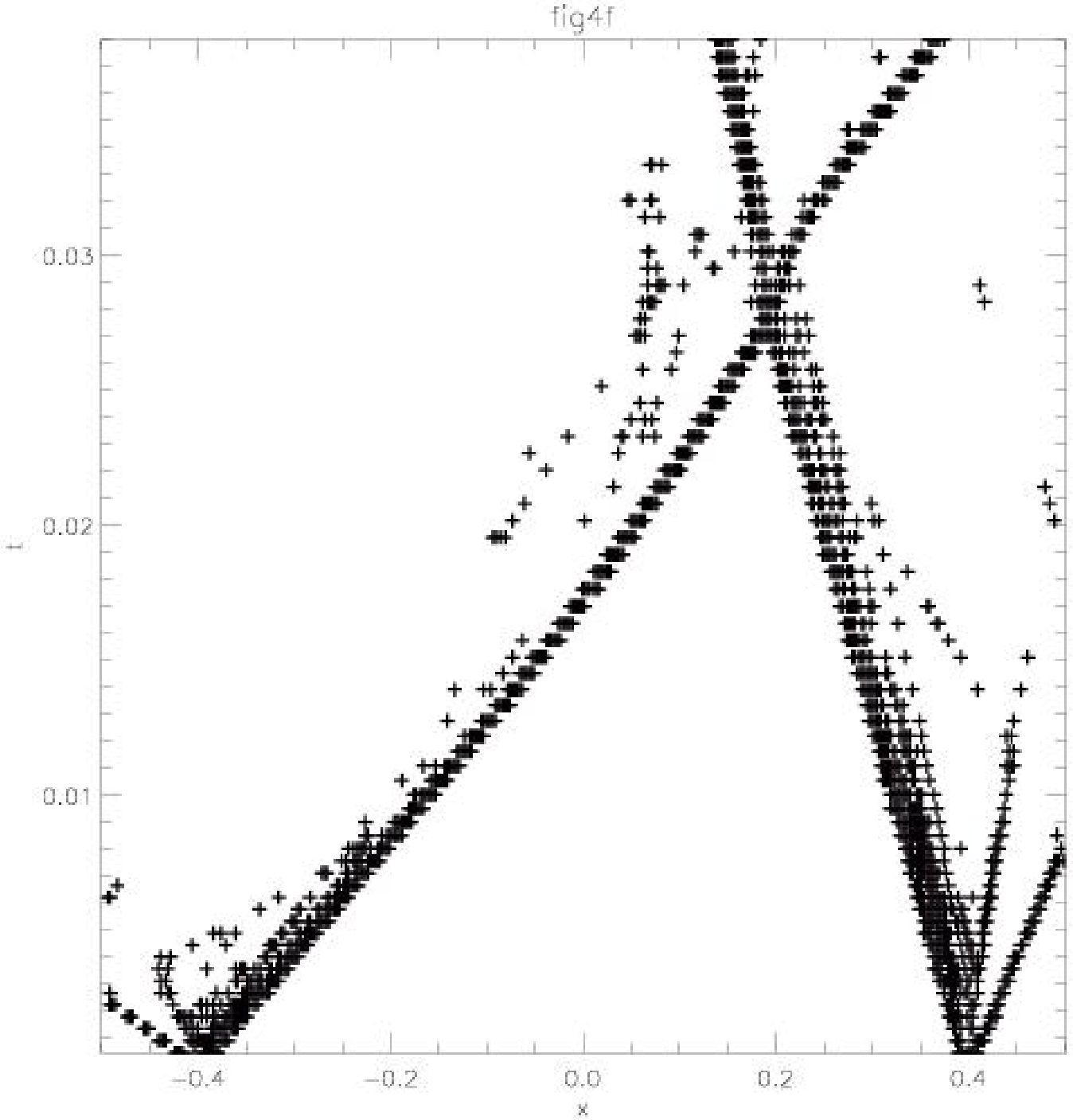

Fig. 6f

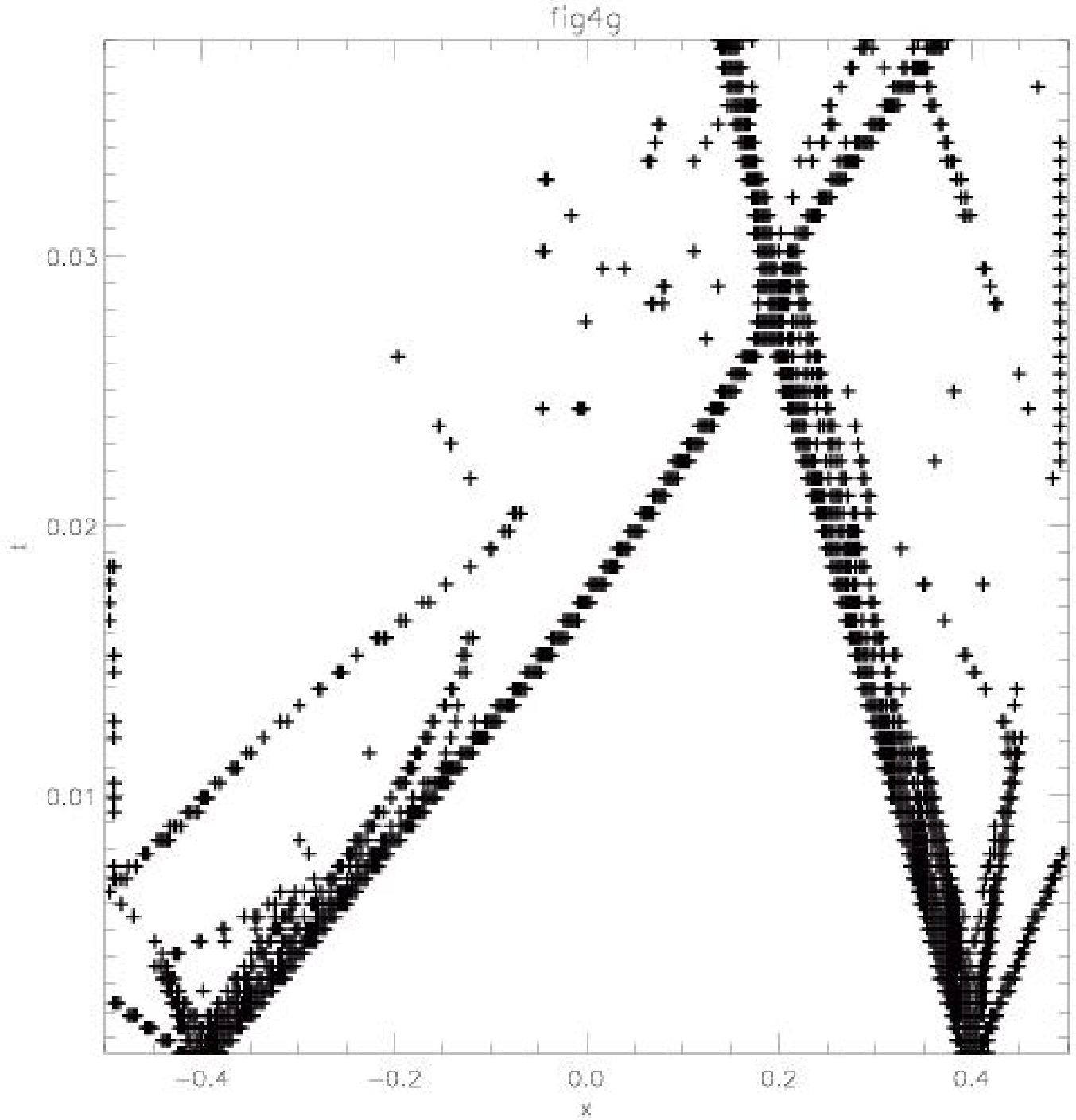

Fig. 6g

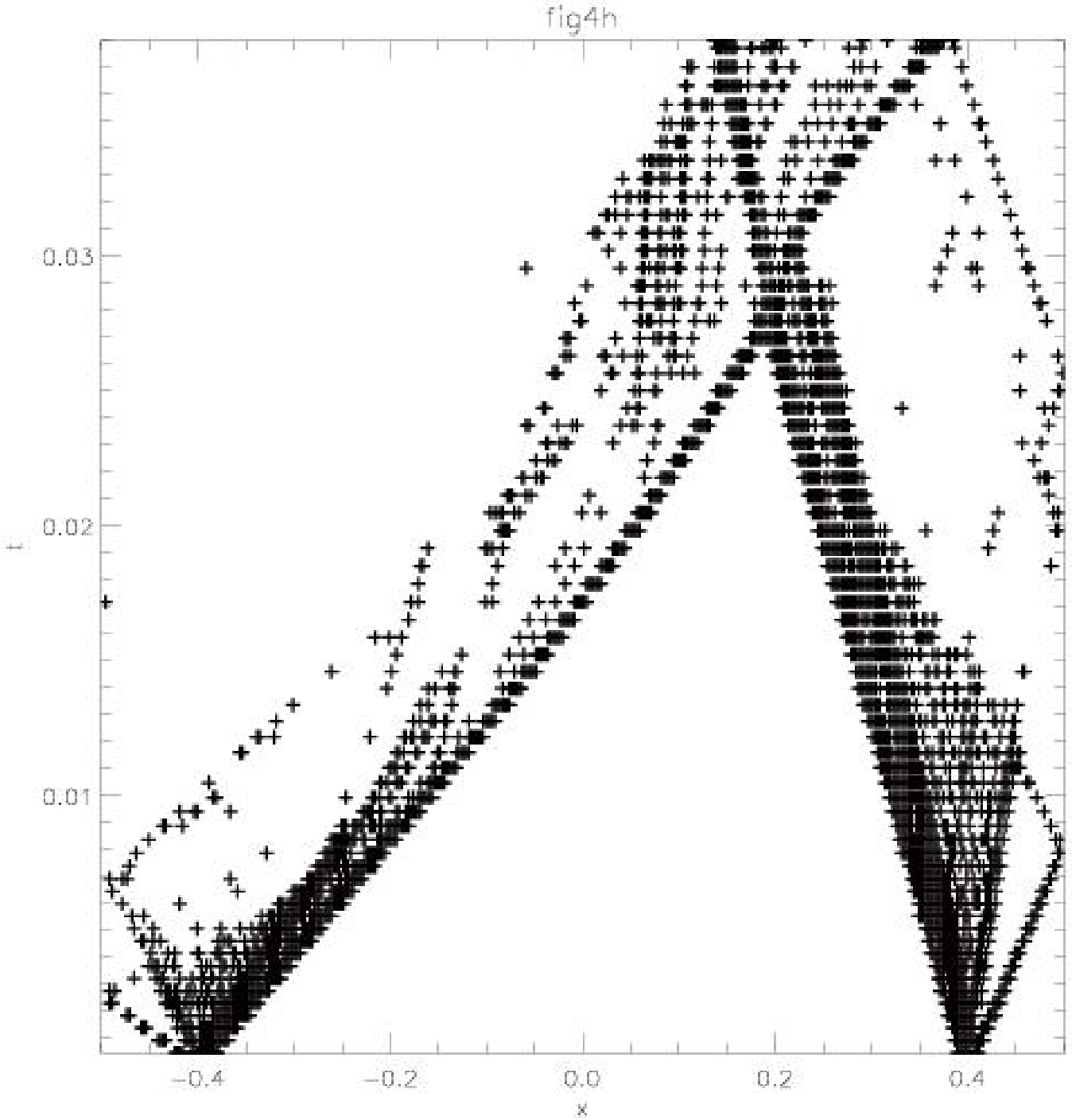

Fig. 6h

# Fig. 7

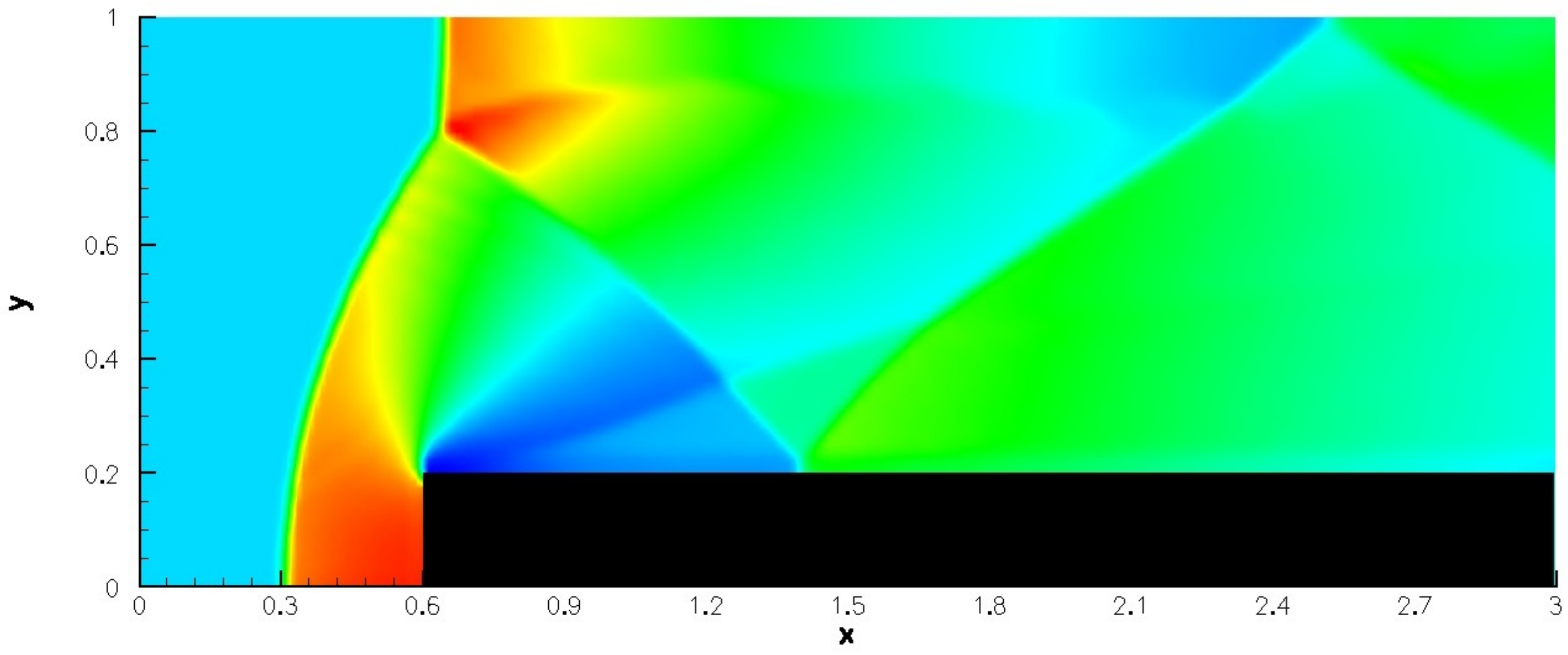